\documentclass{article}

\usepackage{url}
\usepackage{graphicx}
\DeclareGraphicsExtensions{.jpg,.png,.pdf}
\usepackage{amssymb,amsmath}
\DeclareMathAlphabet\mathbfcal{OMS}{cmsy}{b}{n}
\usepackage[normalem]{ulem}

\begin{document}


\title{When topological derivatives met regularized Gauss-Newton
iterations in holographic 3D imaging}

\author{A. Carpio \footnote{Universidad Complutense de Madrid, Madrid 
28040, Spain, and Courant Institute, New York University, NY 10012, USA}, 
T.G. Dimiduk \footnote{Tesla, Palo Alto, CA 94304, USA},
F. Le Lou\"er \footnote{Universit\'e de Technologie de Compi\`egne, 60203 Compi\`egne, France},
M.L.  Rap\'un \footnote{Universidad Polit\'ecnica de Madrid, Madrid 28040, 
Spain}}

\maketitle

{\bf Abstract.}
We propose an automatic algorithm for 3D inverse electromagnetic scattering based on the combination of topological derivatives and regularized Gauss-Newton iterations. The algorithm is adapted to decoding digital holograms. A hologram is a two-dimensional light interference pattern that encodes information about three-dimensional shapes and their optical properties.
The formation of the hologram is modeled using Maxwell theory for light scattering by particles. 
We then seek shapes optimizing error functionals which measure the deviation from the recorded holograms. Their topological derivatives provide initial guesses of the objects. Next, we correct these predictions by regularized Gauss-Newton techniques. In contrast to standard Gauss-Newton methods, in our implementation the number of objects can be automatically updated during the iterative procedure by new topological derivative computations. We show that the combined use of topological derivative based optimization and iteratively regularized Gauss-Newton methods produces fast and accurate descriptions of the geometry of objects formed by multiple components with nanoscale resolution, even for a small number of detectors and non convex components aligned in the incidence direction. The method could be applied in general imaging set-ups involving other waves (microwave imaging, elastography...) provided closed-form expressions for the topological 
and Fr\'echet derivatives are determined.

\section{Introduction}
\label{sec:intro}

\begin{figure}[!t]
\centering
(a) \hskip 5.5cm (b) \\
\includegraphics[width=5cm]{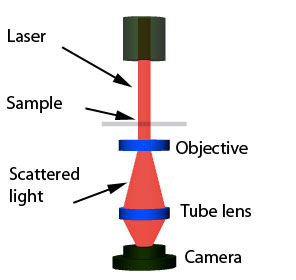} \hskip 1cm
\includegraphics[width=5cm]{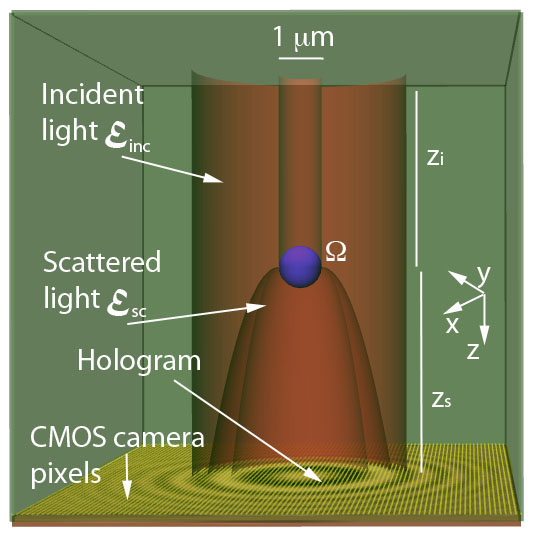}
\caption{\small (a) A typical optical set-up to record in-line holograms.
A collimated laser beam illuminates a sample. The resulting
hologram is recorded on a CMOS camera.
(b) Schematic representation of an in-line holography set-up
used in the algorithms and simulations presented in this paper.
The incident reference wave ${\mathbfcal E}_{\rm inc}$ interacts
with the sample $\Omega$ generating a scattered wave
${\mathbfcal E}_{\rm sc}$.
The interference pattern created by both of them forms the
hologram ${\mathbfcal I}_{\rm meas}=
|{\mathbfcal E}_{\rm inc}+{\mathbfcal E}_{\rm sc}|^2$.
We have superimposed on the CMOS screen the
hologram created by a sphere placed at a distance
$z_s=5\, \mu$m ($1 \mu$m=$10^{-6}$ m) from that
screen, when a $660$ nm  ($1$ nm=$10^{-9}$ m)
laser is emitted.}
\label{fig1}
\end{figure}

Digital in-line holography is a promising tool for high speed three dimensional (3D) imaging of live cells and soft matter \cite{dimidukfitting,holocell}. It can achieve high temporal (microseconds) and spatial (nanometers) resolution while avoiding the usage of toxic stains and fluorescent markers.
Holograms are two-dimensional (2D) light interference patterns that contain information about the 3D positions and optical properties of an object or set of  objects \cite{holo}.
Figure \ref{fig1} illustrates the formation of an in-line hologram from the interference of the light field scattered from a sample and the undiffracted beam \cite{lee}.
In tracking experiments, one expects to infer the position of objects
as a function of time analyzing a time-series of holograms.
Instead, in characterization experiments, one aims to extract the size, shape and refractive index of the particles under study.
Traditional optical reconstructions shine light back through the hologram to produce a 3D image, though this process may introduce a number of artifacts for sizes comparable to the employed light wavelength \cite{pu}.
In contrast, digital holography aims to achieve numerical reconstructions, facing the inherent difficulty of computationally recovering 3D  geometries from the 2D holograms they generate. From the mathematical point of view, finding objects and their optical properties from holograms is an ill-posed inverse scattering problem.

Recent work has relied on scattering theory to analyze
holograms. As demonstrated in \cite{ovryn1,ovryn2}, in-line holograms can be predicted combining Lorenz-Mie scattering theory with a model for the propagation and interference of the light fields in the microscope. Later, spherical colloidal particles were successfully tracked and characterized by fitting scattering models based on the Lorenz-Mie theory to holograms \cite{lee}. More general scattering approximations allow to treat non-spherical particles like rods as well as clusters of spherical particles \cite{dimidukfitting,dimidukdda}
fitting to the data a forward model, that is, a model that can evaluate a hologram based on a theory of scattering and propagation inside the microscope, using solvers such as discrete dipole approximations (DDA),  see for instance \cite{ddatheory,dimidukdda}.
These methods proceed by `least squares fitting': a few parameters
representing radius, orientation, position, refractive index and so on are adjusted iteratively by a Levenberg-Marquardt algorithm to minimize the error when comparing  the synthetic holograms generated by the approximate objects as predicted by the selected forward model and the true measured hologram. Alternative bayesian and machine learning approaches are discussed in \cite{bayesian,grier}.
Successful reconstructions typically require significant a priori knowledge about the objects being imaged, such as their approximate positions in the field of view and a simple parametrization (sphere, cylinder...), for example.

To track and characterize objects without a priori knowledge (other than the optical properties of the ambient medium and the incident light) we may formulate more comprehensive optimization problems. The idea is to optimize the error functionals  with respect to arbitrary shapes and arbitrary functions representing the imaged objects and their optical properties. When illuminating  simple object configurations with time harmonic and polarized light, the problem somewhat simplifies, since we can use scalar Helmholtz approximations  to simulate the polarized component of light \cite{siam2016,siam2018}. Within this simplified framework, Ref. \cite{siam2018} succeeds in producing first guesses of objects and of their optical properties from the holograms they generate by combining topological derivative and gradient based optimization  procedures, provided the size of the objects is of the same order or smaller than the employed light wavelength. Helmholtz equations for the forward problems are solved in \cite{siam2018}
by coupled boundary element (BEM)/finite element (FEM) methods \cite{virginiafem,nedelec}, whereas shapes are constructed by  means of blobby molecule coverings and signed distance functions as in \cite{siam2016}, without imposing any specific parametrization.

Instead of relying on particular scattering theories, here we implement topological derivative based optimization using the 3D vector Maxwell  equations, which requires new closed-form formulas  adapted to the holographic setting. 
We obtain them extending ideas developed in \cite{siam2018} to deal with holograms $I=|{\mathbf E}|^2$ under  scalar approximations for polarized light to derivations for error functionals using complex measured data $\bf E$ and vector Maxwell constraints \cite{masmoudi,ip2008,lelouerrapun,lelouerrapun2}.
As it happens in other imaging problems (acoustics, elastography...), topological derivatives allow us to generate first guesses of holographied objects and to correct the number of boundaries by creating, merging or destroying components.  However,  topological derivative based iterations may get stuck without providing a precise description of the shapes unless data distributed over a wide enough angle for a wide enough range of frequencies or incoming incident directions are available \cite{park,ln2008,ip2008,guzina}. We find that this is the case in holography. Due to the way microscopes are built, only one incident direction and one frequency are usually available and data are measured on a limited screen behind the object \cite{siam2016,lee,dimidukdda}.  Moreover, our tests show that the results do not really improve switching to optimization procedures that deform the approximate object contours along vector fields, such as shape derivative based deformations or level set techniques \cite{caubet,dorn1,osher,masmoudi0}.
In this paper, we overcome this difficulty by combining topological derivative based optimization with iteratively regularized Gauss-Newton methods, see the videos in the supplemental material.  For the true objects we will only require ${\cal C}^2$ regularity. However, those objects will be approximated by star-shaped components during the iterative procedure. Star-shaped objects are defined by rays emerging from one fixed position $\mathbf c$. Their boundary is located at distances of that point which vary with  the angles. Therefore, they can be described using a spherical coordinate  system, as it happens for  ellipsoids and smoothed polyhedra, for instance. A general object formed by several star-shaped components is represented as
$\Omega=\cup_{\ell=1}^{\cal L} \Omega_{\ell}, $
$\partial \Omega_{\ell}= \mathbf c_{\ell} + \mathbf r_\ell (\mathbb S^2),$
where $\mathbf r_\ell: \mathbb S^2 \rightarrow \mathbb R^3$ is a combination of spherical harmonics. Our method identifies automatically the number of defects ${\cal L}$ (i.e., we do not assume ${\cal L}$ to be known), and provides the centers $\mathbf c_{\ell}$ and the radii functions $\mathbf r_\ell$ defining our approximation of the true configuration, which does not need to consist in star-shaped objects. Furthermore, the method could be extended to a broader class of parameterizations dropping the star-shaped constraint, see \cite{Hohage3}.  To simplify, we  assume that the optical properties are known and constant inside each component.

The starting point of our algorithm are star-shaped parametrizations fitted to initial guesses of the holographied objects provided by  topological derivative optimization of the error functional comparing the synthetic hologram generated by any object with the true hologram. Next, we linearize the synthetic hologram contribution in this quadratic error functional about the current parametrization to obtain a nonlinear least squares problem for the next star-shaped parametrization, which is regularized including a Tikhonov  quadratic term. Gauss-Newton iterations are a very powerful tool for finding the best descent direction for minimizing nonlinear least squares  problems  \cite{Bakushinskii,Hohage,Hohage2,Hohage3}. There
is work on coupling topological gradients and Gauss-Newton iterations \cite{FehrenbachMasmoudi} to solve a 2D inverse problem in elasticity, applied to parameter reconstruction instead of shape reconstruction. That study does not consider the iteratively regularized Gauss-Newton method whose regularizing Tikhonov parameter is exponentially decreasing to ensure the convergence of the algorithm  \cite{Hohage,Hohage2} when the stoppping rule is given by Morozov's discrepancy principle,  as we do here. Moreover, the classical application of the iteratively regularized Gauss-Newton method  requires an initial guess formed by the true number of components, while our hybrid method does not: it automatically updates this number by topological derivative computations. This is the key starting point for constructing automatic/smart inverse algorithms.  The hybrid inverse algorithm we are proposing  is therefore new, as it is its application to holography.

Many other approaches have been developed for inverse scattering problems involving time harmonic electromagnetic waves of wavelengths larger than those of light. We may mention qualitative techniques such as  linear sampling, factorization and MUSIC methods \cite{coltonsampling}  and a variety of methods tracking permittivity variations \cite{litman}.
In such cases, the complex amplitude (modulus and phase) of the scattered field is measured.  When working with light, phases are not measurable due to its high frequency, only intensities in the form of interference patterns such as holograms are measurable. Ref. \cite{siam2016} showed that in case the complex amplitude was known the performance of topological methods might improve. Ref. \cite{siam2018} proposes a procedure based on gradient optimization and gaussian filtering to numerically approximate the modulus and the phase from the measured hologram.
However, the applicability of methods exploiting this numerical approximation of the complex field is currently limited by the numerical error and by the constrained use of a single incident wave.

The paper is organized as follows. Section \ref{sec:inverse} recalls the  formulation of the inverse holography problem. Section \ref{sec:priors} adapts topological techniques to generate initial approximations of the holographied objects in the absence of  a priori information, other than the emitted light and the refractive index of the ambient medium. Section \ref{sec:gauss} formulates an iteratively regularized  Gauss-Newton method employing star-shaped parametrizations. This allows to use closed-form expressions of the Fr\'echet derivatives with respect to the parametrization in the linearized operators. It also permits the use of fast spectral solvers for the Maxwell equations.
Section \ref{sec:algorithm} proposes a hybrid algorithm combining topological derivative based iterations to create, merge, or destroy objects with  regularized Gauss-Newton iterations to sharpen their shape. We illustrate numerically the performance of this scheme for configurations containing multiple and non necessarily convex  neither star-shaped components. 
The numerical solution of the auxiliary Maxwell systems and details on the computation of Fr\'echet derivatives and their adjoints are discussed in Section  \ref{sec:numerical}. Finally, Section \ref{sec:conclusions} summarizes our conclusions and comments on possible further developments. A few appendices contain complementary information and technical details. Appendix
\ref{sec:explicitwhole} briefly compares with scalar approximations based on Helmholtz equations. Appendix
\ref{sec:derivatives} establishes formulas for the Fr\'echet, shape and topological derivatives relevant  in holography. Appendix
\ref{sec:sphere} collects some background on the selected spherical harmonics and related Mie expansions for ease of the reader.

\section{The inverse holography problem}
\label{sec:inverse}

The general form of an inverse scattering problem is the following.
An incident wave ${\mathbfcal E}_{\rm inc}$ (electromagnetic, elastic, acoustic, thermal...) interacts with objects contained in a medium. The resulting wave field ${\mathbfcal E}$ is somehow measured at a set of detectors.
Knowing the emitted waves, the data measured at the detectors, and the properties of the background medium, we aim to reconstruct the geometry of the objects and their relevant material properties with regard to the employed wave (permittivities and permeabilities, elastic constants,  thermal diffusivities...).
In other words, we seek objects $\Omega$ with material parameters
$\kappa$ such that, when the emitted waves ${\mathbfcal E}_{\rm inc}$ interact with such objects, the resulting wave field ${\mathbfcal E}_{\Omega,\kappa}$ generates at the detectors data ${\cal D}_{\Omega,\kappa}$ which agree with the measured data ${\cal D}_{\rm meas}.$
In practice, one never knows the exact value of the true field at the
detectors due to errors and noise.
Therefore, it is often enough to find objects $\Omega$ with material parameters $\kappa$ such that the error $J(\Omega,\kappa)$ when comparing ${\cal D}_{\Omega,\kappa}$ and ${\cal D}_{\rm meas}$  at the detectors is as small as possible, that is, we seek objects $\Omega \subset {\mathbb R}^3$ and  parameter functions  $\kappa: \Omega \rightarrow {\mathbb R}$ minimizing the selected cost functional $J(\Omega,\kappa)$.

In holography,  the emitted waves are light beams,  the measured data take the
form of a hologram ${\mathbfcal I}_{\rm meas}$  recorded  at a screen behind the object, and the material parameters of interest  are usually its permittivity or its refractive index. 
The resulting wave field is governed by the linear time dependent Maxwell equations.
When the emitted light beams are time harmonic, that is,
${\mathbfcal E}_{\rm inc}(\mathbf x,t)= {\rm Re}[e^{-\imath \omega t}
{\mathbf E}_{\rm inc}(\mathbf x)]$, the resulting wave fields
also happen to be time harmonic
${\mathbfcal E}_{\Omega,\kappa}(\mathbf x,t) = {\rm Re}[ e^{-\imath \omega t} {\mathbf E}_{\Omega,\kappa}(\mathbf x)]$
and the complex amplitude $\mathbf E_{\Omega,\kappa}(\mathbf x)$ satisfies a stationary version of the time dependent Maxwell equations, the so-called forward problem:
\begin{eqnarray}
\begin{array}{rcl}
\mathbf{curl} \,  ({1\over \mu_e}\mathbf{curl} \,  \mathbf E)
- {\kappa_e^2\over \mu_e} \mathbf E =0  & \mbox{in} &
\mathbb R^3\setminus\overline{\Omega},  \quad   \\ [1ex]
\mathbf{curl} \,  ({1\over \mu_i} \mathbf{curl} \,  \mathbf E)  - {\kappa_i^2\over
\mu_i} \mathbf E =0 & \mbox{in} &  \Omega,  \quad  \\ [1ex]
 \hat{\mathbf n}  \times \mathbf E^- = \hat{\mathbf n} \times \mathbf E^+,
& \mbox{on} &  \partial \Omega,  \quad \quad \\ [1ex]
{1\over \mu_i}  \hat{\mathbf n} \times \mathbf{curl} \,  \mathbf E^- =
{1 \over \mu_e}  \hat{\mathbf n} \times \mathbf{curl} \,  \mathbf E^+,
& \mbox{on} & \partial \Omega, \quad  \\ [1ex]
{\rm lim}_{|\mathbf x| \rightarrow \infty} |\mathbf x|  \big|
\mathbf{curl} \,  (\mathbf E - \mathbf E_{\rm inc}) \times {\mathbf x \over
|\mathbf x|} -\imath \kappa_e (\mathbf E - \mathbf E_{\rm inc}) \big| =0, & &
\end{array} \label{forwarddim}
\end{eqnarray}
where $\mu_i,\varepsilon_i,\kappa_i$ and $\mu_e,\varepsilon_e,\kappa_e$
are the permeabilities, permittivities and wavenumbers
$\kappa^2 = \omega^2 \varepsilon \mu$ of the objects and the ambient
medium, respectively \cite{borhen,coltonkress,kirsch,monk,nedelec}.
In biological media, $\mu_i \sim \mu_e \sim \mu_0$,
$\mu_0$ being the vacuum permeability \cite{lin}.
The signs $+$ and $-$ denote the values from outside and inside 
$\Omega$, respectively.
The vector $ \hat{\mathbf n}$ represents the outer unit normal vector.
We have imposed transmission conditions at the interface $\partial \Omega$
between the objects and the ambient medium, together with the
Silver-M\"uller radiation condition at infinity.  We will consider
incident plane waves polarized in a direction $\hat {\mathbf p}$ orthogonal to the
direction of propagation $\hat {\mathbf d}$,  that is, with 
$\mathbf E_{\rm inc}(\mathbf x)
= E_0 \hat{\mathbf p} e^{\imath \kappa_e \hat{\mathbf d} \cdot \mathbf x}$
where $E_0$ represents the units and magnitude of the incident field. As sketched in Fig. \ref{fig1}(b), in practice the direction of propagation is the $z$ axis,
orthogonal to the hologram  recording screen.

It is well known that system (\ref{forwarddim}) has a unique solution for any real positive $\kappa_e>0$ \cite{nedelec}. Elliptic regularity implies that this solution belongs to  the Sobolev
space $H^{2,0}(\Omega')=\{{\mathbf E}\in H^2(\Omega'),\; {\rm div }\,{\mathbf E}=0\}$ for any smooth bounded domain $\Omega' \subset \mathbb R^3 \setminus \overline{\Omega}$ \cite{gilbart,grisvard}.  Sobolev's embeddings ensure then continuity in
$\Omega'$. 

The solution of system (\ref{forwarddim}) is the forward field ${\mathbf E}_{\Omega,\kappa_i}$.
The total   field ${\mathbf E}_{\Omega,\kappa_i}$ is the sum of the scattered field ${\mathbf E}_{\rm sc,\Omega,\kappa_i}$
and the incident field ${\mathbf E}_{\rm inc}$ outside the objects, but becomes the transmitted  field  ${\mathbf E}_{\rm tr,\Omega,\kappa_i}$ inside. The hologram is $I_{\Omega,\kappa_i}= |{\mathbf E}_{\rm inc}+{\mathbf E}_{\rm sc,\Omega,\kappa_i}|^2 =|{\mathbf E}_{\Omega,\kappa_i}|^2 $  evaluated at detectors placed at  the screen.

Strategies to solve numerically (\ref{forwarddim}) are discussed in Section \ref{sec:numerical}.
For numerical purposes, it is convenient to nondimensionalize the problem. We set
$\mathbf{\tilde x}:=\mathbf{x}/L$,
$\tilde \Omega:=\Omega/L$,
$\tilde {\mathbf E}(\mathbf{\tilde x})=\mathbf E(\mathbf{x}/L)/E_0$,
$\tilde {I}(\mathbf{\tilde x})=  I(\mathbf{x}/L)/E_0^2$,
$\tilde {\mathbf E}_{inc}(\mathbf{\tilde x})= \mathbf E_{inc}(\mathbf{x}/L)/E_0,$ where $L$ is a reference length unit.
Choosing $L$ equal to a typical diameter of the object $\Omega$, for instance, we obtain the dimensionless forward system:
\begin{eqnarray}
\begin{array}{rcl}
\mathbf{curl}_{\tilde{\mathbf x}} \,  (\mathbf{curl}_{\tilde{\mathbf x}} \,
\tilde{\mathbf E})  - k_e^2 \tilde{\mathbf E} =0  & \mbox{in} & \mathbb R^3\setminus\overline{\tilde \Omega},  \quad \\
\mathbf{curl}_{\tilde{\mathbf x}} \,  (\mathbf{curl}_{\tilde{\mathbf x}} \,
\tilde{\mathbf E})  - k_i^2  \tilde{\mathbf E} =0 & \mbox{in} &
\tilde \Omega,  \quad  \\
\tilde{\mathbf n} \times \tilde{\mathbf E}^- =
\tilde{\mathbf n} \times \tilde{\mathbf E}^+,
& \mbox{on} &  \partial \tilde \Omega,  \quad \quad  \\
\beta \tilde {\mathbf n} \times \mathbf{curl}_{\tilde{\mathbf x}} \,
\tilde{\mathbf E}^- =
\tilde {\mathbf n} \times \mathbf{curl}_{\tilde{\mathbf x}} \,
\tilde{\mathbf E}^+,
& \mbox{on} & \partial \tilde \Omega, \quad  \\
{\rm lim}_{|\tilde{\mathbf x}| \rightarrow \infty} |\tilde{\mathbf x}|  \big|
\mathbf{curl}_{\tilde{\mathbf x}} \,  (\tilde{\mathbf E} -
\tilde{\mathbf E}_{\rm inc}) \times {\tilde{\mathbf x} \over |\tilde{\mathbf x}|}
-\imath k_e (\tilde{\mathbf E} - \tilde{\mathbf E}_{\rm inc}) \big| =0, & &
\end{array} \label{forwardadim}
\end{eqnarray}
$k_i= \kappa_i L$ and $k_e= \kappa_e L$ being the `dimensionless wavenumbers' (size parameters in the terminology of \cite{borhen}) inside and outside the object, respectively. This assumes that the permeabilities are constant in the ambient medium and in the connected components of $\Omega$, so that they can be scaled out. We will also assume $\beta={\mu_e \over \mu_i} \sim 1.$
The standard refractive indexes become
$n_i={\kappa_i \lambda \over 2 \pi}={k_i\lambda \over 2 \pi   L}$ and
$n_e={\kappa_e \lambda \over 2 \pi}={k_e\lambda \over 2 \pi   L}$, where
$\lambda$ is the employed light wavelength.
For ease of  notation, in the sequel we will drop the  \,$\tilde{}$ symbol.  In what follows, all the magnitudes are dimensionless.  We wish to image objects using light, ranging from sizes of a few nanometers (viruses, colloidal particles, cell structures) to a few microns (prokaryotic cells). The reference length will be set to $L=1 \mu$m.
We will work with laser lights of wavelengths varying from $405$ nm (violet light) to $660$ nm (red light). For refractive indexes typical of cellular structures we find $k_i$ and $k_e$ in the ranges $12-15$ and $20-25$, respectively, for instance.

In the general framework described above and assuming the ambient refractive index known, the inverse holography problem consists in finding objects $\Omega=\cup_{\ell=1}^{\cal L}\Omega_\ell$ and functions $k_i: \Omega \rightarrow \mathbb R^+$ satisfying the equation
\begin{eqnarray}
I_{\rm meas}(\mathbf x_j) = |{\mathbf E}_{\Omega,k_i}(\mathbf x_j)|^2,
\quad j=1, \ldots, N,
\label{eqinversek}
\end{eqnarray}
where $\mathbf E_{\Omega,k_i}={\mathbf E}_{\rm inc}+{\mathbf E}_{\rm sc,\Omega,k_i}$ is the solution of the forward problem (\ref{forwardadim}) with object $\Omega$ and dimensionless wavenumber $k_i$ (i.e. refractive index $n_i$) whereas $I_{\rm meas}$ represents the hologram measured at the screen points $\mathbf x_j,$ $j=1, \ldots, N.$

To simplify, in this paper we consider that the refractive indexes are constant and known inside each component  of the objects. Therefore, the problem becomes finding objects $\Omega$ such that the equation:
\begin{eqnarray}
I_{\rm meas}(\mathbf x_j) =
|{\mathbf E}_{\Omega}(\mathbf x_j)|^2,
\quad j=1, \ldots, N,
\label{eqinverse}
\end{eqnarray}
is satisfied. Alternatively, we can reformulate this equation as a constrained optimization problem:  Find the global minimum $\Omega$ of
\begin{eqnarray}
J(\mathbb R^3 \setminus \overline{\Omega})= {1\over 2} \sum_{j=1}^N | I_{\Omega}(\mathbf x_j) - I_{\rm meas}(\mathbf x_j)|^2.
\label{costH}
\end{eqnarray}
Here, $I_{\Omega}= |{\mathbf E}_{\Omega}|^2 $ and $\mathbf E_{\Omega}$ is the solution of the dimensionless forward system (\ref{forwardadim}). The object $\Omega$ is the design variable. The stationary Maxwell system (\ref{forwardadim}) is the constraint. The true objects are a global minimum at which the cost functional vanishes.

In practice, we never know the true hologram $I_{\rm meas}$
but a hologram $I_{\rm meas}^{\eta}= I_{\rm meas} + I_{\eta}$ affected by noise $I_{\eta}$ of  magnitude $\eta$, meaning that
\begin{eqnarray}\begin{array}{l}
\| I_{\rm meas} - I_{\rm meas}^{\eta} \|_{2} = \big(\sum_{j=1}^N |
I_{\rm meas}(\mathbf x_j) - I_{\rm meas}^{\eta}(\mathbf x_j) |^2 \big)^{1/2} \leq \eta \, \| I_{\rm meas} \|_{2},\end{array} 
\label{eta}
\end{eqnarray}
where $\delta= \| I_{\rm meas} -  I_{\rm meas}^{\eta} \|_{2}$ is the noise level.
Thus, $I_{\rm meas}$ is replaced in (\ref{eqinverse}) and (\ref{costH}) by $I_{\rm meas}^{\eta}$.
In the next section we explain how to approximate the number, size
and location of holographied objects $\Omega$ from a hologram
using the topological derivative of the cost functional (\ref{costH}).

\section{Topological derivative based imaging}
\label{sec:priors}

In the absence of any information on the holographied objects, other
than the measured hologram and the nature of the ambient medium,
some key features can be captured by a topological sensitivity analysis
of the cost functional (\ref{costH}).

\subsection{First guesses}
\label{sec:1st}

The topological derivative of functional
(\ref{costH}) measures its sensitivity to including and removing points
in reconstructions of an object. Given a region ${\cal R}$ and a point
$\mathbf x \in {\cal R}$, we have the expansion
\begin{equation}\label{expansion}
J({\cal R} \setminus \overline{B_\varepsilon (\mathbf{x})})=
J({\cal R})+{4\over 3} \pi \varepsilon^3 D_T(\mathbf{x},{\cal R})
+o(\varepsilon^3),\qquad \varepsilon\to 0,
\end{equation}
for any ball $B_\varepsilon (\mathbf{x})=B(\mathbf x, \varepsilon)$
centered at $\mathbf x$ with radius $\varepsilon$. The
coefficient $D_T(\mathbf{x},{\cal R})$ is the topological derivative of
the functional at ${\mathbf x}$ \cite{sokolowski}. When
$D_T(\mathbf{x},{\cal R})$ is negative,  the cost functional decreases
for $\varepsilon>0$ small, that is,
$J({\cal R}\setminus \overline{B_\varepsilon (\mathbf{x})})<J({\cal R})$.
This suggests that the cost functional decreases by forming
objects $\Omega_{\rm ap}$ with points where the topological derivative
is negative and large \cite{ip2008,feijoo,masmoudi}:
\begin{equation}
    \Omega_{\rm ap}:=\{{\bf x}\in {\cal R} \,|\, D_T({\bf x}, {\cal R}) <-C \}.
    \label{initialguesstd}
\end{equation}
For large enough $C>0,$ we expect $J({\cal R} \setminus
\overline{\Omega}_{\rm ap})<J({\cal R}).$ In practice, to select a first approximation of the holographied objects we set ${\cal R}=\mathbb R^3$ 
and evaluate the topological derivative in a bounded region ${\cal R}_{\rm obs}\subset 
\mathbb R^3$ where objects are assumed to be located, i.e.,
\begin{equation}\label{initialguesstd_impl}
\Omega_{\rm ap}:=\{{\bf x}\in  {\cal R}_{\rm obs} \,|\, D_T({\bf x},\mathbb R^3)
    <(1-C_0)\min_{{\bf y}\in {\cal R}_{\rm obs}}D_T({\bf y},\mathbb R^3)\},
\end{equation}
with $0<C_0<1$ arbitrarily chosen (in most of our examples we will set $C_0=0.15$ or $C_0=0.2$).

Computing the topological derivative using the expansion (\ref{expansion}) is too costly from the computational point of view. We use explicit expressions in terms of adequate forward and adjoint fields instead. For the cost functional (\ref{costH}) and ${\cal R}=\mathbb R^3$:
\begin{eqnarray}
D_T(\mathbf x, \mathbb R^3)= 3 \, {\rm Re}
\left[
{k_e^2  (k_e^2  - k_i^2 \beta) \over   (k_i^2 \beta +2k_e^2 )}
\mathbf E(\mathbf x) \cdot \overline{\mathbf P}(\mathbf x) -
{1 -\beta \over 1 + 2 \beta}
\mathbf{curl} \,  \mathbf E(\mathbf x) \cdot \mathbf{curl} \,
\overline{\mathbf  P}(\mathbf x)
\right] \label{DT}
\end{eqnarray}
where $\beta={\mu_e \over \mu_i}$, see Appendix \ref{sec:tdcost}.
Notice that $\beta \sim 1$ for biological samples \cite{lin}, therefore we will neglect the second term. The forward $\mathbf E$  and adjoint $\mathbf P$ fields are solutions of
\begin{eqnarray} 
\begin{array}{r}
\mathbf{curl} \,  (\mathbf{curl} \,  \mathbf E)  - k_e^2 \mathbf E = 0
\quad \mbox{in $\mathbb R^3$},  \\
{\rm lim}_{|\mathbf x| \rightarrow \infty} |\mathbf x|  \big|
\mathbf{curl} \,  (\mathbf E - \mathbf E_{\rm inc}) \times \hat{\mathbf x}
-\imath k_e (\mathbf E - \mathbf E_{\rm inc}) \big| =0, 
\end{array} \label{forwardempty} \\[1ex]
\begin{array}{r}
\mathbf{curl} \,  (\mathbf{curl} \,  \mathbf P)  - k_e^2 \mathbf P =
2 \sum_{j=1}^N (I_{\rm meas} - |\mathbf E|^2) \mathbf E \,
\delta_{\mathbf x_j}
\quad \mbox{in $\mathbb R^3$}, \\ [1ex]
{\rm lim}_{|\mathbf x| \rightarrow \infty} |\mathbf x|  \big|
\mathbf{curl} \,  \mathbf P \times \hat{\mathbf x}
+\imath k_e \mathbf P \big| =0,
\end{array}  \label{adjointempty}
\end{eqnarray}
where $\delta_{\mathbf x_j}$ are Dirac masses concentrated at the detectors
$\mathbf x_j$, $j=1,...,N,$ and $ \hat{\mathbf x} = {\mathbf x \over |\mathbf x|}.$
As shown in Appendix \ref{sec:tdcost}, the equation governing the forward electric field $\mathbf E$ and the boundary conditions at the interface of the object determine the expression of the derivative. The specific  dependence on $\mathbf E$ of the cost functional influences the source of the adjoint problem instead.
Expressions of the form (\ref{DT}) were first established in \cite{masmoudi} 
with a different adjoint  field to account for a different cost functional.

For an incident plane wave $\mathbf E_{\rm inc}(\mathbf x)
= \hat{\mathbf p} e^{\imath k_e  z},$
polarized in a direction $\hat {\mathbf p}$ orthogonal to the
direction of propagation $(0,0,1)$,
the solution of (\ref{forwardempty}) is $\mathbf E = \mathbf E_{\rm inc}$.
The conjugate of the solution of (\ref{adjointempty}) is
\begin{eqnarray}
\label{adjointexplicit}
\overline{\mathbf P}(\mathbf x) ={1\over k_e^2}
\sum_{j=1}^N \mathbf{curl}_{\mathbf x} \mathbf{curl}_{\mathbf x}
\left( G_{k_e}(\mathbf x \!-\! \mathbf x_j) \, 2
(I_{\rm meas}(\mathbf x_j)\!-\!|\mathbf E_{\rm inc}(\mathbf x_j)|^2)\,
\overline{\mathbf E_{\rm inc}(\mathbf x_j)} \right),
\end{eqnarray}
where $G_{k_e}(\mathbf x) = {e^{\imath k_e |\mathbf x|} \over 4 \pi |\mathbf x|}$
is the outgoing Green function of Helmholtz equation, that is, the solution of
$-\Delta G - k_e^2 G = \delta$ satisfying an outgoing radiation condition.
\ref{sec:explicitwhole} quantifies the deviation of these 3D formulas for the
topological derivatives of the holography cost functional with Maxwell
constraints from the scalar Helmholtz approximations employed
in \cite{siam2018}.


\begin{figure}[!h]
\centering
\hskip -0.5cm (a) \hskip 3.5cm (b) \hskip 3cm (c)  \\
\hskip -2mm
\includegraphics[height=3.4cm]{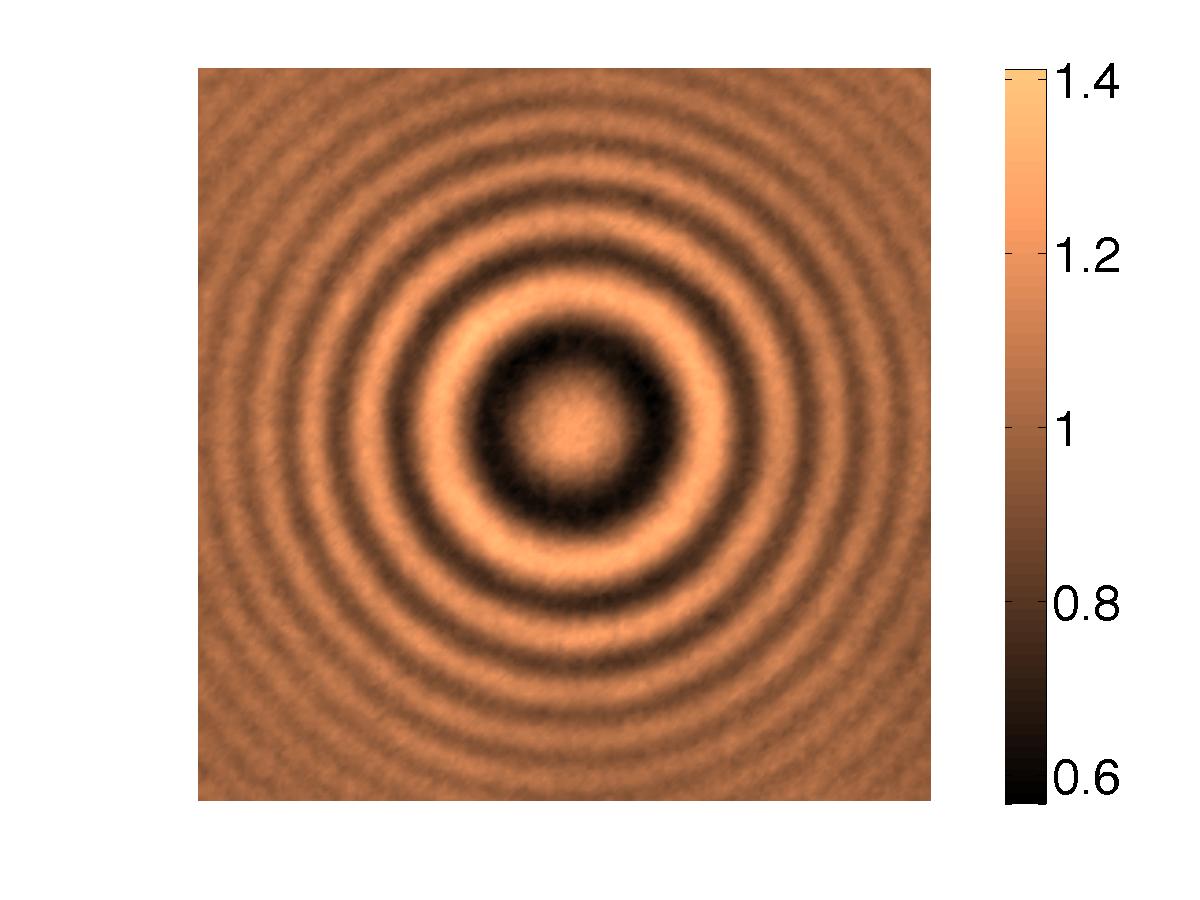} \hskip -4mm
\includegraphics[height=3.4cm,width=3.8cm]{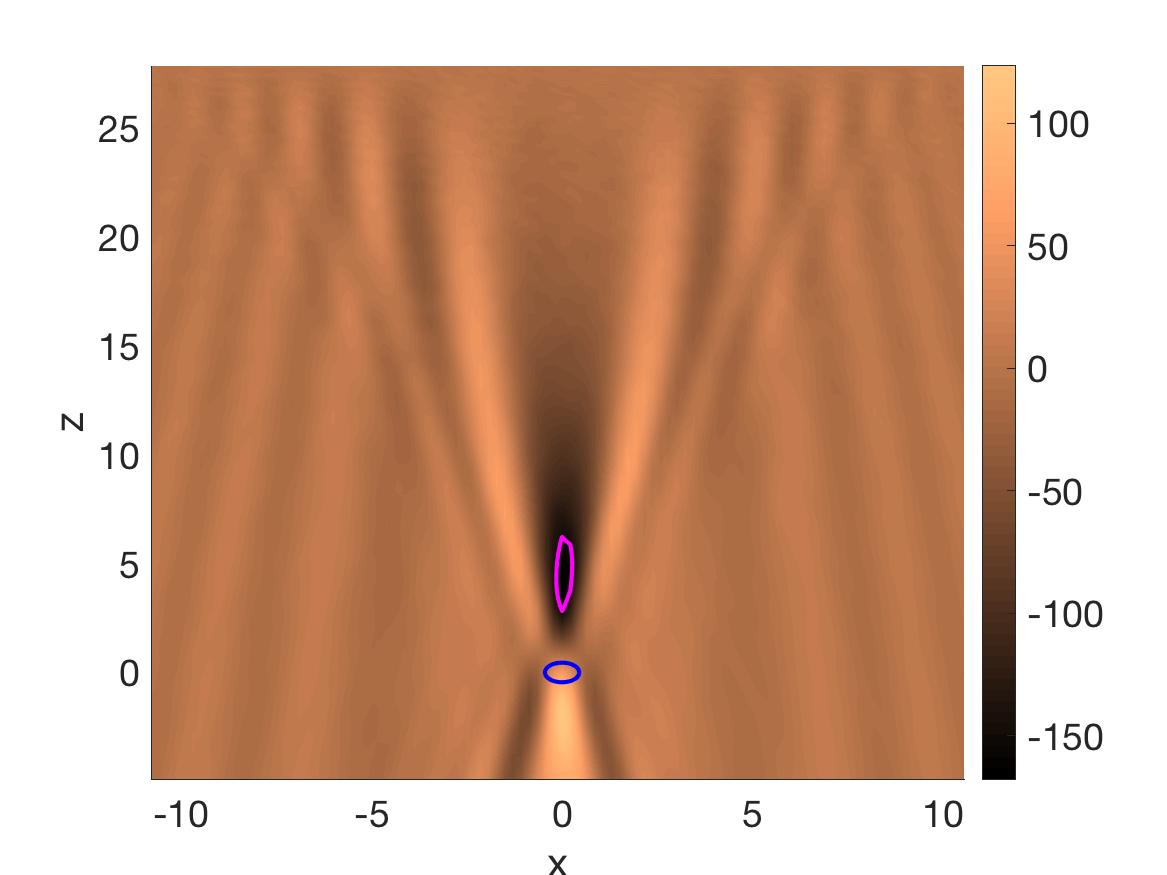} \hskip -3mm
\includegraphics[height=3.4cm,width=3.8cm]{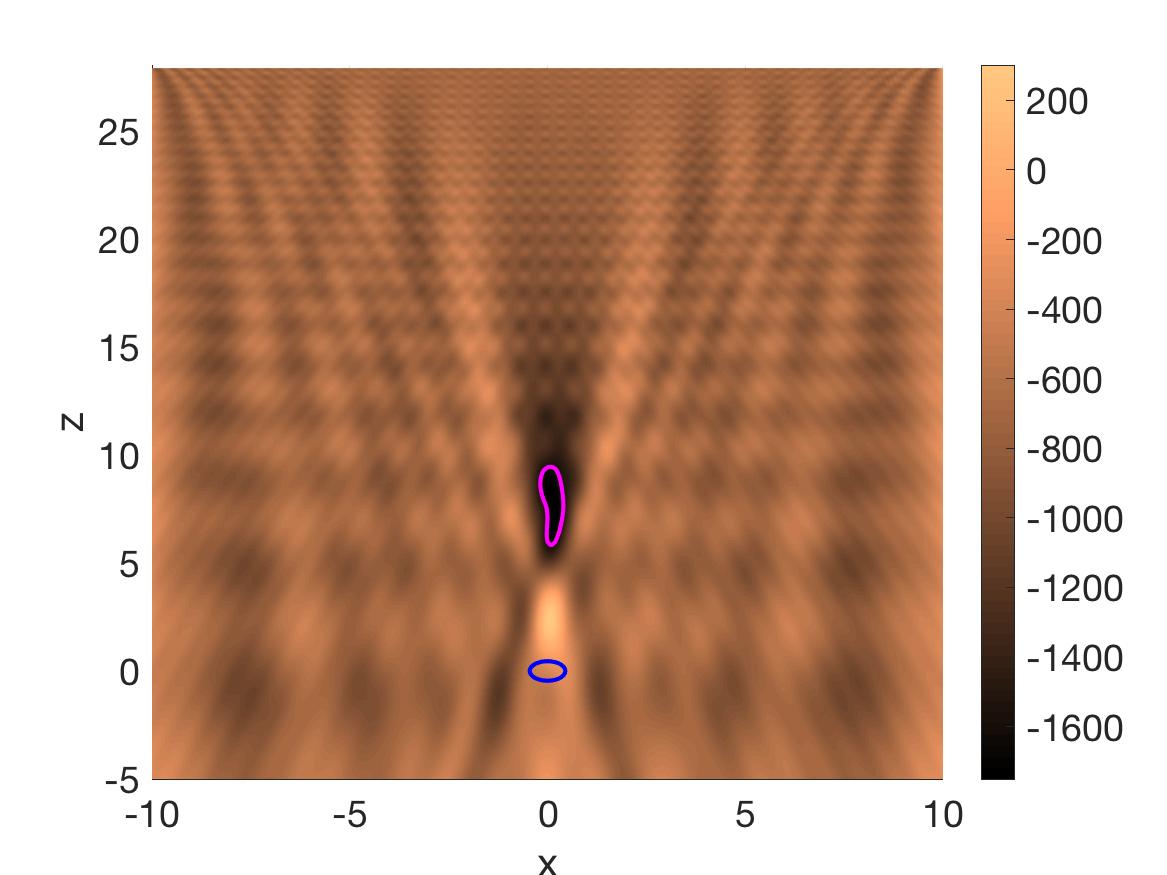} \hskip -3mm
\includegraphics[height=3.2cm,width=0.8cm]{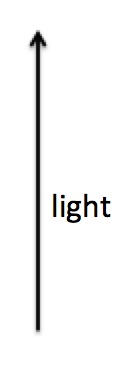}
\caption{ \small
(a) Experimentally measured hologram ${\mathbfcal I}_{\rm meas}$
generated by a sphere
of radius $0.45 \mu$m,  illuminated by polarized light of wavelength $520$nm and
placed at a distance $28 \mu$m of a CMOS screen of size $21.45
\mu{\rm m} \times 21.45 \mu{\rm m}$ with a pixel grid of step
$0.1078  \mu{\rm m}$.
The  dimensionless wavenumbers are $k_e=16.07$ and $k_i=19.33$.
Slice $y=0$ of (b) the topological derivative computed using expression (\ref{DT})
with forward and adjoint fields given by (\ref{forwardempty})-(\ref{adjointexplicit}),
and (c) the scalar approximation of the topological derivative in \ref{sec:explicitwhole}.
The blue contour represents the true object, while the lighter magenta contour
represents an approximation using the topological derivative.
}
\label{fig2}
\end{figure}

Figure \ref{fig2} shows results for an experimentally recorded hologram.
Panel (b) represents the topological derivative constructed using the
hologram in panel (a) in the vector Maxwell framework. The region where
large negative values are attained would indicate the approximate location
of an object. We can appreciate that this prediction is elongated and shifted
towards the screen. Panel (c) displays the topological derivative obtained using
the scalar formulas in \ref{sec:explicitwhole}.
This latter prediction improves noticeably and becomes comparable
when we replace the true hologram by a synthetic hologram consistently
generated using the scalar approximation for the polarized component.
The increased shift is the result of neglecting the nonpolarized
components, which contribute to the experimentally measured hologram. 
All the subsequent numerical tests in this paper use synthetically generated
holograms, and devices of smaller size to reduce the computational cost of 
the evaluation of the involved fields in 3D regions. The distance to the hologram 
recording screen will be about $5 \mu$m and the size of the screen 
$10 \mu$m $\times$ $10 \mu$m in the set-up sketched in Fig. \ref{fig1}(b).

\begin{figure}[!h] \centering
\hskip -0.5cm (a) \hskip 3.25cm (b) \hskip 3.25cm (c) \\
\hskip -0.0 cm \includegraphics[width=4cm]{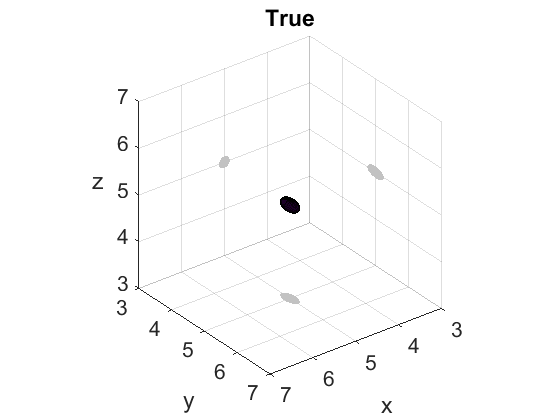} \hskip -5mm
\includegraphics[width=4cm]{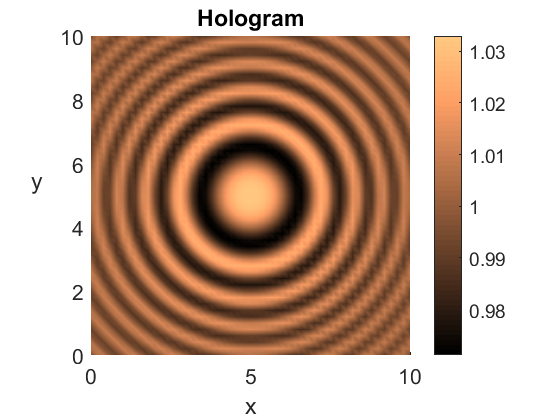} \hskip -5mm
\includegraphics[width=4cm]{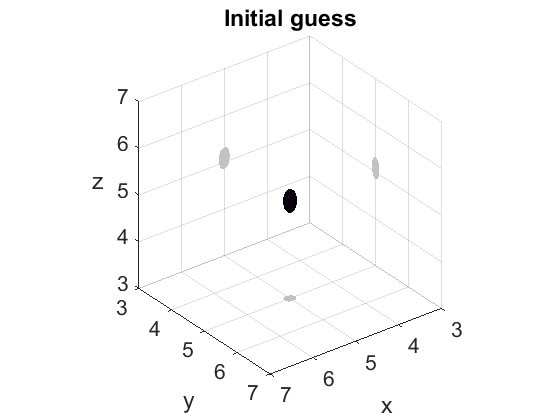} 
\\
\hskip -0.5cm (d) \hskip 3.25cm (e) \hskip 3.25cm (f) \\
\includegraphics[width=4cm]{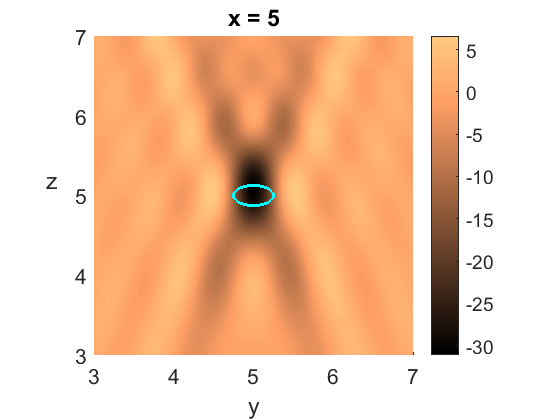} \hskip -2mm
\includegraphics[width=4cm]{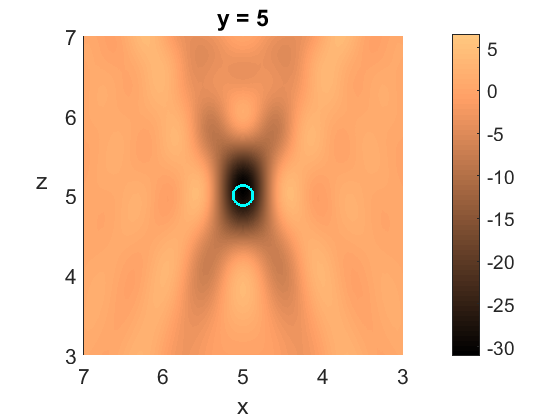} \hskip -2mm
\includegraphics[width=4cm]{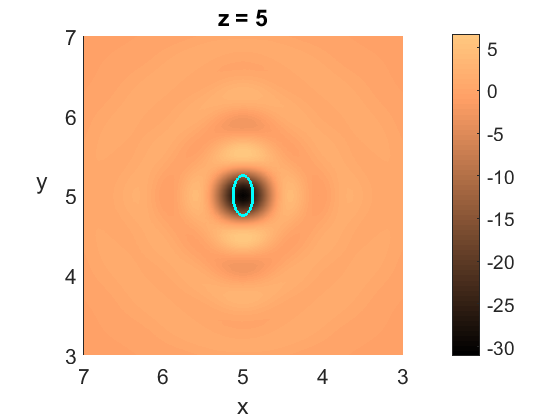}
\caption{ \small
Detection of an ellipsoid using topological  derivatives in the set-up depicted in Fig. \ref{fig1}(b) with $k_e=12.56$, $k_i=15.12$ and $\beta=1$.
(a) True geometry:  The ellipsoid is centered at $(5,5,5)$ and oriented along the $y$ axis, with  semi-axes  $a=0.125$, $b=0.25$, $c=0.125$.
(b) Synthetic hologram recorded on the screen $z=10$, on a grid with $51 \times 51$ detectors located at the points ${\bf x}_{k\ell}=(0.2k,0.2\ell,10)$, $k,\ell=0,\dots,50$.
The direction of the incident red light is $(0,0,1)$, the polarization vector is (1,0,0).
(c) Initial guess defined by (\ref{initialguesstd_impl}) and (\ref{DT})-(\ref{adjointexplicit}) with $C_0=0.15$ when we add to the hologram $2\%$ noise, 
i.e. $\eta=0.02$ in (\ref{eta}).
The proposed object is elongated in the incidence direction and its center slightly shifted towards the screen. (d)-(f) Slices of the topological derivative: (d) $x=5$, (e) $y=5$, (f) $z=5$. Cyan contours represent sections of the true object.}
\label{fig3}
\end{figure}

Figure \ref{fig3} illustrates the procedure for an ellipsoid 
when using red light of $660$ nm. The estimation of the 
object location improves as we place
it closer to the recording screen and reduce its size. However, the elongation in
the predicted shape remains and the true orientation of the ellipsoid is missed
in the initial guess.  The same behavior is observed for violet
light of $405$ nm.

Fixing the hologram size, the distance to the hologram recording screen and
the light wavelength, we have noticed that oscillations may appear
in the topological derivative as the size of the object grows. Then, the objects
are more clearly located tracking the peaks of a companion field
\cite{tenergy}, the topological energy
$E_T(\mathbf x, \mathbb R^3)= |\mathbf E|^2 |\overline{\mathbf P}|^2,$
where $\mathbf E$ and $\mathbf P$ are the forward and adjoint fields
governed by (\ref{forwardempty}) and (\ref{adjointempty}). This issue
is further discussed in  \cite{siam2016,ln2008,guzina}.
Here, we focus on improving the reconstructions of objects whose sizes are
similar to or smaller than the light wavelength to avoid this transition.

\subsection{Correction of the number of components}
\label{sec:tditeration}

Once we have an initial reconstruction of the objects $\Omega_{\rm ap}$ we may
improve it by a topological derivative based iteration.
We construct a new approximation $\Omega_{\rm new}$ from $\Omega_{\rm ap}$ as follows  \cite{siam2016,ip2008}:
\begin{eqnarray}
    \Omega_{\rm new}:= \{{\bf x}\in \Omega_{\rm ap}  \;|\;
    D_T({\bf x},{\mathbb R}^3\setminus \overline{\Omega}_{\rm ap})< (1\!-\!c_1)\max_{{\bf y}\in\Omega_{\rm ap}}D_T({\bf y},\mathbb R^3\setminus\overline{\Omega}_{\rm ap}) \}   \hskip 1cm
    \label{updatedguess2td} \\
\cup \, \{{\bf x}\in  {\cal R}_{\rm obs}\setminus\overline{\Omega}_{\rm ap}   \;|\; D_T({\bf x},{\mathbb R}^3\setminus \overline{\Omega}_{\rm ap})< (1\!-\!C_{1})
\min_{{\bf y}\in {\cal R}_{\rm obs}\setminus\overline{\Omega}_{\rm ap}}
\hskip -4mm D_T({\bf y},\mathbb R^3\setminus\overline{\Omega}_{\rm ap}) \}.
\hskip 5mm
\nonumber
\end{eqnarray}
The positive constants $C_{1}$, $c_{1}$ in (\ref{updatedguess2td}) are
selected to ensure a decrease in the shape functional:
$J({\mathbb R}^3 \setminus\overline{\Omega}_{\rm new})<J({\mathbb R}^3 \setminus\overline{\Omega}_{\rm ap})$. While the points where the updated topological derivative is negative and large are added to the former reconstructions,  the points where it is positive and large are removed.   In principle, we will set $C_1=c_1=C_0$ but they may be automatically reduced in case the cost functional does not decrease. This descent strategy is suggested by expansion
(\ref{expansion}) for $\mathbf x \in {\cal R}= \mathbb R^3 \setminus\overline{\Omega}$ and its equivalent
\begin{equation}\label{expansion2}
\begin{array}{ll}
J(\mathbb R^3 \setminus (\overline{ \Omega \setminus
\overline{B_\varepsilon (\mathbf{x})} })) & =
J((\mathbb R^3  \cup B_\varepsilon (\mathbf{x})) \setminus \overline{\Omega})\\[1ex]
& = J( \mathbb R^3\setminus\overline{\Omega} )-{4\over 3} \pi \varepsilon^3 D_T(\mathbf{x},\mathbb R^3 \setminus \overline{\Omega}) +o(\varepsilon^3)
\end{array}
\end{equation}
for $\mathbf x \in \Omega.$
Notice the change of sign when compared with expansion (\ref{expansion}) for 
$\mathbb R^3 \setminus \overline{\Omega}.$ This is a matter of choice. Keeping the same sign in both we should add and remove regions with large and negative topological derivative. Changing the sign we achieve a simpler and more visual interpretation, which is easier to implement. Also, this choice ensures that the global expression of the topological derivative when $\beta=1$ is continuous in $\mathbb R^3$ for scalar problems  \cite{siam2018,ip2008}.

\begin{figure}
\centering
\hskip 0.0cm (a) \hskip 3.25cm (b) \hskip 3.25cm (c) \\
\includegraphics[width=4cm]{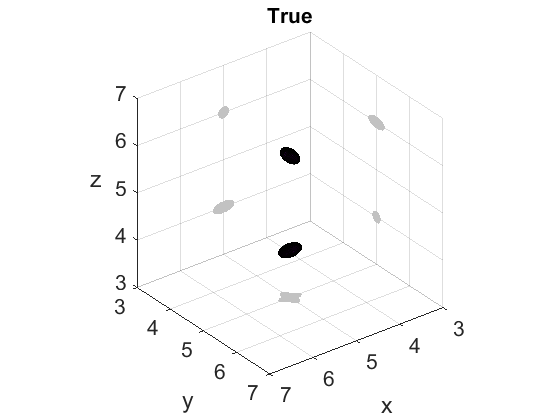} \hskip -2mm
\includegraphics[width=4cm]{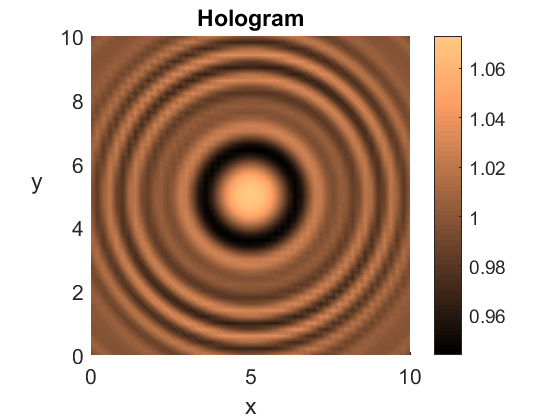}\hskip -2mm
\includegraphics[width=4cm]{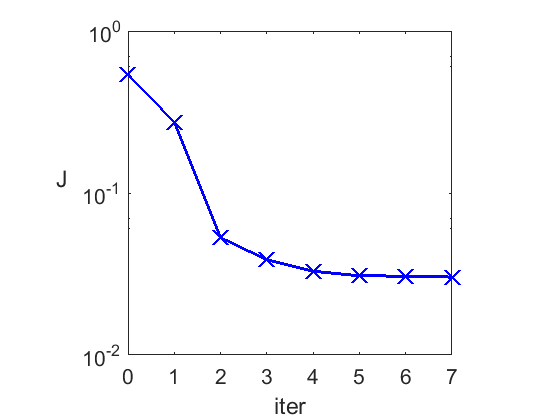} \\
\hskip 0.0cm (d) \hskip 3.25cm (e) \hskip 3.25cm (f) \\
\includegraphics[width=4cm]{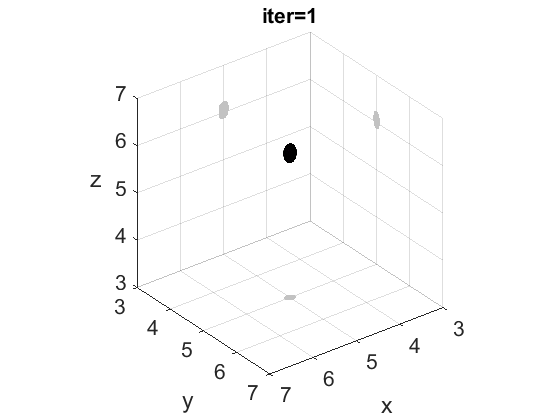} \hskip -2mm
\includegraphics[width=4cm]{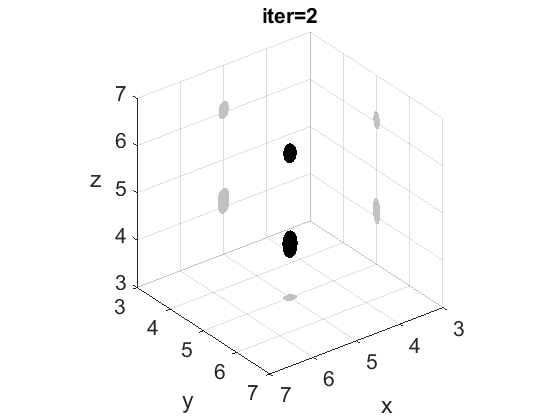} \hskip -2mm
\includegraphics[width=4cm]{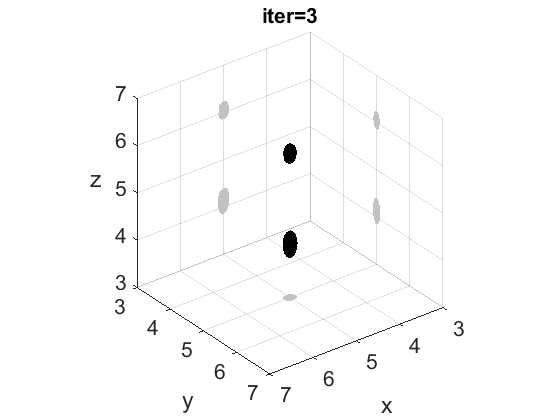} \\
\hskip 0.0cm (g) \hskip 3.25cm (h) \hskip 3.25cm (i) \\
\includegraphics[width=4cm]{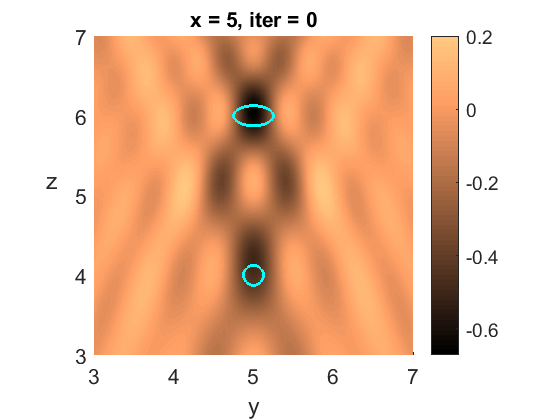} \hskip -2mm
\includegraphics[width=4cm]{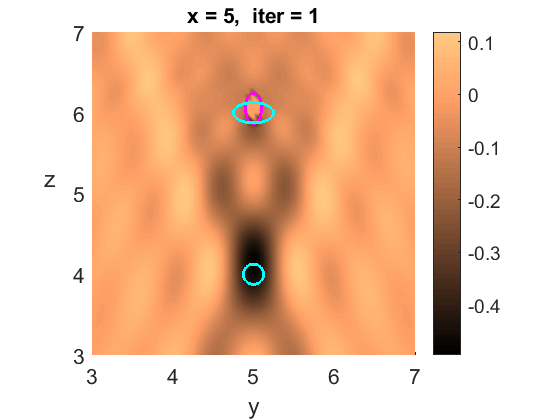} \hskip -2mm
\includegraphics[width=4cm]{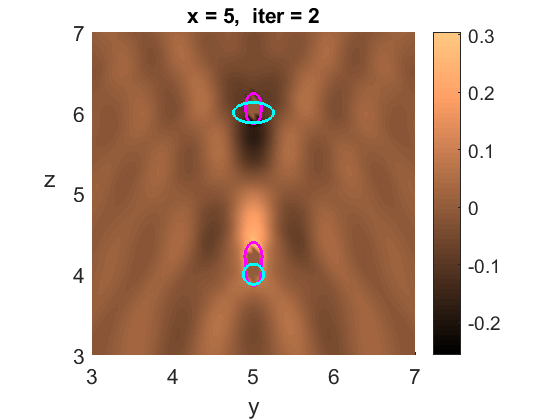} \\
\hskip 0.0cm (j) \hskip 3.25cm (k) \hskip 3.25cm (l) \\
\includegraphics[width=4cm]{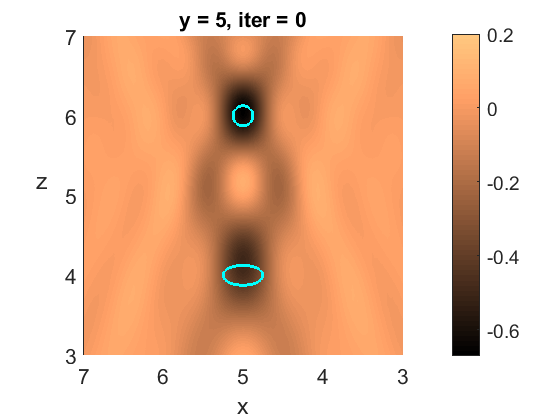} \hskip -2mm
\includegraphics[width=4cm]{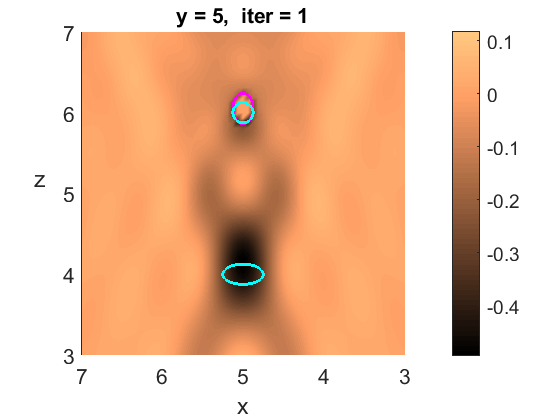} \hskip -2mm
\includegraphics[width=4cm]{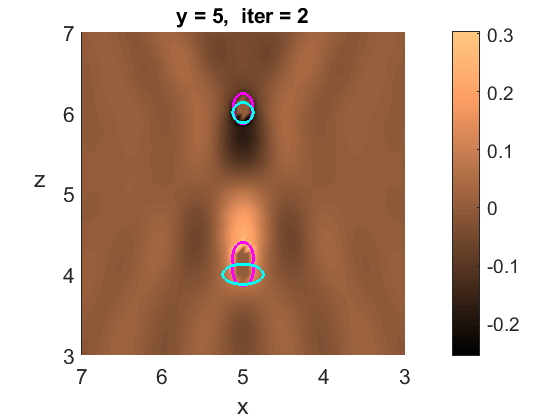}
\caption{ \small
Detection of two ellipsoids using topological  derivatives in the set-up depicted in Fig. \ref{fig1}(b) with $k_e=12.56$, $k_i=15.12$ and $\beta=1$.
(a) True geometry:  One ellipsoid is centered at $(5,5,6)$ and oriented along the $y$ axis, with  semi-axes $a=0.125$, $b=0.25$, $c=0.125$. The second one is centered at $(5,5,4)$ and oriented along the $x$ axis, with  semi-axes $a=0.25$, $b=0.125$, $c=0.125$.
(b) Synthetic hologram recorded on the screen $z=10$, on a grid with $51\times 51$ detectors. The direction of the incident red light is $(0,0,1)$ and the polarization vector is (1,0,0). (c) Evolution of the cost functional.
(d) Initial guess defined by (\ref{initialguesstd_impl}) and (\ref{DT})-(\ref{adjointexplicit}) with $C_0=0.15$. Only one object is detected, which is elongated and shifted in the incidence direction.
(e) and (f) Approximate objects  in the 2nd and 3rd iterations defined
by (\ref{updatedguess2td}) and (\ref{forwardomega})-(\ref{adjointomega})
with $C_1=c_1=0.15$. The second object is detected and the location of the center improves.  However, the elongation remains and the orientation is missed. Cyan contours plot sections of the true objects whereas magenta contours represent the approximations.
Slices of the topological derivative up to the 2nd iteration (g),(h),(i) $x=5$  and (j),(k),(l) $y=5$. No significative changes are observed until the process stops at the 7th iteration.}
\label{fig4}
\end{figure}

In the presence of an object $\Omega= \Omega_{\rm ap}$, the topological derivative is given by
\begin{eqnarray}
D_T(\mathbf x, \mathbb R^3 \setminus \overline{\Omega})=
\left\{ \begin{array}{l}
3 \, {\rm Re}
\big[ {k_e^2  (k_e^2  - k_i^2 ) \over   (k_i^2 +2k_e^2 )}
\mathbf E(\mathbf x) \cdot \overline{\mathbf P}(\mathbf x)
\big],  \quad \mathbf x \in \mathbb R^3 \setminus\overline{\Omega}, \\
3 \, {\rm Re}
\big[ {k_i^2  (k_e^2  - k_i^2 ) \over   (k_e^2 +2k_i^2 )}
\mathbf E(\mathbf x) \cdot \overline{\mathbf P}(\mathbf x)  \big],
 \quad \mathbf x \in \Omega,
\end{array} \right.
\label{DTsimple}
\end{eqnarray}
when $\beta={\mu_e \over \mu_i}\sim 1$, see Appendix \ref{sec:tdcost},
with forward and conjugate adjoint fields satisfying transmission Maxwell 
problems with object $\Omega=\Omega_{\rm ap}$
\begin{eqnarray}  
\begin{array}{rcl}
\mathbf{curl} \,  (\mathbf{curl} \,  \mathbf E)  - k_e^2 \mathbf E =0,
& \mbox{in} & \mathbb R^3\setminus\overline{\Omega},   \\
\mathbf{curl} \,  (\mathbf{curl} \,  \mathbf E)  - k_i^2 \mathbf E =0, &
\mbox{in} & \Omega,   \\
\hat{ \mathbf n}  \times \mathbf E^- =
\hat{ \mathbf n} \times \mathbf E^+,
& \mbox{on} & \partial \Omega,  \\
\beta \hat{ \mathbf n} \times \mathbf{curl} \,  \mathbf E^- =
\hat{ \mathbf n} \times \mathbf{curl} \,  \mathbf E^+,
& \mbox{on} & \partial \Omega,    \\
{\rm lim}_{|\mathbf x| \rightarrow \infty} |\mathbf x|  \big|
\mathbf{curl} \,  (\mathbf E - \mathbf E_{\rm inc}) \times \hat{\mathbf x} -
\imath k_e (\mathbf E - \mathbf E_{\rm inc}) \big| =0, &&
\end{array} \label{forwardomega} \\[1ex]
\begin{array}{rcl}
\mathbf{curl} \,  (\mathbf{curl} \,  \overline{\mathbf P})  - k_e^2
\overline{\mathbf P} =
2 \sum_{j=1}^N (I_{\rm meas} - |\mathbf E|^2)
\overline{\mathbf E} \, \delta_{\mathbf x_j}
& \mbox{in} & \mathbb R^3\setminus\overline{\Omega},  \\
\mathbf{curl} \,  (\mathbf{curl} \,  \overline{\mathbf P})  - k_i^2
\overline{\mathbf P} = 0,
& \mbox{in} & \Omega,    \\
\hat{ \mathbf n}  \times \overline{\mathbf P}^- =
\hat{ \mathbf n} \times \overline{\mathbf P}^+,
& \mbox{on} & \partial \Omega,  \\
\beta \hat{ \mathbf n} \times \mathbf{curl} \,  \overline{\mathbf P}^- =
\hat{ \mathbf n} \times \mathbf{curl} \,  \overline{\mathbf P}^+,
& \mbox{on} & \partial \Omega,    \\
{\rm lim}_{|\mathbf x| \rightarrow \infty} |\mathbf x|  \big|
\mathbf{curl} \,  \overline{\mathbf P} \times \hat{\mathbf x}
- \imath k_e \overline{\mathbf P} \big| =0,
\end{array}  \label{adjointomega}
\end{eqnarray}
where $\hat{\mathbf n}$ is the unit normal and
$ \hat{\mathbf x} = {\mathbf x \over |\mathbf x|}.$

The procedure described in (\ref{updatedguess2td}) allows for the creation of new components forming the objects (if missed in the previous iteration),  for merging existing components (if they are close and intermediate points are indentified), for the destruction of proposed components (if they turn out to be spurious), and for the creation of holes inside them \cite{ip2008}. It does not rely on any object parametrization and can be used to reconstruct non convex and multiple objects \cite{siam2016,siam2018}.

A delicate step in the process is the fitting of surfaces to the objects defined by  (\ref{initialguesstd_impl}) and (\ref{updatedguess2td}),  in such a way that the corresponding forward and adjoint fields can be computed numerically. Here we resort to star-shaped parametrizations and spectral solvers, as described in Sections \ref{sec:algorithm} and \ref{sec:numerical}. When we are only interested in determining the number and location of the objects, we may just fit spheres to each component and use explicit Mie series solutions (see Appendix \ref{sec:sphere}) as done
in \cite{siam2018} for scalar problems. In general, we may avoid choosing specific parametrizations by fitting to (\ref{updatedguess2td}) smooth surfaces defined by means of blobby molecules and signed distance functions, and employing coupled BEM-FEM solvers as in \cite{siam2016}. We could also work directly with (\ref{updatedguess2td}) and DDA solvers \cite{ddatheory,dimidukdda}.

Figure \ref{fig4} illustrates the performance of the  iterative method based on (\ref{updatedguess2td}) to detect two ellipsoids aligned in the incidence direction for the red light. The synthetic hologram is generated here solving numerically the forward problem (\ref{forwardomega}) when the scatterers $\Omega$ are the true ellipsoids and adding $2\%$ random noise, fnamely, when $\eta=0.02$ in (\ref{eta}).  The two objects are detected, with adequate approximate locations and sizes, but the process stagnates without recovering the correct orientation. The same happens if we modify the light, increase the number of detectors or change the thresholds $C_0$ and/or $C_1$ and $c_1$.

\section{Iteratively regularized Gauss-Newton approach}
\label{sec:gauss}

Topological derivative-based methods yield predictions of the number, location and
size of  holographied objects. However,  the objects appear to be displaced and elongated towards the hologram recording screen and their global 3D shape is not really recognizable, though some $xy$ slices may suggest it. We will show that this information can be refined by iteratively regularized Gauss-Newton methods (IRGNM)  to obtain accurate reconstructions of the holographied shapes, provided we can rely on some kind of parametrization.  
However, IRGNM can only improve shapes, not alter the number of predicted object contours. This section adapts the IRGNM to holographic settings. A
hybrid algorithm combining topological derivatives to allow for variations in the
number of contours and IRGNM to correct shapes will be presented in Section \ref{sec:algorithm}.

\subsection{General framework}
\label{sec:irgngeneral}

We choose an algorithm studied in \cite{Hohage,Hohage2} and based on the
IRGN method introduced in \cite{Bakushinskii}, which  enjoys some convergence properties  \cite{Bakushinskii,Hohage}.  Let us recall the main ideas. Given two Hilbert spaces $X$, $Y$
and a Fr\'echet differentiable operator  ${\cal F}:  D({\cal F}) \subset X \longrightarrow Y$, we assume that the exact data $y \in Y$ are attainable, that is, there is 
$x \in X$ such that ${\cal F}(x)=y,$ but only noisy data  $y^{\delta}$ satisfying 
$\| y^\delta -y \|_Y \leq \delta$ are available. Starting from an initial guess $x_0$, the IRGN method \cite{Bakushinskii} generates a sequence
$x_{k+1}^\delta$ as follows.  At each step, we linearize the equation
at $x_k^\delta$, solve ${\cal F}(x_k^\delta) + {\cal F}'(x_k^\delta) \xi = y^\delta$
by means of the minimization problem:
\begin{eqnarray} \begin{array}{l}
\xi_{k+1} = {\rm Argmin}_{\xi \in X} \|{\cal F}(x_k^\delta) + {\cal F}'(x_k^\delta) \xi -
y^\delta \|_Y^2 + \alpha_k \|x_k^\delta + \xi  - x_0  \|_{X}^2,
\end{array} \label{tikhonov1} \end{eqnarray}
and set $x_{k+1}^\delta = x_{k}^\delta + \xi_{k+1}.$ The penalty term
$ \alpha_k \|\xi  \|_{X} $ would lead to the Levenberg-Marquart algorithm
\cite{leven,marq}. The Tikhonov regularizing term
$\alpha_k \|x_k^\delta + \xi  - x_0  \|_{X}^2 $ has additional regularizing
properties and facilitates convergence for specific choices of $\alpha_k$
and of initial guesses $x_0$ \cite{Hohage}. From the theory
of linear Tikhonov regularization, the unique solution of (\ref{tikhonov1})
is given by
\begin{eqnarray} \begin{array}{l}
\xi_{k+1} = - ({\cal F}'(x_k^\delta)^* {\cal F}'(x_k^\delta) + \alpha_k I)^{-1}
[ {\cal F}'(x_k^\delta)^* ({\cal F}(x_k^\delta)-y^\delta) + \alpha_k
(x_k^\delta - x_0)],
\end{array} \label{tikhonov2}
\end{eqnarray}
where ${\cal F}'(x_k^\delta)^*$ denotes the adjoint of the Fr\'echet derivative
${\cal F}'(x_k^\delta)$, see \ref{sec:frechet}.

\subsection{Variant adapted to inverse holography}
\label{sec:irgnadapted}

We introduce next a procedure based on star-shaped parameterizations inspired in \cite{Hohage3,lelouerspectral}.  We assume that the holographied object $\Omega_{\rm true}$  can be approximated by a finite number of star-shaped objects. This happens, of course, when $\Omega_{\rm true}$ is the union of star-shaped objects, but also for a wide range of 3D objects that are not star-shaped, like some ``bean'' or ``peanut''-like objects. More complicated shapes, for instance ``dolphin'' or ``aeroplane''-like ones cannot be well approximated by star-shaped parameterizations and the forthcoming method would fail. However, the ideas can be adapted to more general parameterizations \cite{Hohage3} to deal with this kind of objects. This extension is out of the scope of the paper.

Let us first assume that $\Omega_{\rm true}$ consists in just one component which can be reasonably well-approximated by a star-shaped object $\Omega_{\rm star}$. The extension to multiple components is straightforward and will be discussed at the end of this section. Points on the surface of $\Omega_{\rm star}$ are located at rays emerging from a point $\mathbf c_{\rm star}$, at distances of this point  which vary with  the angles and can be referred to a spherical coordinate  system by $\mathbf r_{\rm star}(\hat {\mathbf x}) = r_{\rm star}(\hat {\mathbf x}) \hat {\mathbf x}, $ $\hat {\mathbf x} ={\mathbf x \over |\mathbf  x|}\in \mathbb S^2$. The spherical harmonics \cite{coltonkress,kirsch} furnish a basis to expand functions defined in $L^2(\mathbb S^2)$ which we use to approximate $ r_{\rm star}(\mathbb S^2)$, see \ref{sec:sphere}. Let $\mathbb H_{n_{max}}$ be the finite dimensional space spanned by scalar spherical harmonics  with degree up to $n_{max}\in \mathbb N$.
Given a star-shaped approximation $\Omega_{\mathbf q_{\rm ap}}$ of
$\Omega_{\rm star}$ (namely, of $\Omega_{\rm true}$), we can describe its boundary as
\begin{eqnarray}\begin{array}{ll}
\mathbf q_{\rm ap}= \mathbf c_{\rm ap}+\boldsymbol \xi_{\rm ap}, &
\quad \mathbf c_{\rm ap}\in\mathbb R^3, \;
\boldsymbol \xi_{\rm ap}(\hat{\mathbf x})=  r_{\rm ap}(\hat{\mathbf x})
\hat{\mathbf x}, \\[1ex] & \quad
r_{\rm ap}(\hat{\mathbf x})=  \sum_{n=0}^{n_{max}}\sum_{m=-n}^{n}\gamma_{n,m}^{\rm ap} Y_{n,m}(\hat{\mathbf x}),
\label{starshaped}
\end{array}\end{eqnarray}
for $\hat{\mathbf x} \in \mathbb S^2.$ Our goal is to find $\mathbf c_{\rm ap}$ and $r_{\rm ap}$ in such a way that equation (\ref{eqinverse}) is satisfied by the star-shaped object $\Omega=\Omega_{\rm \mathbf q_{\rm ap}}$ defined by it. Gradient methods would seek to correct the approximate
parametrization $\mathbf q_{\rm ap}$ in the direction in which some kind of domain derivative of the functional (\ref{costH})  is negative. Gauss-Newton methods instead
aim to correct the parametrization by solving equation (\ref{eqinverse}) linearized at
the available approximation $\mathbf q_{\rm ap}$.
Both strategies may undergo stagnation in the direction of small  gradient.
To allow for convergence avoiding this artifact, regularizing terms are added  to
Gauss-Newton methods, as we have discussed in Section \ref{sec:irgngeneral}.

In our framework, starting from an approximate parametrization $\mathbf q_0$, 
the IRGNM solves the linearized equation
$$ {\cal I}(\mathbf q_{k})+{\cal I}'(\mathbf q_{k})\boldsymbol
\xi \approx I^{\eta}_{\rm meas} $$
by minimizing the regularized nonlinear least squares problem
\begin{eqnarray}\begin{array}{l}
\boldsymbol \xi_{k+1}=\underset{\boldsymbol \xi}{\operatorname{Argmin}}\big\| I^{\eta}_{\rm meas}-{\cal I}(\mathbf q_{k})-{\cal I}'(\mathbf q_{k})\boldsymbol \xi\big\|^2_{\rm 2}+\alpha_k \big\| \mathbf q_{k}+\boldsymbol \xi-\mathbf
q_{\rm ap}\big\|^2_{H^s(\mathbb S^2)},
\end{array} \label{xik+1} \end{eqnarray}
where $ \| \boldsymbol \xi \|^2_{H^s(\mathbb S^2)} = \sum_{n=0}^{n_{max}} \sum_{m=-n}^n (1+n^2)^s |\gamma_{n,m}|^2,$ with $\gamma_{m,n}$ as in (\ref{starshaped}). 
We choose $\alpha_k = \alpha_0 ({2 \over 3})^k$ to ensure logarithmic convergence \cite{Hohage},   and fix $\alpha_0=0.1$.
We  update the object shape by setting $\mathbf q_{k+1}=\mathbf q_{k}+\boldsymbol \xi_{k+1}$.  

As indicated in Section \ref{sec:irgngeneral}, the minimizer of
(\ref{xik+1}) is the unique solution of
\begin{eqnarray} \begin{array}{l}
\big( {\cal I}'(\mathbf q_{k})^* {\cal I}'(\mathbf q_{k}) + \alpha_k I \big)
\boldsymbol \xi_{k+1} =
 {\cal I}'(\mathbf q_{k})^* (I^{\eta}_{\rm meas} - {\cal I}(\mathbf q_{k})) +
 \alpha_k (\mathbf q_{\rm ap} - \mathbf q_{k}),
\end{array} \label{xik+1bis}
\end{eqnarray}
where ${\cal I}'(\mathbf q_{k})^*$ represents the adjoint of the Fr\'echet derivative
${\cal I}'(\mathbf q_{k})$.

Let us detail explicit expressions for both operators.
Admissible perturbations $\mathbf q \in {\cal Q}$ of a given spherical
parametrization $\mathbf q_{\rm ap}$ of an initial guess star-shaped with
respect to a fixed center $\mathbf c_{\rm ap}$ take the form:
\begin{eqnarray} \begin{array}{l}
\mathbf q(\boldsymbol \xi) =  \mathbf q_{\rm ap} + \boldsymbol \xi, \quad
\boldsymbol \xi(\hat{\mathbf x})  =   r (\hat{\mathbf x})
\hat{\mathbf x},
\end{array} \label{admissible}
\end{eqnarray}
and define deformed domains $\Omega_{\mathbf q(\boldsymbol \xi)}$ such that $\partial\Omega_{\mathbf q(\boldsymbol \xi)}:= \mathbf c_{\rm ap}+\boldsymbol \xi_{\rm ap}(\mathbb{S}^2) + \boldsymbol \xi(\mathbb{S}^2).$ In this way, the function $\boldsymbol\xi$ is uniquely determined by $\partial\Omega_{\mathbf q(\boldsymbol \xi)}$. As argued in Section \ref{sec:frechet}, the Fr\'echet derivative of the hologram admits the explicit formula
\begin{eqnarray}
{\cal I}'(\mathbf q_{\rm ap}) \boldsymbol \xi = \begin{pmatrix}2 {\rm Re} \left[
\overline{\mathbf  E_{\Omega_{\rm ap}}(\mathbf x_1)} \cdot
\dot{\mathbf E}(\mathbf x_1)\right]\\\vdots\\2 {\rm Re} \left[
\overline{\mathbf  E_{\Omega_{\rm ap}}(\mathbf x_N)} \cdot
\dot{\mathbf E}(\mathbf x_N)\right]\end{pmatrix},
 \label{frechetholo}
\end{eqnarray}
where  ${\bf x}_1,\dots,{\bf x}_N$ are the screen detectors, and $\dot {\mathbf E}$ is the solution of
\begin{eqnarray} \label{characterization}
\begin{array}{ll}
\mathbf{curl} \,  (\mathbf{curl} \,  \dot{\mathbf E})  - k_e^2 \dot{\mathbf E} =0 & \quad \mbox{in $\mathbb R^3\setminus\overline{\Omega}$}, \\
\mathbf{curl} \,  (\mathbf{curl} \,  \dot{\mathbf E})  - k_i^2 \dot{\mathbf E} =0 & \quad \mbox{in $\Omega$}, \\
\hat{\mathbf n}  \times \dot{\mathbf E}^+
- \hat{\mathbf n} \times \dot{\mathbf E}^-
=g_D & \quad\mbox{on $\partial \Omega$},\\
 \hat{\mathbf n} \times \mathbf{curl} \,  \dot{\mathbf E}^+ -
\beta \, \hat{\mathbf n} \times \mathbf{curl} \,  \dot{\mathbf E}^-
=g_N & \quad\mbox{on $\partial \Omega$}, \\
{\rm lim}_{|\mathbf x| \rightarrow \infty} |\mathbf x|  \big|
\mathbf{curl} \,  \dot{\mathbf E} \times {\mathbf x \over |\mathbf x|}
-\imath k_e \dot{\mathbf E} \big| =0, &
\end{array}
\end{eqnarray}
with object $\Omega=\Omega_{\rm ap}$ and transmission data
\begin{eqnarray}
g_D&=& - ({\boldsymbol \xi} \cdot \hat{\mathbf n}) \left(
\hat{\mathbf n} \times \mathbf{curl} \,  \mathbf E^+
-  \hat{\mathbf n} \times \mathbf{curl} \,  \mathbf E^- \right)
\times \hat{\mathbf n} \nonumber \\
&& - \hat{\mathbf n} \times \nabla  \left( ({\boldsymbol \xi}
 \cdot \hat{\mathbf n}) (\hat{\mathbf n} \cdot \mathbf E^+
- \hat{\mathbf n} \cdot \mathbf E^-)  \right), \label{gD} \\
g_N&=& - ({\boldsymbol \xi} \cdot \hat{\mathbf n})
\left( k_e^2  \hat{\mathbf n} \times  \mathbf E^+
- k_i^2 \beta \, \hat{\mathbf n} \times \mathbf E^- \right) \times \hat{\mathbf n} \nonumber \\
&& - \hat{\mathbf n} \times \nabla \left( ({\boldsymbol \xi} \cdot \hat{\mathbf n})
\hat{\mathbf n} \cdot (\mathbf{curl} \,  \mathbf E^+ -
\beta  \, \mathbf{curl} \,  \mathbf E^-)
\right), \label{gN}
\end{eqnarray}
$\mathbf E= \mathbf E_{\Omega_{\rm ap}}$ being the solution of
(\ref{forwardadim}) with $\Omega= \Omega_{\rm ap}$ and $\hat{\mathbf n}$ the outer unit normal. Notice that this solution is continuously differentiable in a classical sense far from the obstacles \cite{ola}.

Let us characterize now the adjoint. Let $\boldsymbol h$ be a real-valued  function defined
on the screen detectors {$\mathbf x_{1},\hdots,\mathbf x_{N}$. The $L^2$-adjoint operator of ${\cal I}'(\mathbf q_{\rm ap})$ is defined  by
\begin{eqnarray}
\boldsymbol\xi^*(\hat{\mathbf x})=  r^* (\hat{\mathbf x}) \hat{\mathbf x}={\cal I}'(\mathbf q_{\rm ap})_{|L^2}^*\boldsymbol h, \label{adjoint1}
\end{eqnarray}
and
\begin{eqnarray}
\begin{aligned} r^*=\hat{\mathbf x}\cdot{\cal I}'(\mathbf q_{\rm ap})_{|L^2}^*\boldsymbol h=&\; r^2_{\rm ap}{\rm Re}\left[-(1-\beta^{-1})\hat{\mathbf n} \times \mathbf{curl}_{\mathbf x} \,  \overline{\mathbf E^+_{\boldsymbol h}}\cdot \hat{\mathbf n} \times \mathbf{curl}_{\mathbf x} \,  \overline{\mathbf E^+}\right.\\
&\left. \hspace{1cm}
+(\frac{1}{k_{e}^2}-\frac{1}{\beta k_{i}^2}){\rm div}_{\partial\Omega}\big(\hat{\mathbf n} \times \mathbf{curl} \,  \overline{\mathbf E^+_{\boldsymbol h}}\big)\, {\rm div}_{\partial\Omega}\big(\hat{\mathbf n} \times \mathbf{curl} \,  \overline{\mathbf E^+}\big)\right. \\
& \left.\hspace{1cm}
-({k_{e}^2}-{\beta k_{i}^2})\hat{\mathbf n} \times  \overline{\mathbf E^+_{\boldsymbol h}}\cdot \hat{\mathbf n} \times \, \overline{ \mathbf E^+}\right. \\
&\left.\hspace{1cm}+(1-\beta){\rm div}_{\partial\Omega}\big( \hat{\mathbf n} \times  \overline{\mathbf E^+_{\boldsymbol h}}\big) \,{\rm div}_{\partial\Omega}\big( \hat{\mathbf n} \times \, \overline{ \mathbf E^+}\big)\right] \circ\mathbf q_{\rm ap}, \end{aligned} \label{adjoint2}
\end{eqnarray}
where ${\mathbf E^+_{\boldsymbol h}}$ is the solution of the transmission problem with object $\Omega=\Omega_{\rm ap}$ and incident field
\begin{eqnarray}
\mathbf E^{inc}_{\boldsymbol h}(\mathbf x)=
\frac{1}{k_e^2}
\mathbf{curl}_{\mathbf x}\, \mathbf{curl}_{\mathbf x} \sum_{j=1}^N
\frac{e^{\imath k_e |\mathbf x-\mathbf x_j|}}{4\pi |\mathbf x- \mathbf x_j|}
2\boldsymbol h(\mathbf x_j) \overline{\mathbf E_{\Omega_{\rm ap}}(\mathbf x_j)}.
\label{adjoint2bis}
\end{eqnarray}
The IRGN method has never been used with intensities in electromagnetism before. Formula (\ref{adjoint2}) is the same as that derived in \cite{lelouerspectral}
when working with measurements of the full field.  However, the incident adjoint field 
$\mathbf E^{inc}_{\boldsymbol h}$ changes and it is given by (\ref{adjoint2bis}). To obtain the adjoint of $ {\cal I}'(\mathbf q_{\rm ap})$ in $H^s$ it suffices to compose ${\cal I}'(\mathbf q_{\rm ap})_{|L^2}^*$ with the diagonal operator
$$j_{L^2\to H^s}( \boldsymbol \xi^*)=\sum_{n=0}^{n_{max}}\sum_{m=-n}^n(1+n^2)^s\langle r^*,Y_{nm}\rangle_{L^2}Y_{nm}\,\hat{\mathbf x},$$
where $\langle \phi_{1},\phi_{2}\rangle_{L^2}=\int_{\partial\Omega}\phi_{1}\cdot\overline{\phi_{2}}$. We then have
\begin{eqnarray}
{\cal I}'(\mathbf q_{\rm ap})^*=j_{L^2\to H^s}\circ {\cal I}'(\mathbf q_{\rm ap})^*_{|L^2},
\label{adjoint3}
\end{eqnarray}
see Appendix \ref{sec:frechet} for details. 
Formulas (\ref{gD})-(\ref{gN}) can also be simplified by using only $\mathbf E^+$, as done in \cite{lelouerspectral}, Remark 5, thanks to the transmission conditions.
These explicit expressions for the Fr\'echet derivatives and their adjoints
allow us to solve numerically equation (\ref{xik+1bis})  in $H^s(\mathbb{S}^2)$
as explained in Section \ref{sec:numerical}.

This idea can be generalized to multiple simply connected domains just by considering a set of parametrizations $\mathbf q = (\mathbf q_\ell)_{\ell=1}^{\mathcal L_{\rm ap}}$ for the union of the boundaries of the components $\Omega_{\rm ap, \ell}$ of $\Omega_{\rm ap}.$
Notice that the procedure we have just described fixes the approximate center of the object $\mathbf c_{\rm ap}$ and seeks for an adequate spherical parametrization about it. We
correct the centers by a simple procedure embedded in the algorithm described in the next section.

\section{Hybrid Topological derivative/Gauss-Newton algorithm}
\label{sec:algorithm}

In general, we ignore the number of components of the holographied object 
a priori. Therefore a scheme should allow to create, 
merge and destroy contours at certain stages.
In this section, we propose a hybrid scheme combining topological derivative based optimization with iteratively regularized Gauss-Newton iterations for inverting holographic data which achieves that goal.  The algorithm proceeds in the following steps. \\

\noindent{\it Step 1 - Observation region. } We define a meshed region $\mathcal{R}_{\rm obs}$ where we seek the objects and evaluate the different fields. In our
set-up, we assume the screen is the square $[0,10]\times[0,10]\times \{10\}$.
Considering $\mathcal{R}_{\rm obs} =  [d,10-d] \subset
[0,10]^3$,  the 3D mesh is formed by the $(M_d+1)^3$ points $\mathbf x = (x_{1,m_1},x_{2,m_2},x_{3,m_3})$, where  $x_{i,m_i}= d + m_i {{10-2 d} \over M_d} $, $m_i=0,\ldots,M_d$, $i=1,2,3$.  \\

\noindent{\it Step 2 - Initial Guess.}
\begin{enumerate}
\item[(i)] Choose $0<C_{0}<1$ and define $\Omega_{\rm ap}$  as in (\ref{initialguesstd_impl}) where $D_{T}(\mathbf x,\mathbb R^3)$ is given by (\ref{DT}) with $\beta=1$ and (\ref{forwardempty})-(\ref{adjointexplicit}). Unless otherwise stated, in all our tests we set $C_0=0.15$ for the red light and $C_0=0.2$ for the violet one.

\item[(ii)] For all simply connected components $\Omega_{{\rm ap},\ell}$ of
$\Omega_{\rm ap}$, compute the centroid $\mathbf c_{{\rm ap},\ell}\in\mathbb R^3$ and the boundary points $\mathbf c_{{\rm ap},\ell}+ \mathbf r_{{\rm ap}, \ell}$.

\item[(iii)] Determine  star-shaped parametrizations for
$\mathbf r_{{\rm ap},\ell}(\hat{\mathbf x})= r_{{\rm ap},\ell}(\hat{\mathbf x}) \hat{\mathbf x}$ finding the coefficients $\gamma_{n,m}= \int_{\mathbb S^2} r_{{\rm ap},\ell}(\hat{\mathbf x}) \overline{Y_{n,m}(\hat{\mathbf x})} dS_{\hat{\mathbf x}} $
in the expansion (\ref{starshaped}) by means of the exact quadrature rule (2.42)
in \cite{ganesh} after interpolating $r_{{\rm ap},\ell}(\hat{\mathbf x})$ to obtain its values at the Gauss quadrature points, for each component.
\end{enumerate}

We can store the result of checking the condition (i) on the topological derivative at each grid point as a binary  vector ${\tt M}$ of  $(M_d+1)^3$  entries. In practical implementations, this vector will be longer. After obtaining $\Omega_{\rm ap}$ as described in Step 2(i), we consider a small cube around each simple connected component. This cube is again meshed and the topological derivative (TD) is evaluated at these points. In the end, the domain $\Omega_{\rm ap}$ is defined by the set of all the points in this finer grid where the TD satisfies (\ref{initialguesstd_impl}). 

For (ii), the MATLAB command {\tt CC = bwconncomp(M)} extracts the number of connected components and labels the
points within them. The command {\tt boundary} extracts the boundary points of each component. The property {\tt centroid} of the command {\tt regionprops} provides the centroids. See also \cite{lelouerrapun} for simple MATLAB instructions regarding (iii). \\

As a result, we obtain an approximation $\Omega_{0}=\Omega_{\rm ap}$ with ${\mathcal L_{\rm ap}}$  components $\Omega_{0,\ell}=\Omega_{\rm ap,\ell}$ parameterized by $\mathbf q_{0,\ell}
= \mathbf q_{\rm ap,\ell} = \mathbf c_{\rm ap,\ell} + \mathbf r_{\rm ap,\ell}
= \mathbf c_{0,\ell} + \mathbf r_{0,\ell}$, $\ell=1,\ldots,  {\mathcal L_{\rm ap}}$ (in principle, ${\cal L_{\rm ap}}$ does not coincide with the true number of objects ${\mathcal L}$). Then, for each iteration $k \geq 1$: \\

\noindent{\it Step $2k+1$ - Shape Correction.}
\begin{enumerate}
\item[(i)] Update the shape of each connected component $\Omega_{k,\ell}$
of $\Omega_{k}$ by computing the solution $\boldsymbol \xi_{k+1,\ell}$ of
(\ref{xik+1})-(\ref{xik+1bis}) and setting $\mathbf q_{k+1,\ell}=  \mathbf q_{k,\ell}
+ \boldsymbol \xi_{k+1,\ell}$ for $\ell=1,\ldots,{\mathcal L}_{\rm ap}.$
\item[(ii)] If any of the centers ${\bf c}_{k,\ell}$ is far from the gravity center
of the corresponding component  $\Omega_{k+1,\ell}$ (i.e. the distance
is larger than $10\%$ of the diameter), then we set ${\bf c}_{k+1,\ell}$ equal
to the gravity center and replace $\mathbf q_{k+1,\ell}$ with the parametrization
that uses that center, as indicated in Step 2 (ii)-(iii). Otherwise, set ${\bf c}_{k+1,\ell}
= {\bf c}_{k,\ell}.$
\item[(iii)] Solve the forward problem (\ref{forwardomega}) with $\Omega=\Omega_{k+1}$ and evaluate the solution $\mathbf E_{\Omega_{k+1}}$ and the corresponding
hologram $ I_{\Omega_{k+1}}$ at the detectors $\mathbf x_{1},\hdots,\mathbf x_{N}$.
\end{enumerate}

\noindent{\it Step $2k+2$ - Stop or modify contours.}
\begin{enumerate}
\item[(i)]  We estimate the noise in the measured hologram
$\delta= \|I_{\rm meas}^{\eta} - I_{\rm meas} \|_{2}$ and select as stopping criterion the discrepancy principle. If
\begin{eqnarray}
\|I_{\Omega_{k+1}}-I^{\eta}_{\rm meas}\|_{2}
\leq \tau \delta,
\label{stop}
\end{eqnarray}
we end the algorithm. To ensure an accurate approximation preventing early stops,
we set $\tau=1.01$.
\item[(ii)] If the cost functional stagnates without fulfilling the
stopping criteria, i.e.,
$$ \left| \sqrt{J(\mathbb R^3\backslash\overline{\Omega_{k+1}})} - \sqrt{J(\mathbb R^3\backslash\overline{\Omega_{k}})} \right|<\frac{\delta}{5} \quad\text{ whereas }
\quad\|I_{\Omega_{k+1}}-I^{\eta}_{\rm meas}\|_{2}> 5  \delta$$
then we  automatically create  and/or merge and/or destroy components using the topological derivative:
\begin{enumerate}
\item[(a)] Solve the adjoint problem (\ref{adjointomega}) with
$\Omega=\Omega_{k+1}$, $\mathbf E=\mathbf E_{\Omega_{k+1}}$  and
compute $D_{T}({\bf x},\mathbb R^3 \setminus \overline{\Omega}_{k+1})$ using (\ref{DTsimple}) for all ${\bf x}\in\mathcal{R}_{\rm obs}\setminus\overline{\Omega}_{k+1}$.
\item[(b)] Replace $\Omega_{k+1}$ by
\begin{eqnarray}\begin{array}{lll}
\Omega_{\rm new} &= &\Omega_{k+1}\cup \Big\{\mathbf x\in \mathcal{R}_{\rm obs}\setminus\overline{\Omega}_{k+1}\, \big|
\, D_{T}(\mathbf x,\mathbb R^3\backslash\overline{\Omega}_{k+1}) <  \\
&& \hskip 14mm (1-C_{0})\min_{\mathbf y \in \mathcal{R}_{obs} 
\setminus\overline{\Omega}_{k+1}}
D_{T}(\mathbf y,\mathbb R^3\backslash\overline{\Omega}_{k+1})\Big\}.
\end{array} \label{updatedguess2td_impl}
\end{eqnarray}
If we wish to allow for the destruction of existing components
too, then we use (\ref{updatedguess2td}) to generate
$\Omega_{\rm new}$ from $\Omega_{\rm ap}=\Omega_{k+1}.$
\item[(c)] Go to Step 2 (ii) and repeat the procedure with
$\Omega_{\rm ap} = \Omega_{\rm new}.$
\end{enumerate}
\item[(iii)] Otherwise, go to Step $2k+1$ (i) and repeat both Steps now for $\Omega_{k+1}$.
\end{enumerate}
We solve the problem (\ref{xik+1}) for the  parameterization correctors and the forward and adjoint problems for the topological derivatives (\ref{forwardomega}) and (\ref{adjointomega}) using the methods described in Section \ref{sec:numerical}.

We have tested this algorithm in the set-up depicted in Fig. \ref{fig1}(b).
After non-dimensionalizing as indicated in Section \ref{sec:inverse}, the detectors
are placed on the screen $[0,10]\times[0,10]\times \{10\}$. The holograms are synthetically generated on that screen on a grid with $51\times 51$ detectors by solving numerically 
the corresponding forward problems and adding  random noise of magnitude $2 \%$, i.e.,
such that (\ref{eta}) holds for $\eta=0.02$.
Detectors are located at the points ${\bf x}_{k\ell}=(0.2k,0.2\ell,10)$, $k,\ell=0,\dots,50$.
The direction of the incident light  is $(0,0,1)$  and the polarization vector is (1,0,0). We fix the refractive indexes of the objects, so that $k_e=12.56$ and $k_i=15.12$ for a red incident light of $660$ nm and  $k_e=20.6$, $k_i=24.79$ for a violet incident light of $405$ nm.
As said earlier, $\beta=1$. We keep the default values for the parameters $C_0=0.15$ (red light), $0.2$ (violet light), $C_1=c_1=C_0$, $\alpha_0=0.1$ and $\tau=1.01$ indicated in the description of the algorithm, unless stated otherwise. We have fixed in all our tests $n_{max}=8$ for the definition of the parameterizations (\ref{starshaped}). 

\begin{figure}[h!]
\centering
\hskip 0.0cm (a) \hskip 3.25cm (b) \hskip 3.25cm (c) \\
\includegraphics[width=4cm]{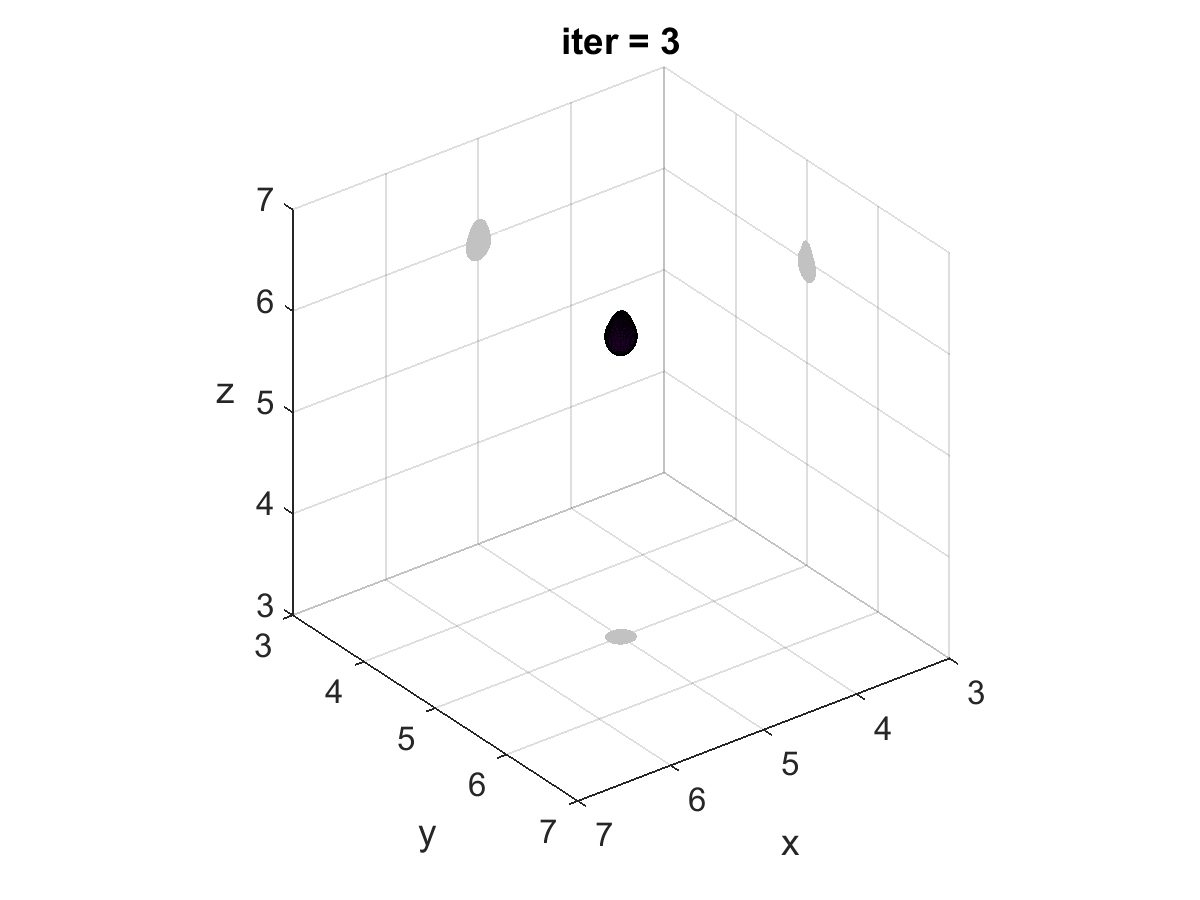}\hskip -2mm
\includegraphics[width=4cm]{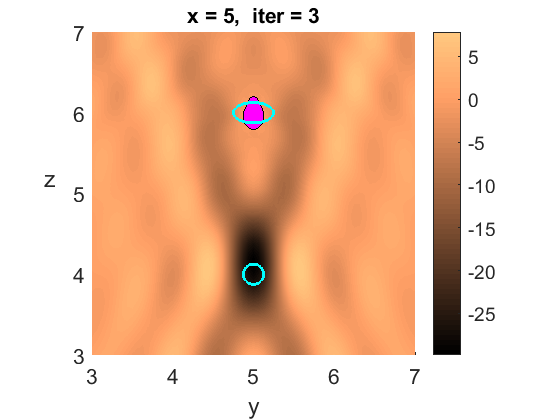}\hskip -2mm
\includegraphics[width=4cm]{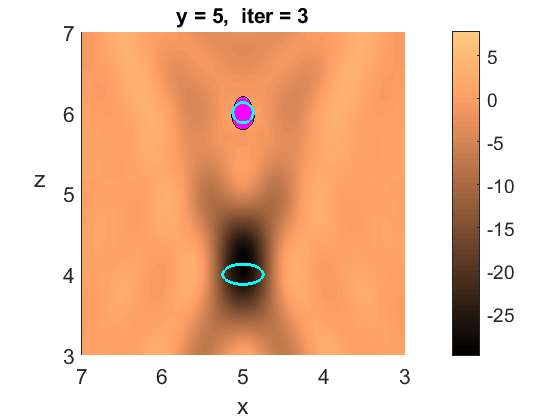} \\
\hskip 0.0cm (d) \hskip 3.25cm (e) \hskip 3.25cm (f) \\
\includegraphics[width=4cm]{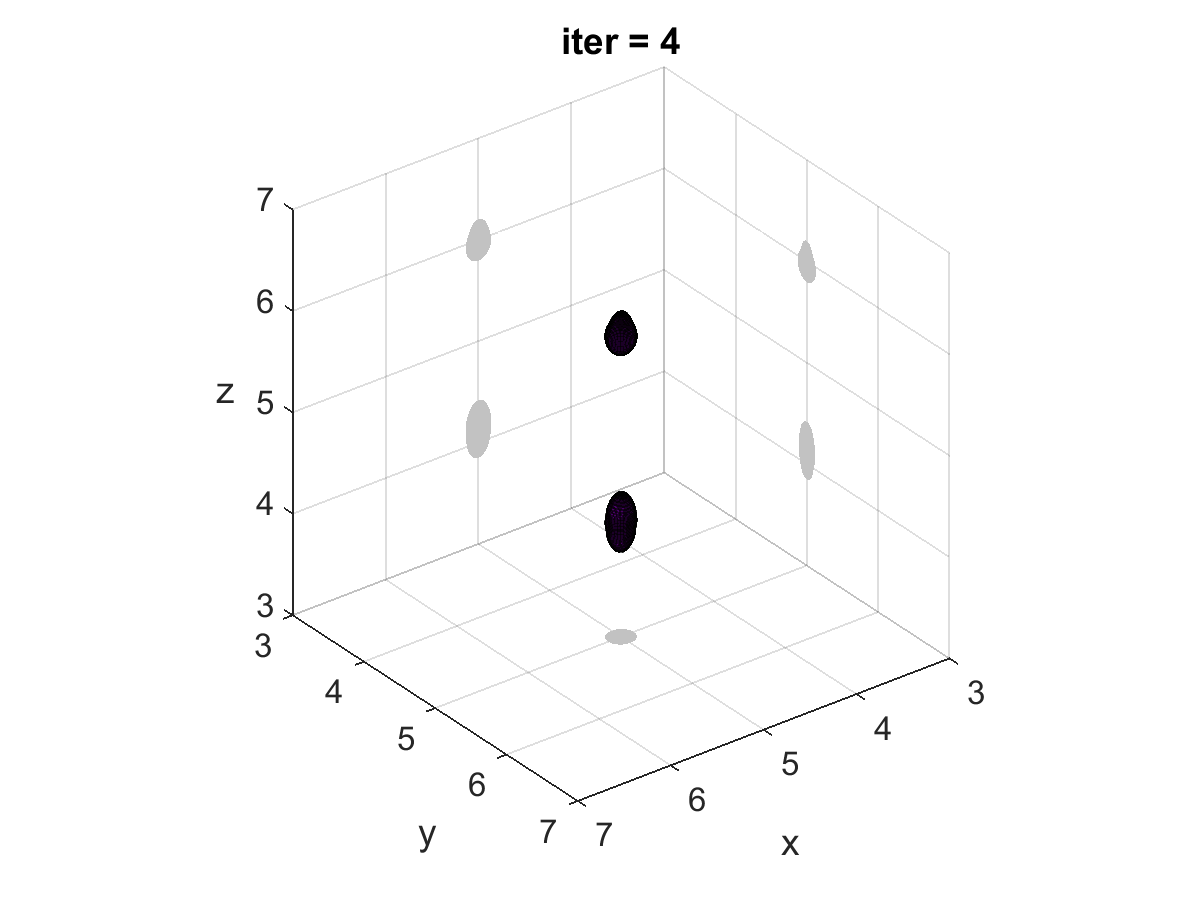} \hskip -2mm
\includegraphics[width=4cm]{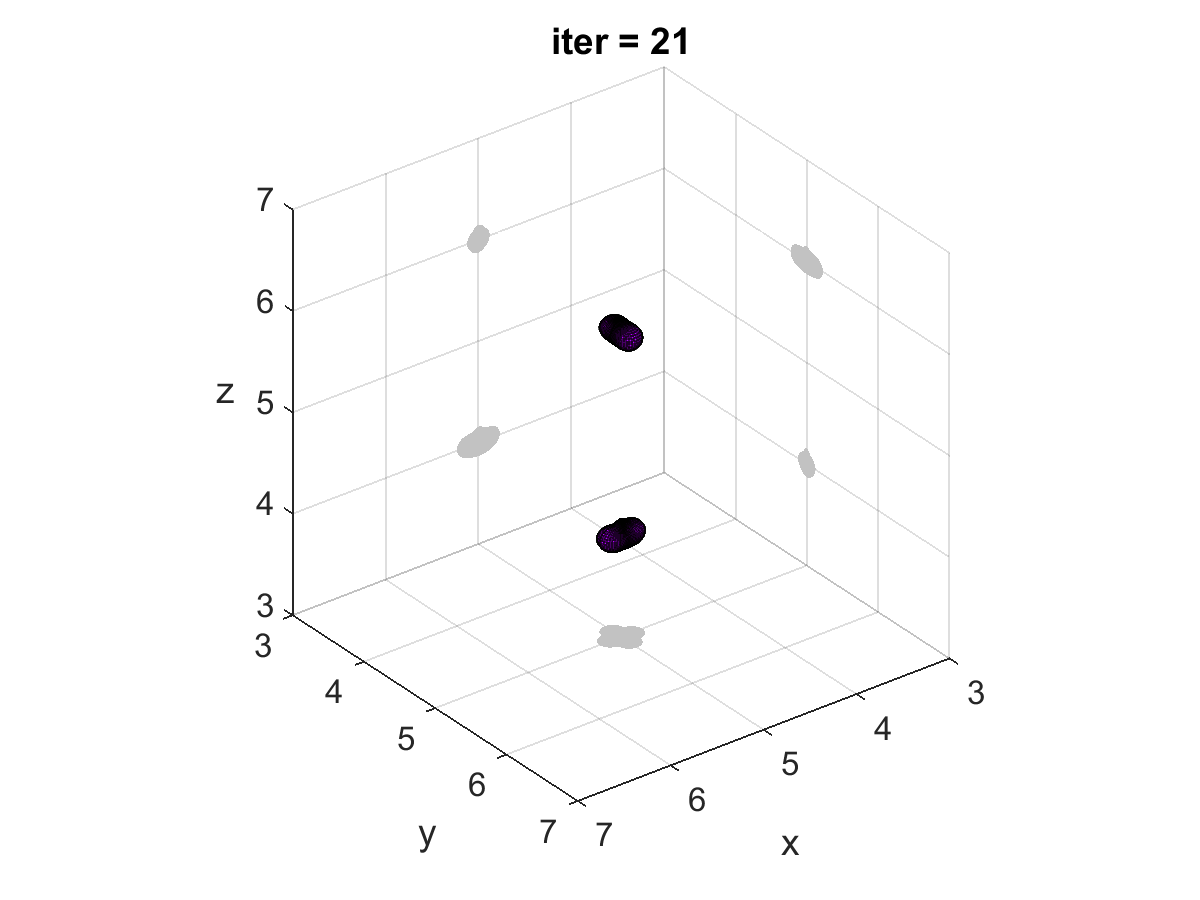} \hskip -2mm
\includegraphics[width=4cm]{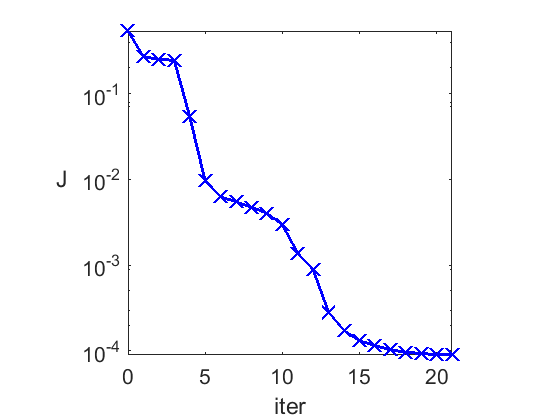}
\caption{ \small
Evolution of the initial guess represented in Fig. \ref{fig4}(d) for the same data
using the hybrid TD/IRGN algorithm described in Section  \ref{sec:algorithm}:
(a) Approximate object $\Omega_3$ obtained by the IRGNM.
(b) and (c) Slices $x=5$ and $y=5$ of the topological derivative (\ref{DTsimple}) with forward and adjoint fields given by (\ref{forwardomega})-(\ref{adjointomega}) when $\Omega=\Omega_3$. Cyan contours represent the true object, whereas the approximate object section is shown in magenta.
(d) Approximate object $\Omega_4$ obtained from $\Omega_3$ using the
TD through (\ref{updatedguess2td_impl}). A new component is detected.
(e) Final approximate object $\Omega_{21}$ generated by the IRGNM.
The true objects are two ellipsoids depicted in Fig. \ref{fig4}(a).
The orientation of the ellipsoids, the location of the center and the length of the
semi-axes are captured.
(f) Evolution of the error functional (\ref{costH}) during the iterative procedure plotted in logarithmic scale. The stagnation at iterations $1$-$3$  disappears when
the second component is introduced at iteration $4$ allowing for convergence.}
\label{fig5}
\end{figure}

\begin{figure}[h!]
\centering
\hskip 0.0cm (a) \hskip 3.25cm (b) \hskip 3.25cm (c) \\
\includegraphics[width=4cm]{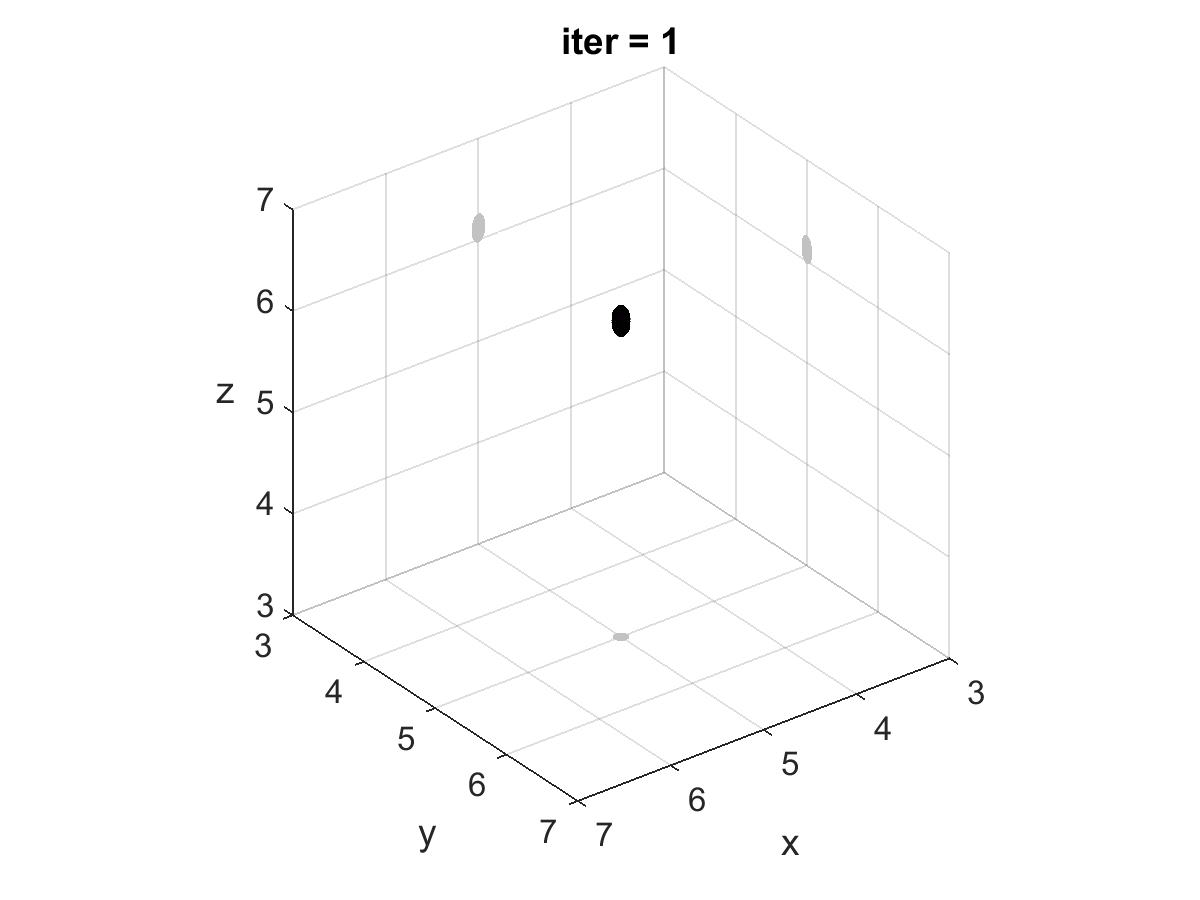}\hskip -2mm
\includegraphics[width=4cm]{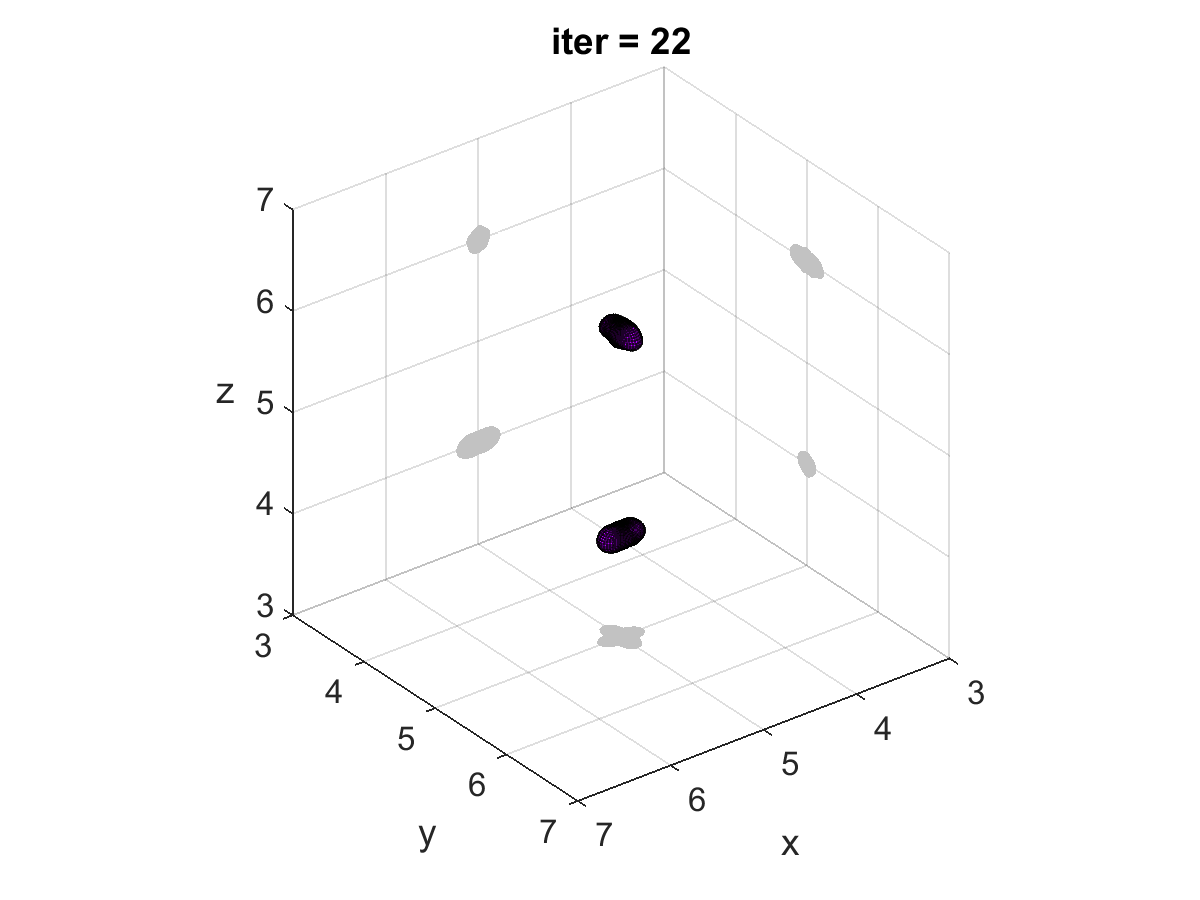}\hskip -2mm
\includegraphics[width=4cm]{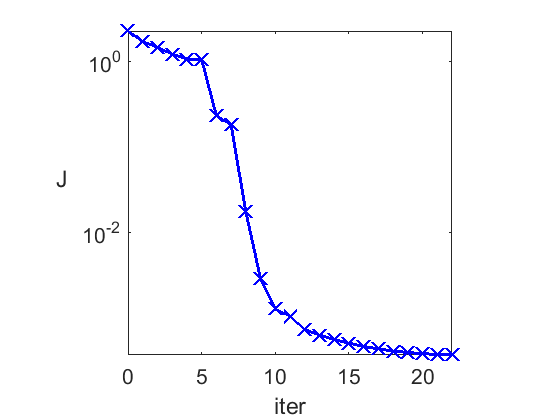} \\
\hskip 0.0cm (d) \hskip 3.25cm (e) \hskip 3.25cm (f) \\
\includegraphics[width=4cm]{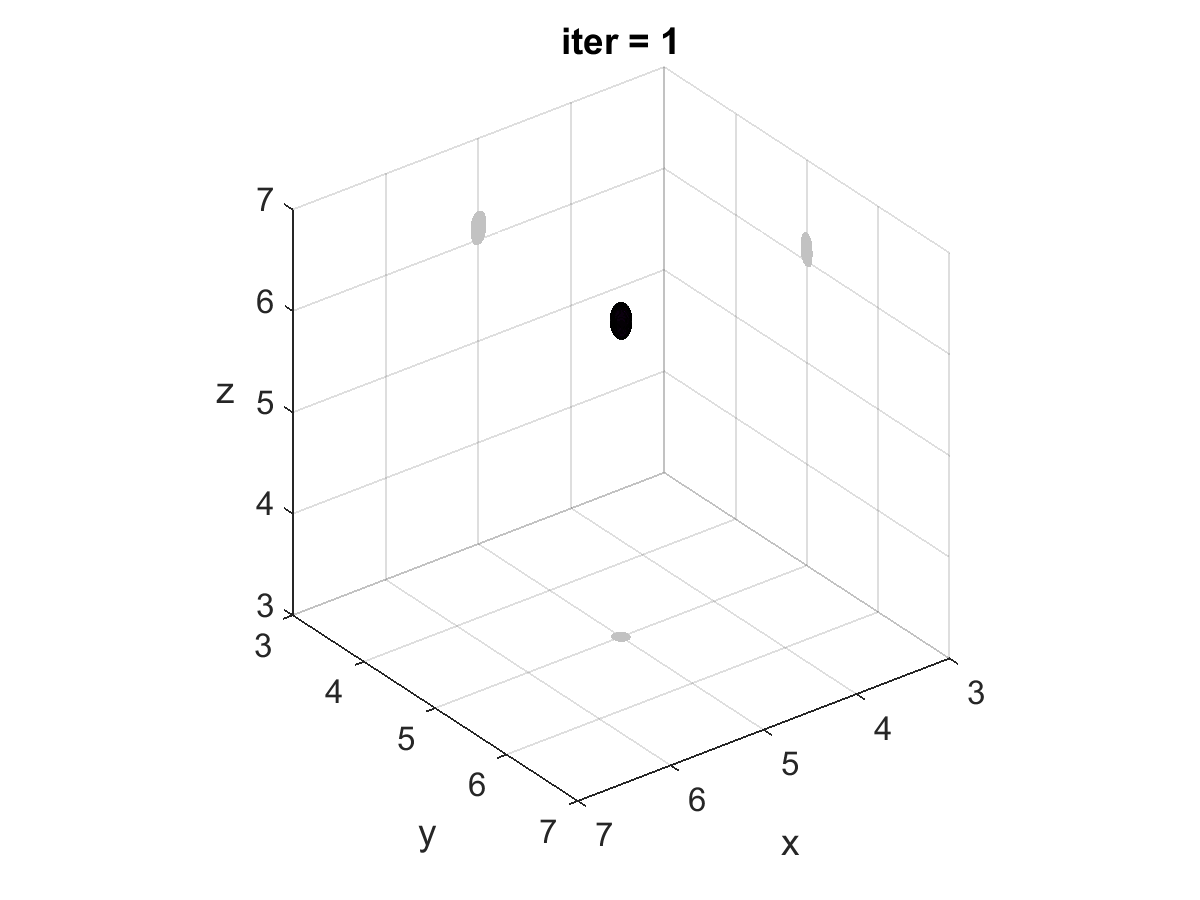} \hskip -2mm
\includegraphics[width=4cm]{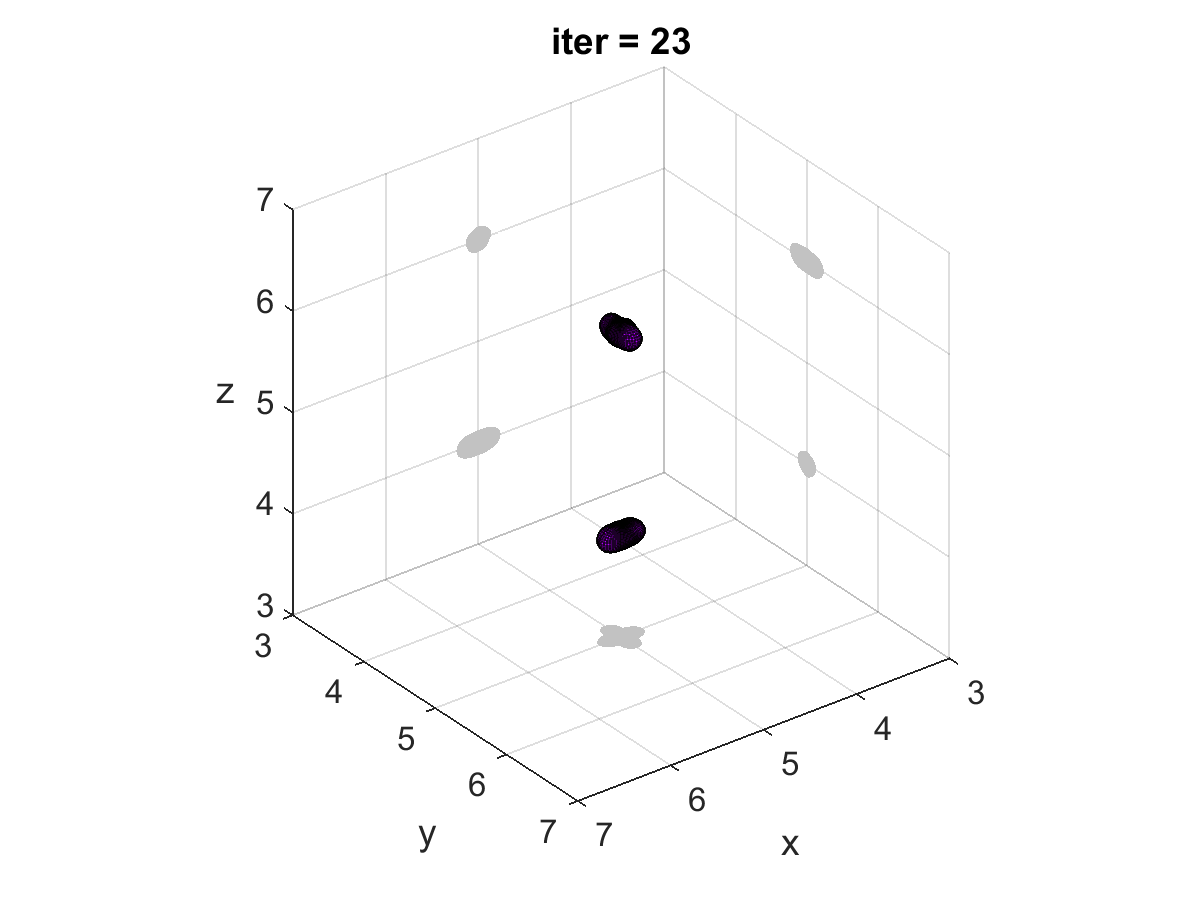} \hskip -2mm
\includegraphics[width=4cm]{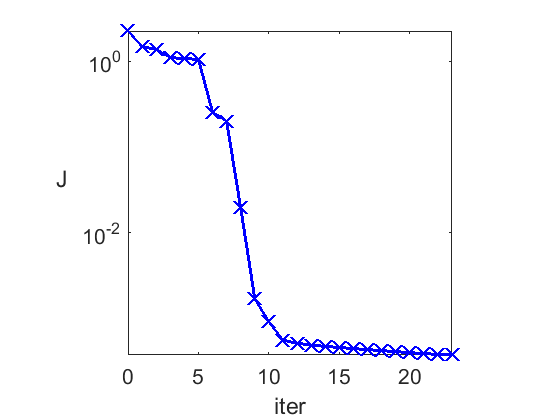} \\
\hskip 0.0cm (g) \hskip 3.25cm (h) \hskip 3.25cm (i) \\
\includegraphics[width=4cm]{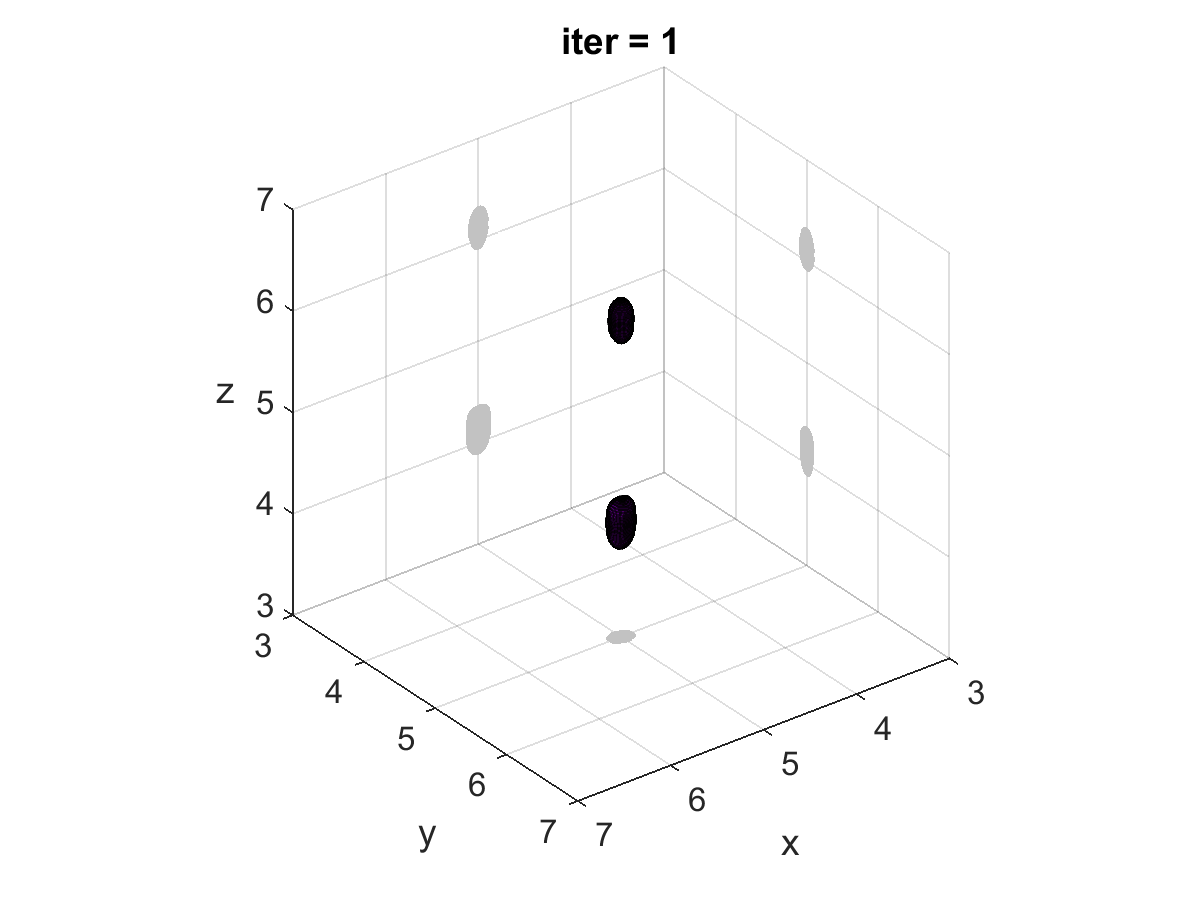} \hskip -2mm
\includegraphics[width=4cm]{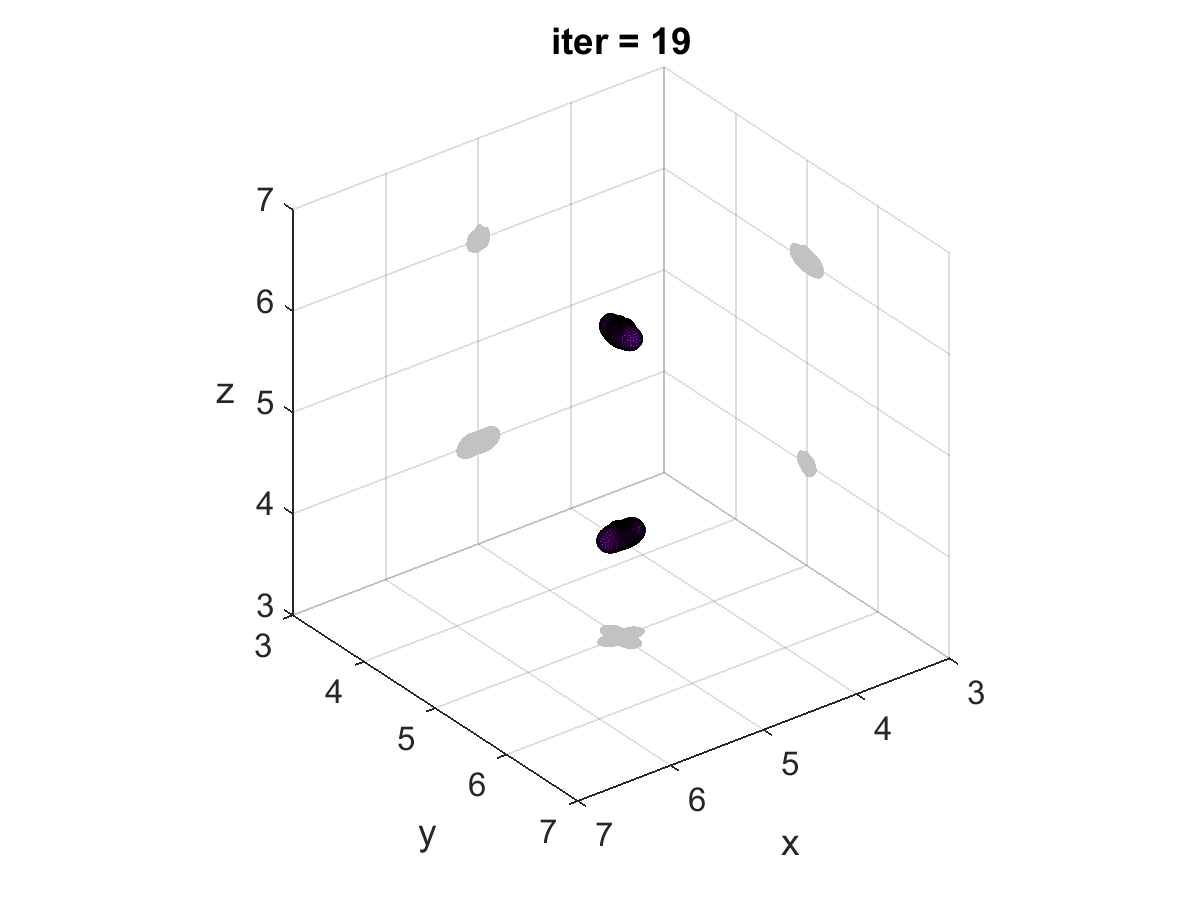} \hskip -2mm
\includegraphics[width=4cm]{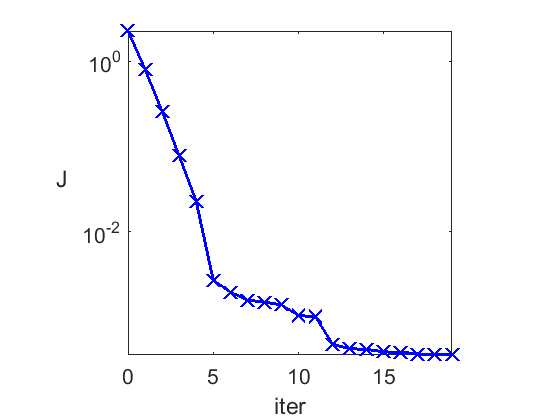}
\caption{ \small
Application of the hybrid TD/IRGN algorithm to the set-up shown in Fig. \ref{fig4}(a)
switching to violet light ($k_e=20.6$ and $k_i=24.79$) and varying the constant
defining the initial guesses: (a,b,c) $C_0=0.15$, (d,e,f) $C_0=0.2$, (g,h,i) $C_0=0.3$. 
The first column represents initial guesses, the second column displays final approximations
and the third column plots the evolution of the cost functional.}
\label{fig5_bis}
\end{figure}

Figure \ref{fig5} revisits the approximation obtained in Figure \ref{fig4} with this technique keeping the parameters for red light. The location, size and shape of the two object components is now recovered with accuracy, setting $\Omega_{\rm ap}$ equal to the initial approximation in Fig. \ref{fig4}(d). The cost functional (\ref{costH}), which stagnated during the iterative procedure proposed in Section \ref{sec:tditeration}, decreases now after the introduction of the second component, as the location, size and orientation of both components improves. Video 1 in the Supplementary Material reproduces the whole sequence of approximations. In Figure \ref{fig5_bis} we repeat the tests for the violet light. We observe that our method is very robust with respect to the choice of the threshold $C_0$  in (\ref{initialguesstd_impl}) and (\ref{updatedguess2td_impl}). It is able to find very accurate reconstructions even when the initial guess has only one component which is rather small (panels (a,b,c)), when it has only one component comparable in size with the objects to be found (panels (d,e,f)), and when the initial guess has the correct number of contours, but sizes are overestimated and orientations are incorrect (panels (g,h,i)). Comparing the decay of the cost functional for the three selected values, we find that for $C_0=0.3$ the cost functional decays rapidly during the first iterations because the initial guess has the correct number of components, and the number of iterations is slightly smaller because stagnation only occurs at the end of the procedure. Remarkably, when starting with a wrong number of components (panels (c,f)),  the cost functional stagnates at the 5th iteration and the number of components is updated.  At about the 10th iteration, the value of the cost function is almost the same for the three situations. Moreover, we have observed that a further increase in $C_0$ promotes initial guesses for which the cost functional increases, so that the method automatically reduces such constants. In the sequel we will set $C_0=0.2$ for the violet light.

\begin{figure}
\centering
\hskip 0.0cm (a) \hskip 3.25cm (b) \hskip 3.25cm (c) \\
\includegraphics[width=4cm]{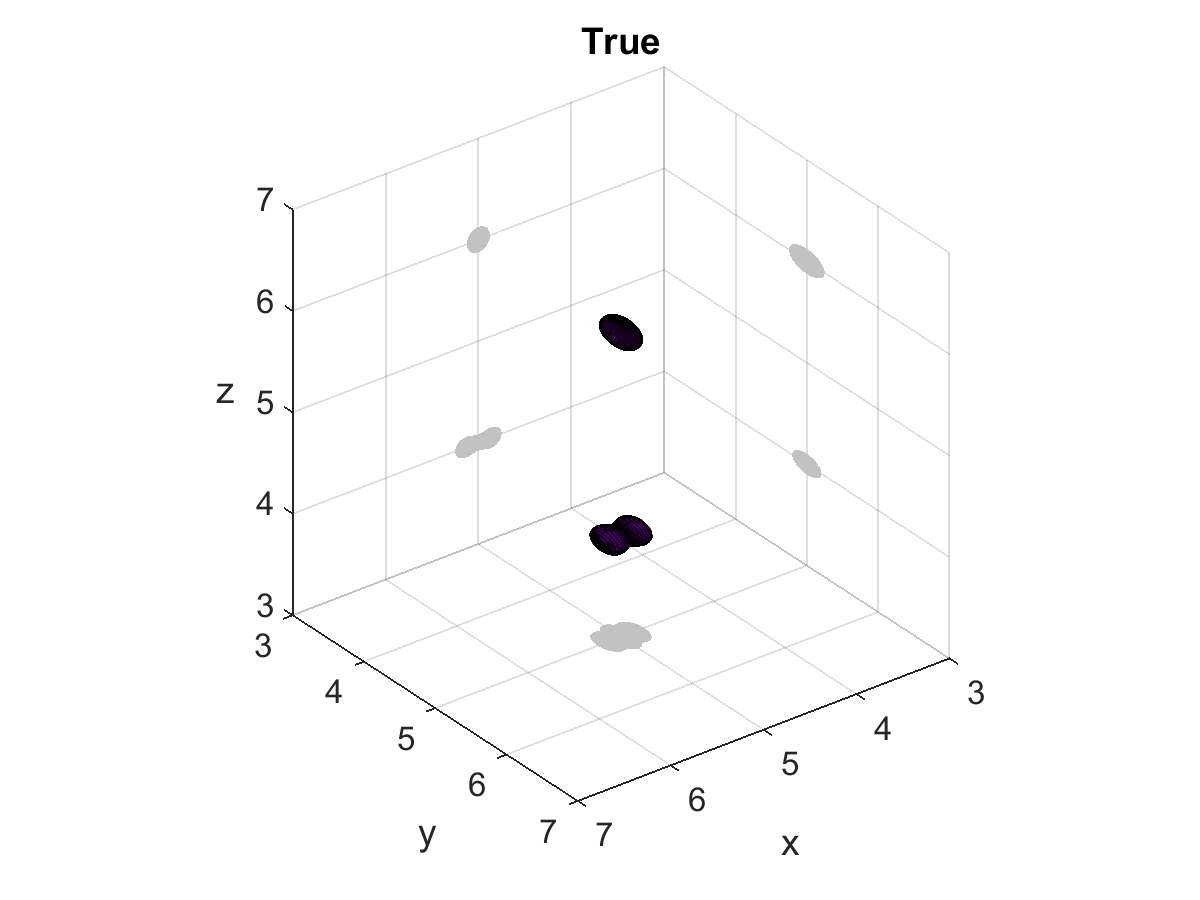} \hskip -2mm
\includegraphics[width=4cm]{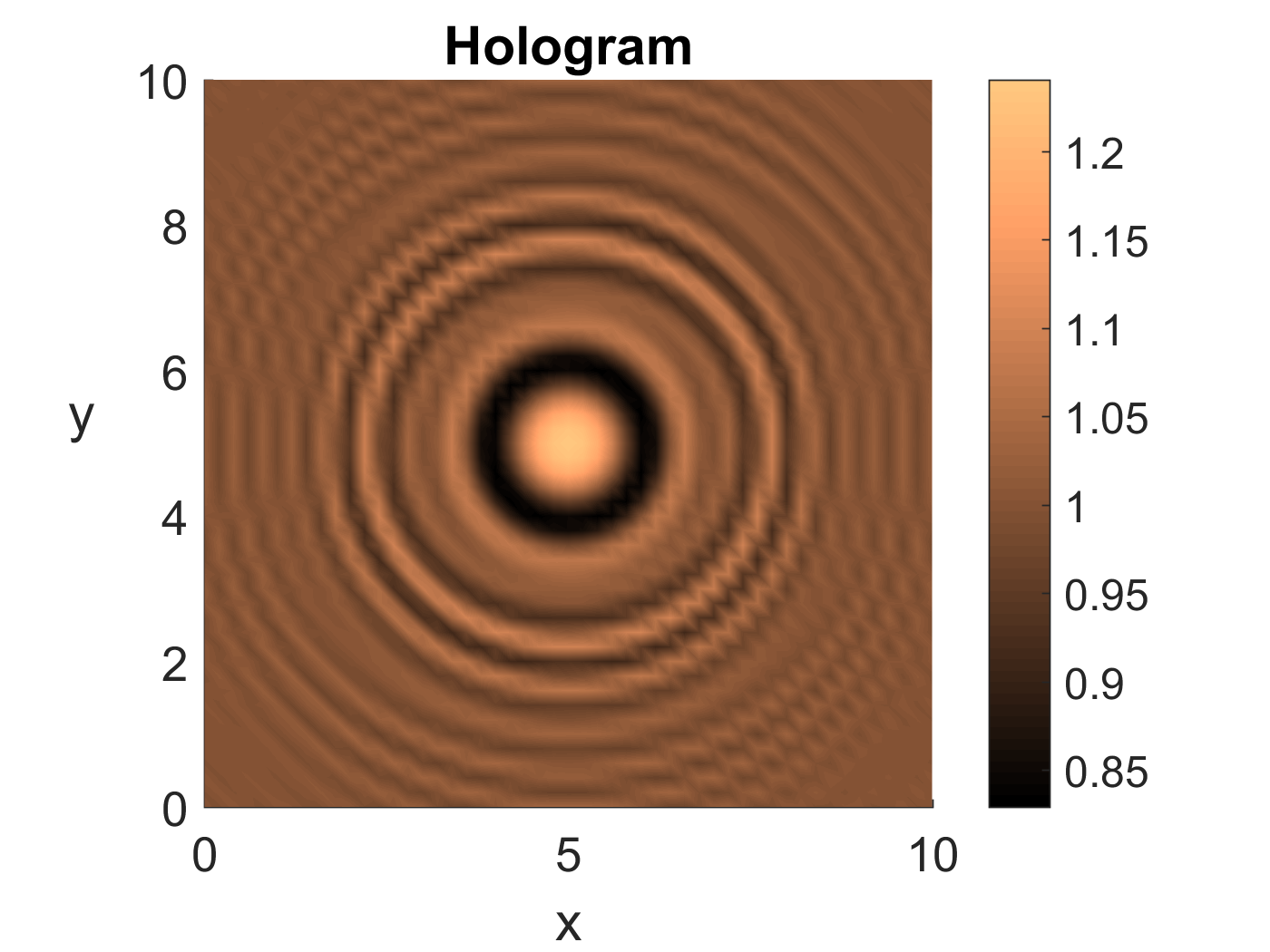} \hskip -2mm
\includegraphics[width=4cm]{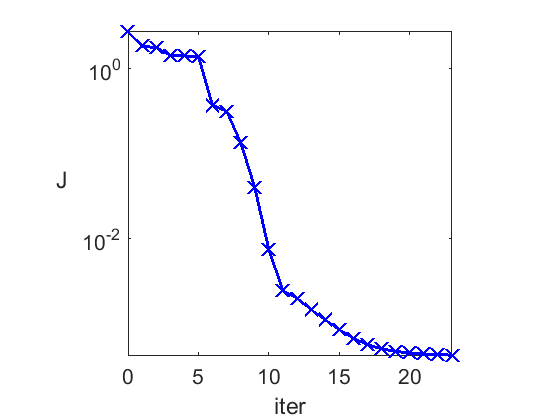} \\
\hskip 0.0cm (d) \hskip 3.25cm (e) \hskip 3.25cm (f) \\
\includegraphics[width=4cm]{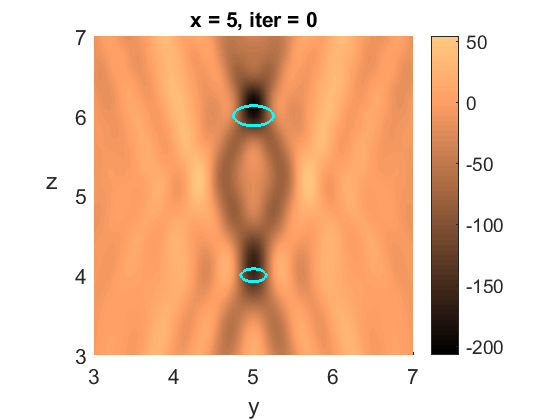} \hskip -2mm
\includegraphics[width=4cm]{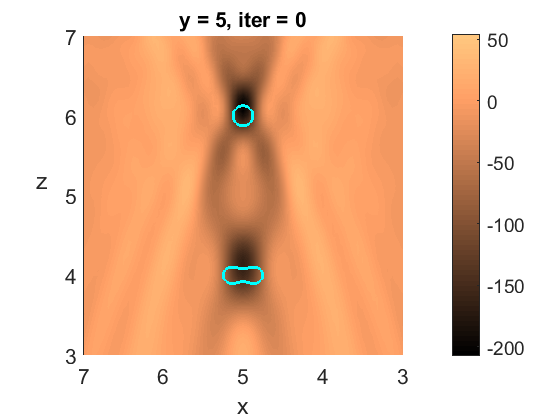} \hskip -2mm
\includegraphics[width=4cm]{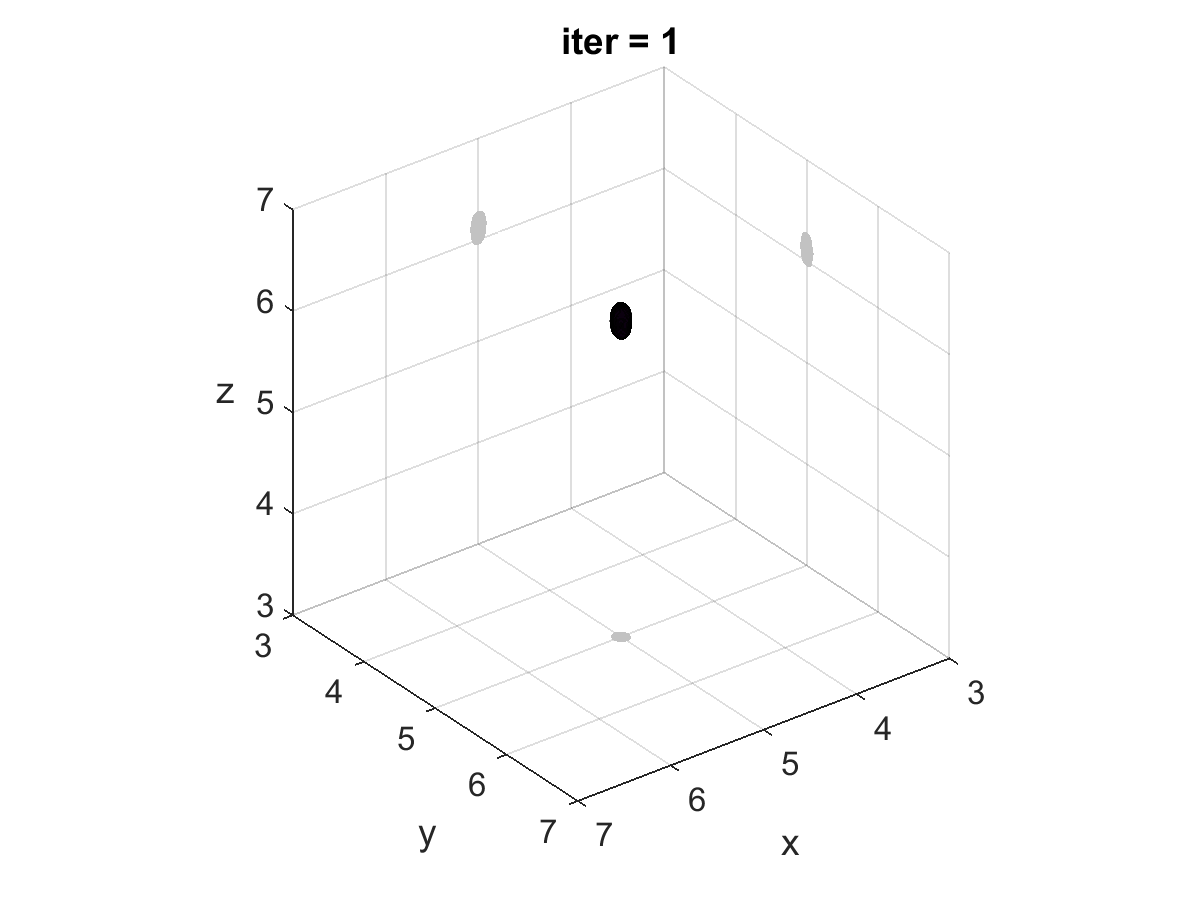} \\
\hskip 0.0cm (g) \hskip 3.25cm (h) \hskip 3.25cm (i) \\
\includegraphics[width=4cm]{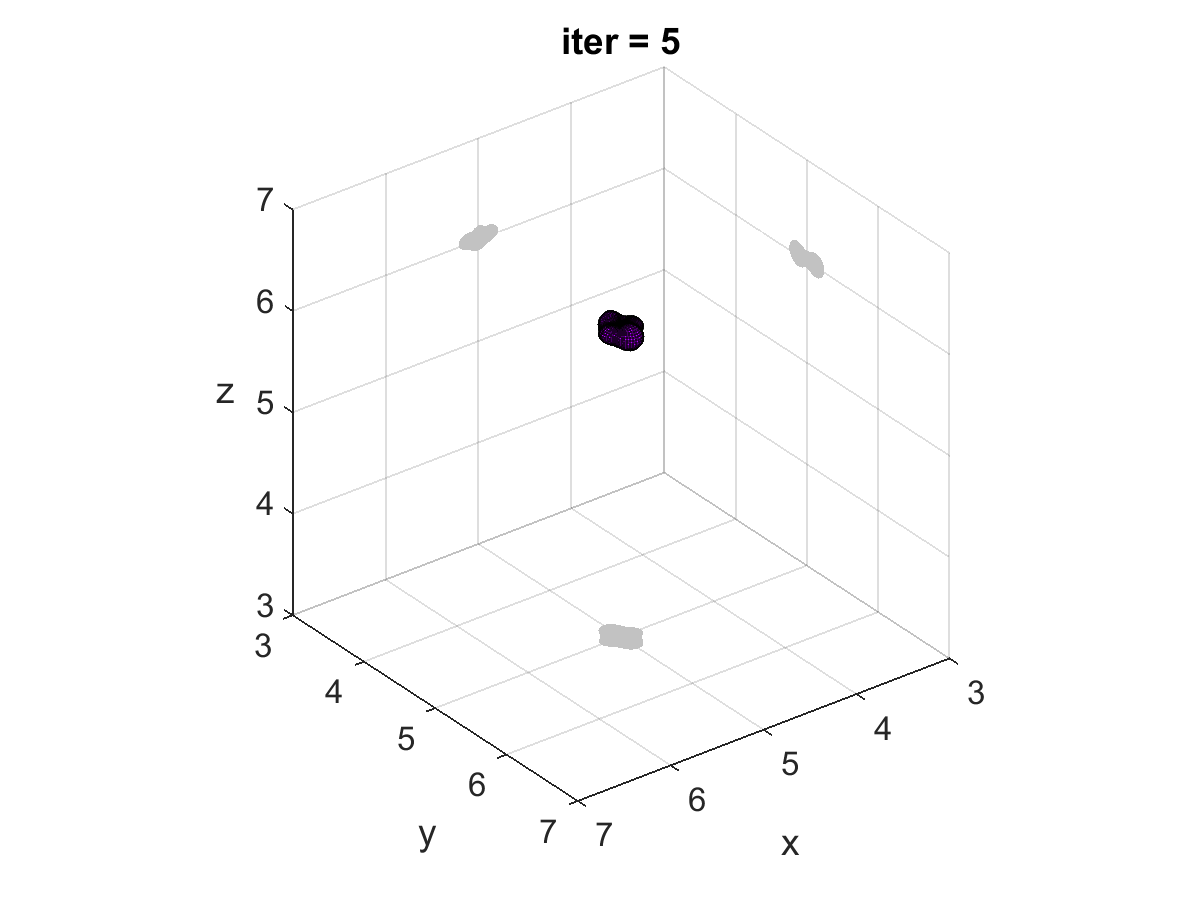} \hskip -2mm
\includegraphics[width=4cm]{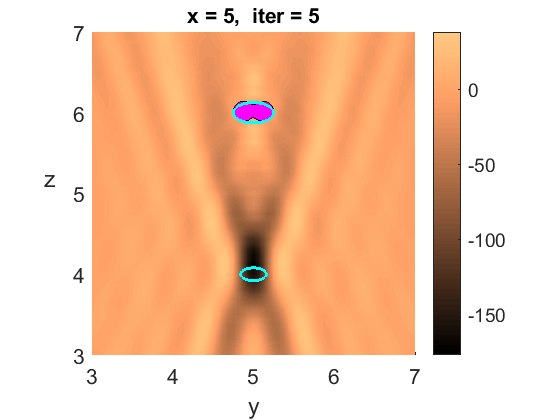} \hskip -2mm
\includegraphics[width=4cm]{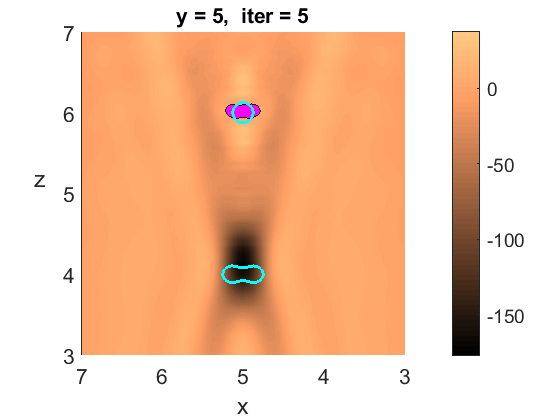} \\
\hskip 0.0cm (j) \hskip 3.25cm (k) \hskip 3.25cm (l) \\
\includegraphics[width=4cm]{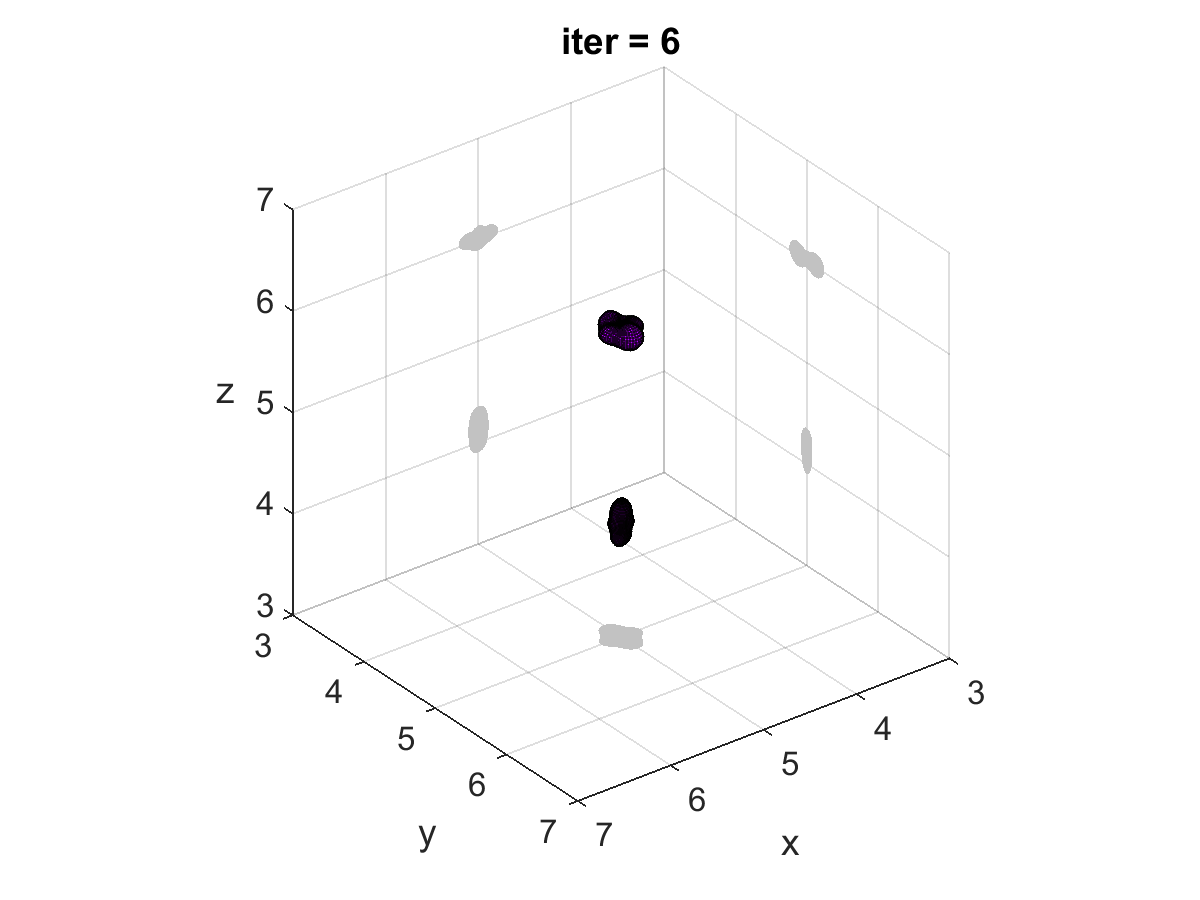} \hskip -2mm
\includegraphics[width=4cm]{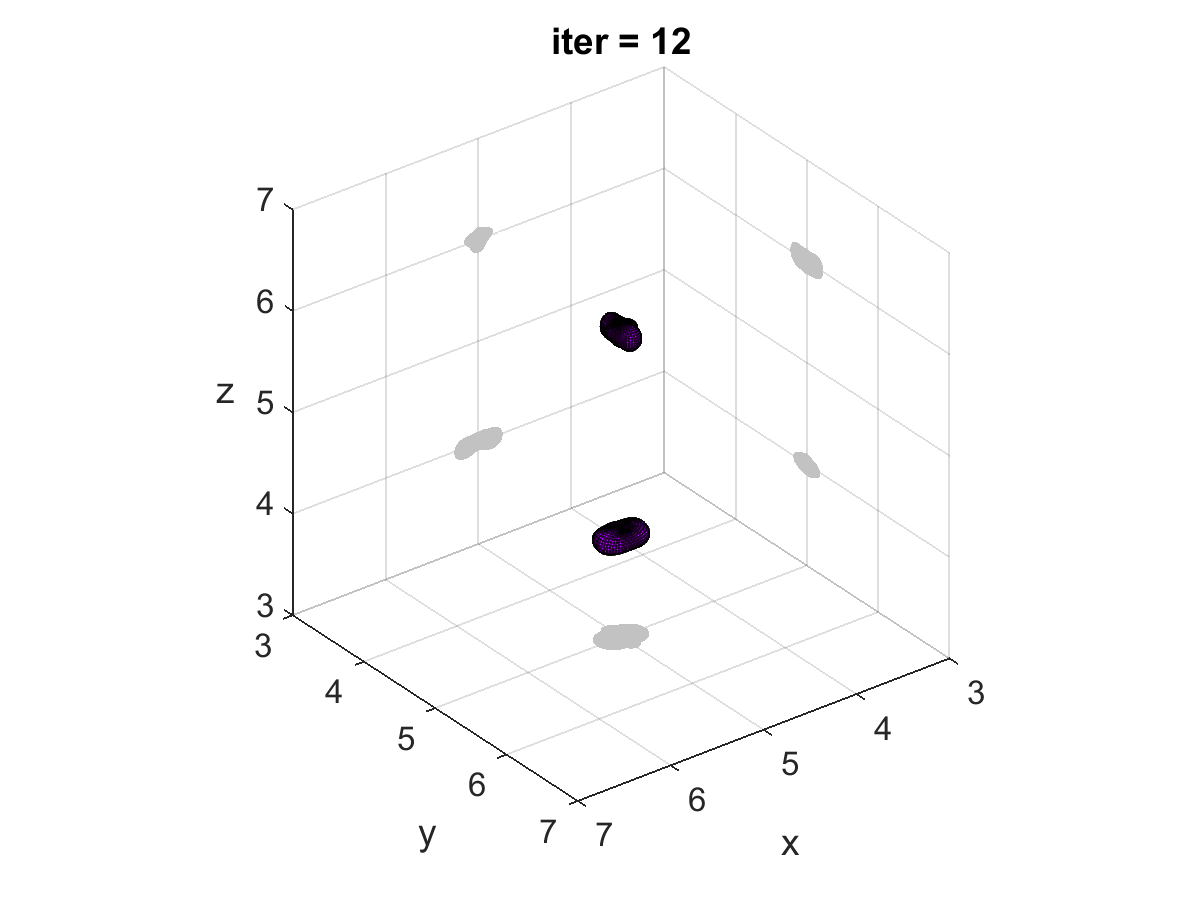} \hskip -2mm
\includegraphics[width=4cm]{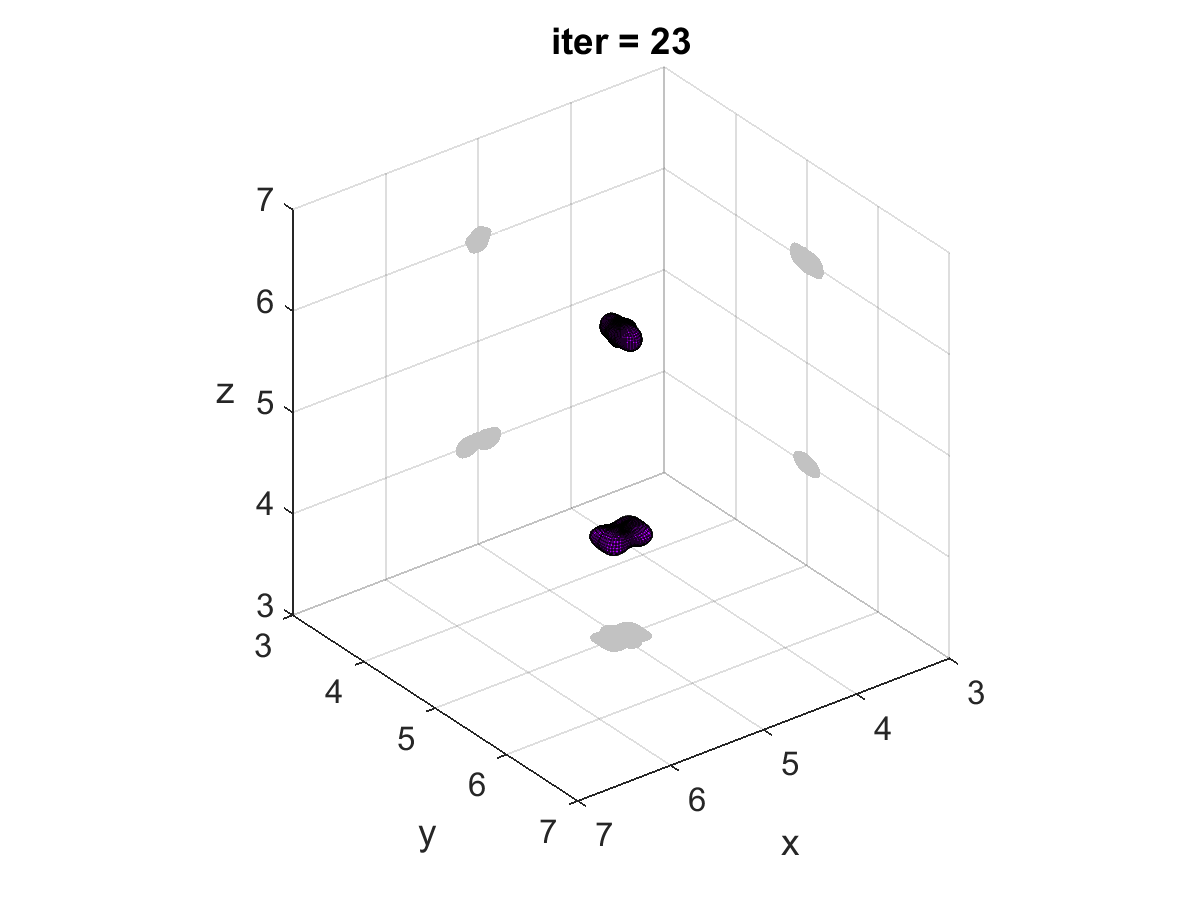}
\caption{\small
Detection of one ellipsoid and one peanut using the hybrid TD/IRGN algorithm 
with violet light.
(a) True geometry: We keep the ellipsoid centered at $(5,5,6)$ of Fig. \ref{fig4}, whereas the second ellipsoid is replaced by the peanut (\ref{parametrizacionpeanut}), with the same center and orientation. (b) Hologram.
(c) Evolution of the cost functional during the optimization process.
(d) and (e) Slices $x=5$ and $y=5$ of the topological derivative (\ref{DT})-(\ref{adjointexplicit}). 
(f) Initial guess $\Omega_1$ defined by (\ref{initialguesstd_impl}). Only one
object is detected, elongated and shifted towards the screen.
(g) Approximate object $\Omega_5$ obtained by the IRGNM. 
The object is overestimated in the $x-$direction.
(h) and (i) Slices $x=5$ and $y=5$ of the topological derivative (\ref{DTsimple}), (\ref{forwardomega})-(\ref{adjointomega}) when $\Omega=\Omega_5$.
Cyan contours represent the true object, whereas the approximated object section is shown in magenta.
(j) Approximate object $\Omega_6$ obtained from $\Omega_5$ using
(\ref{updatedguess2td_impl}). A new component is detected.
(k,l) Approximate objects $\Omega_{12}$ and $\Omega_{23}$ obtained by the IRGNM. 
The location, size and shape of both objects is accurately recovered.}
\label{fig6}
\end{figure}

\begin{figure}
\centering
\hskip 0.0cm (a) \hskip 3.25cm (b) \hskip 3.25cm (c) \\
\includegraphics[width=4cm]{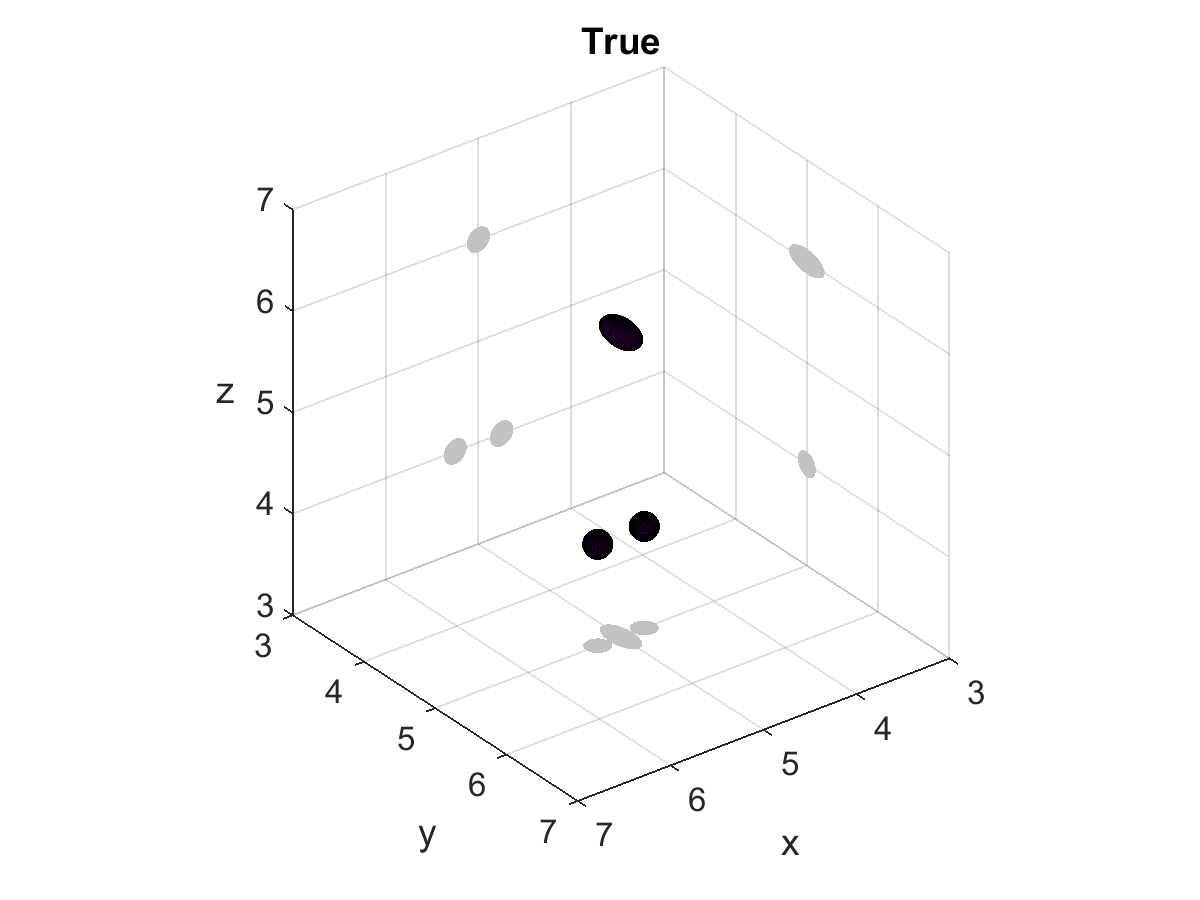}\hskip -2mm
\includegraphics[width=4cm]{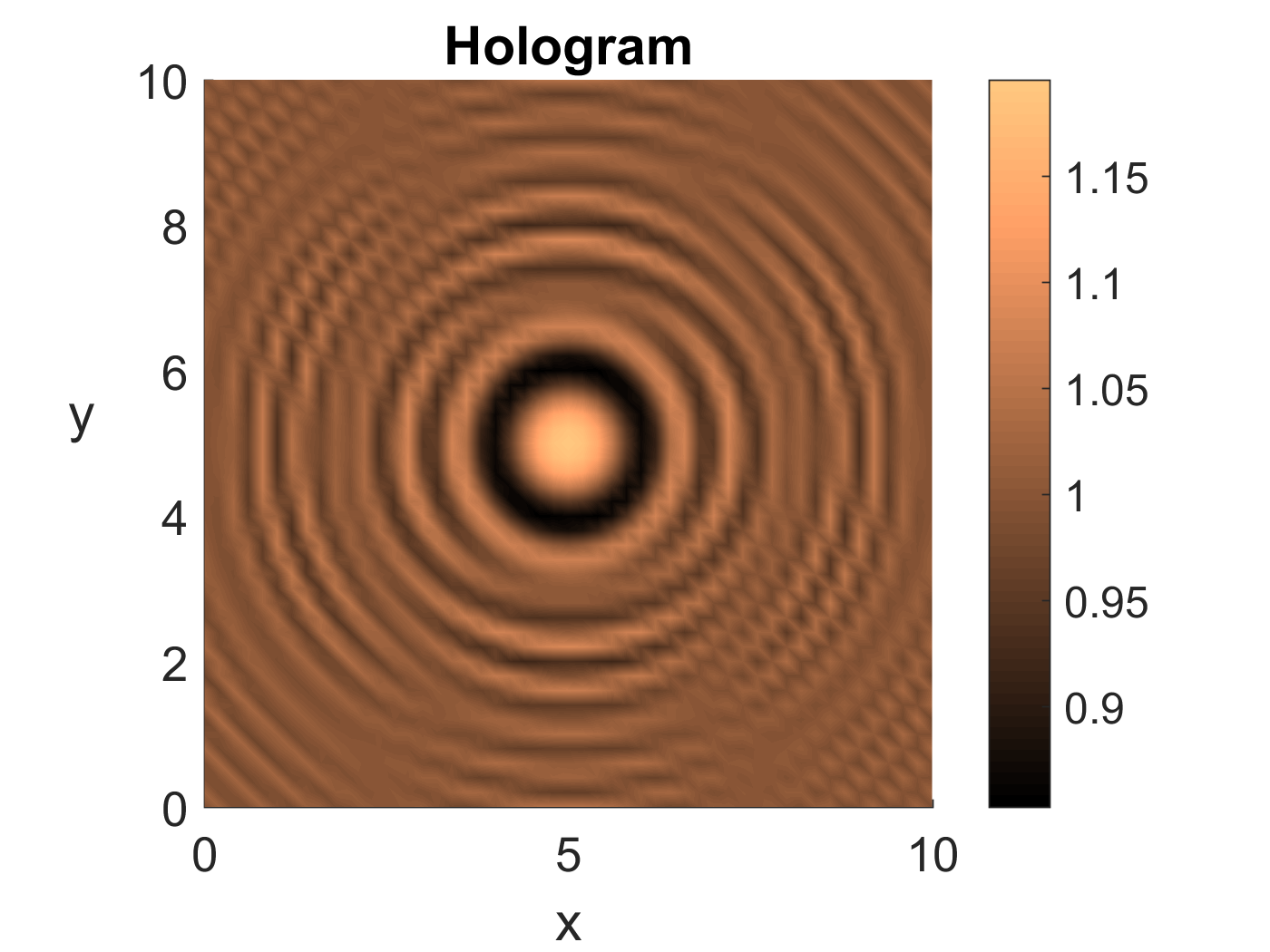}\hskip -2mm
\includegraphics[width=4cm]{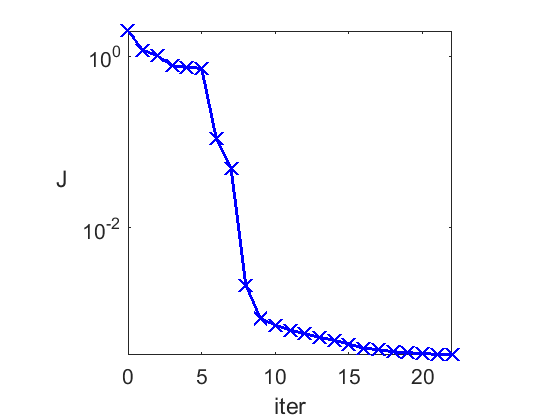} \\
\hskip 0.0cm (d) \hskip 3.25cm (e) \hskip 3.25cm (f) \\
\includegraphics[width=4cm]{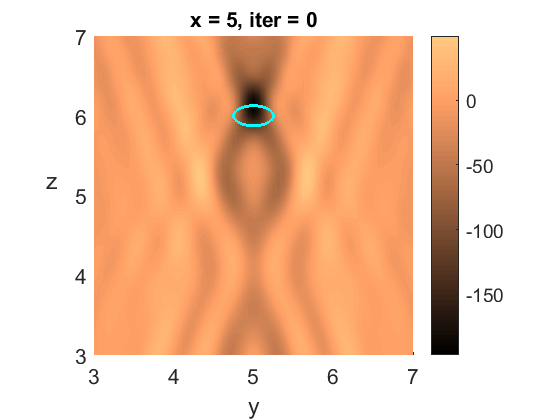}\hskip -2mm
\includegraphics[width=4cm]{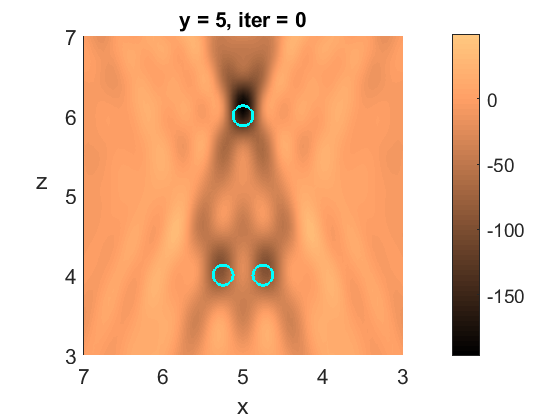}\hskip -2mm
\includegraphics[width=4cm]{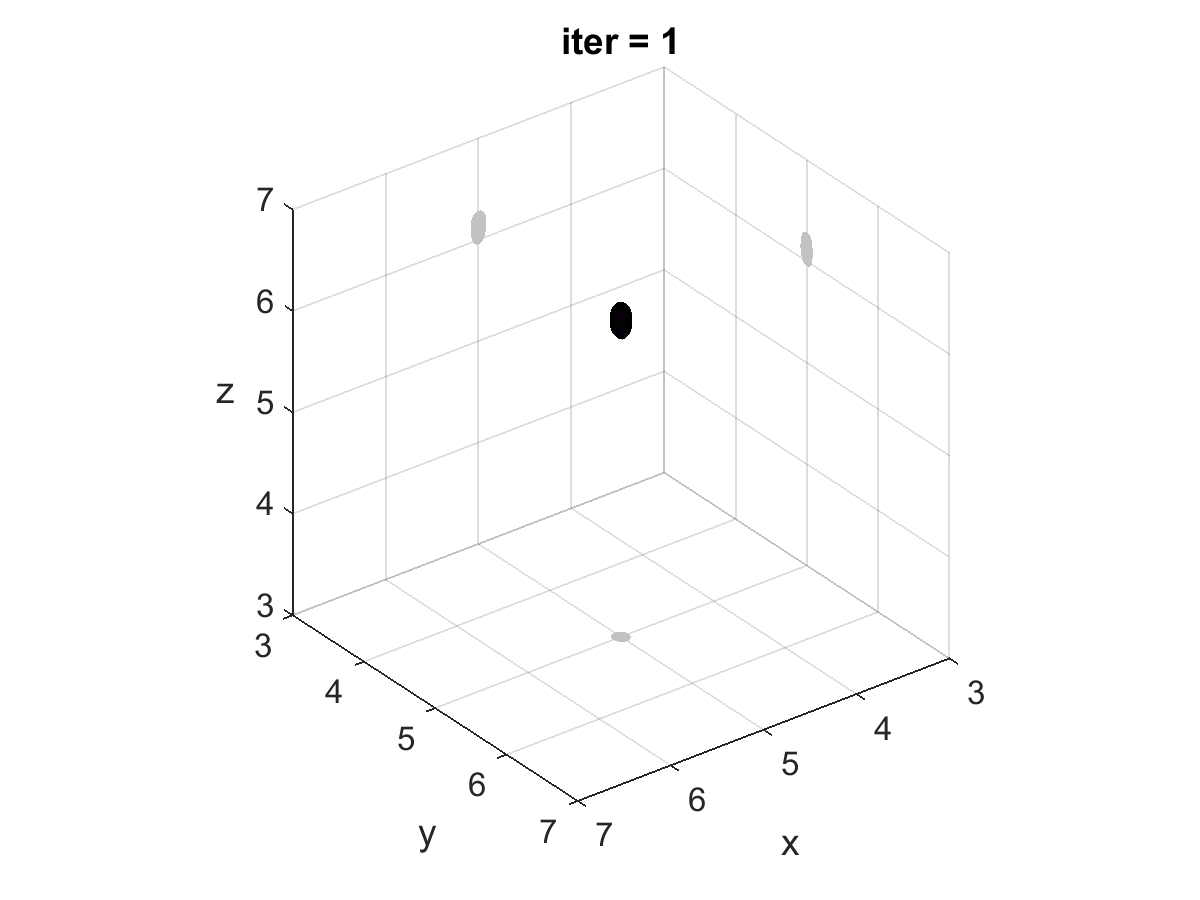} \\
\hskip 0.0cm (g) \hskip 3.25cm (h) \hskip 3.25cm (i) \\
\includegraphics[width=4cm]{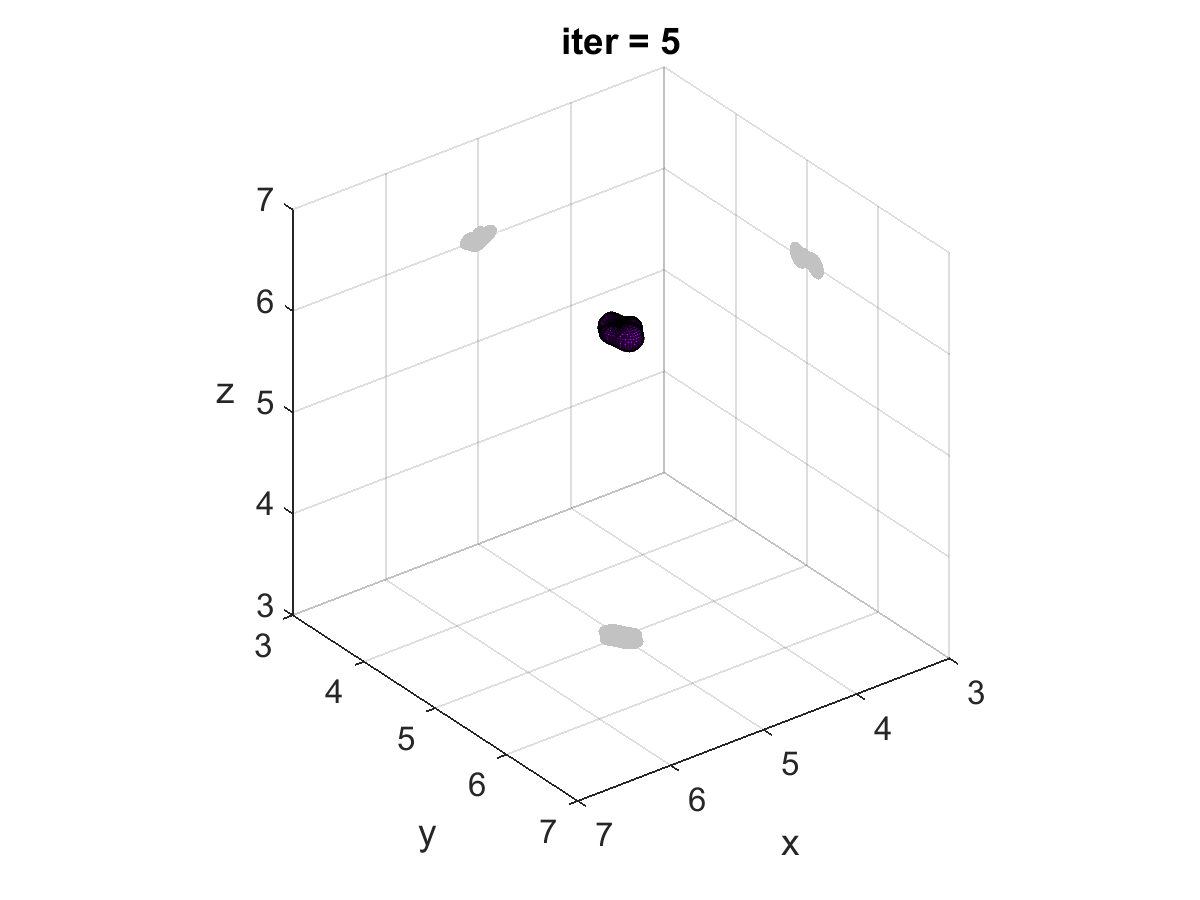}\hskip -2mm
\includegraphics[width=4cm]{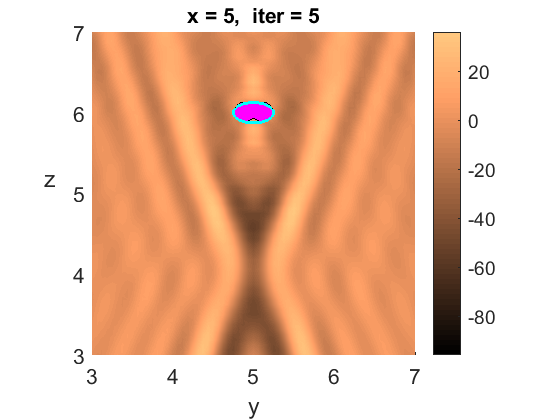}\hskip -2mm
\includegraphics[width=4cm]{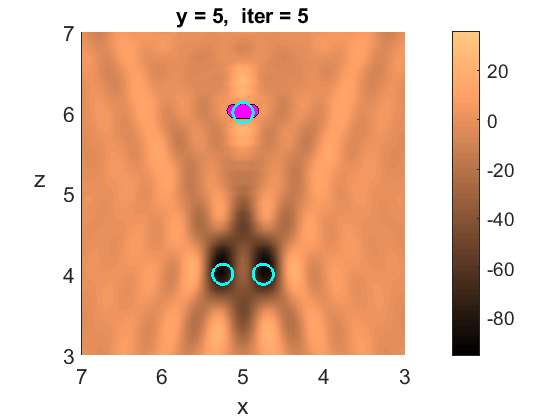} \\
\hskip 0.0cm (j) \hskip 3.25cm (k)  \hskip 3.25cm (l)\\
\includegraphics[width=4cm]{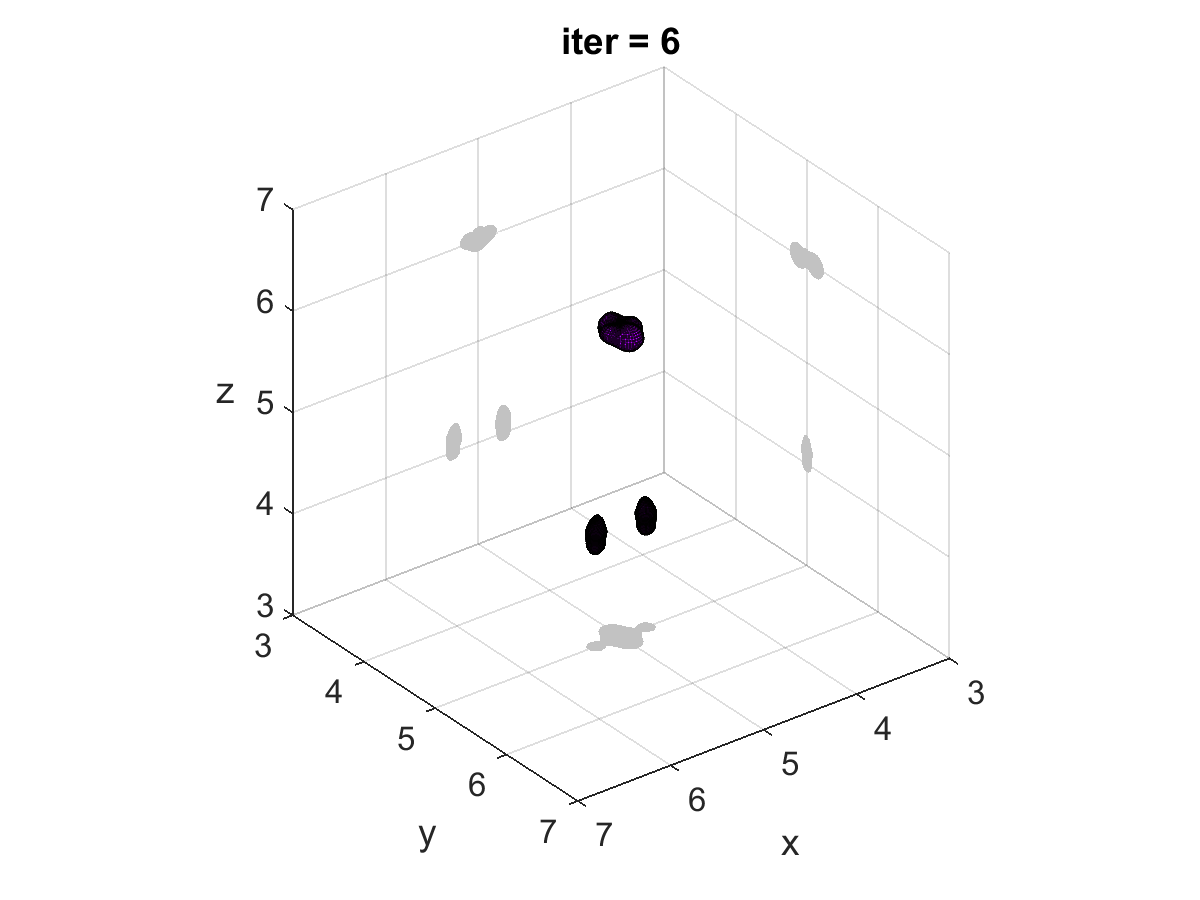}\hskip -2mm
\includegraphics[width=4cm]{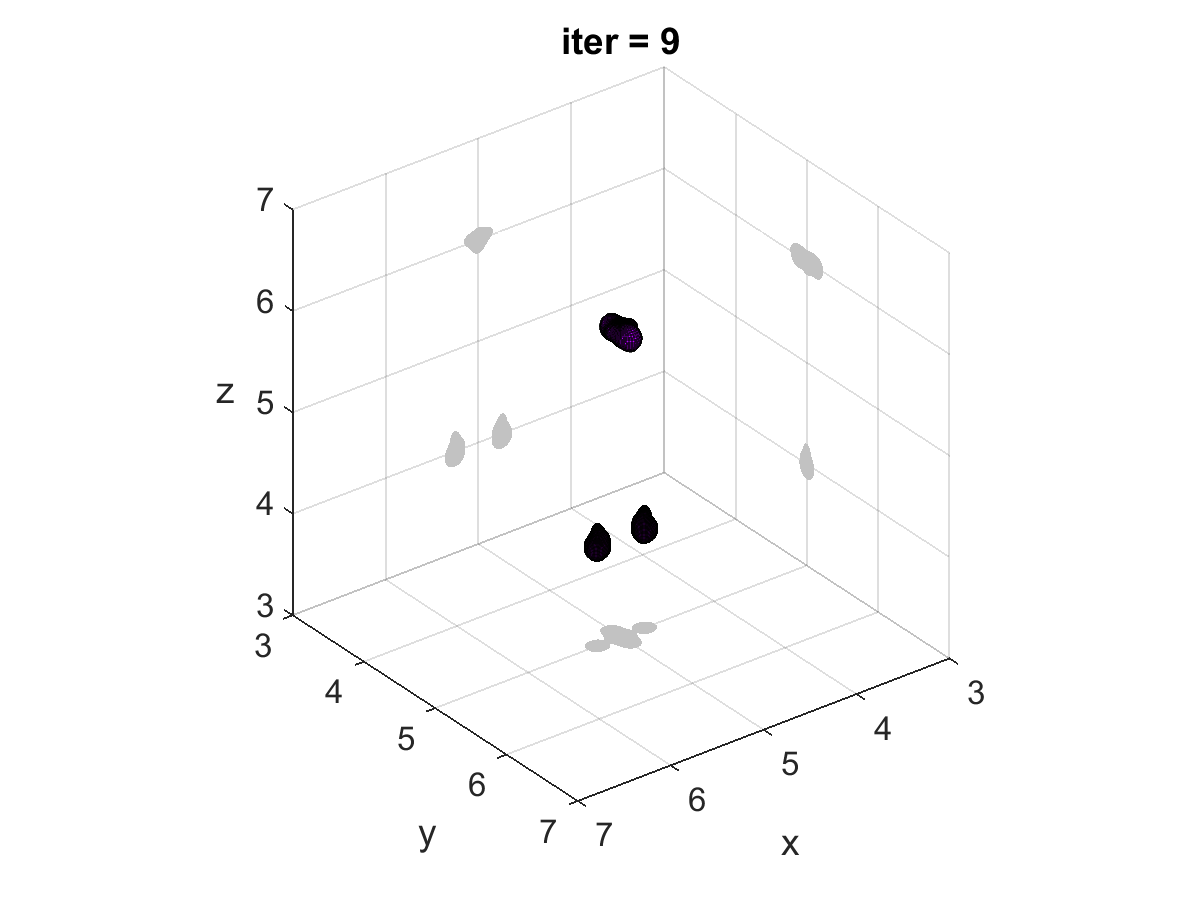}\hskip -2mm
\includegraphics[width=4cm]{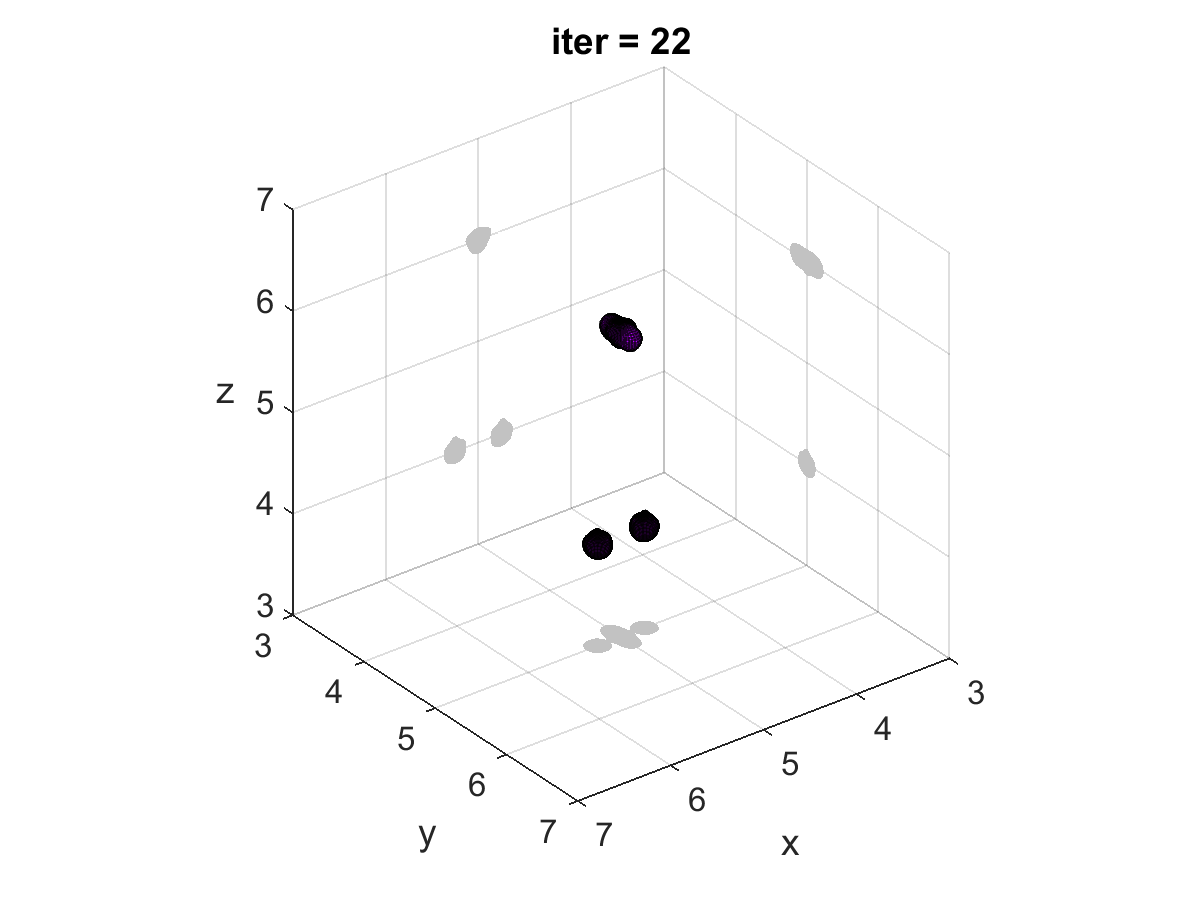}
\caption{ \small Detection of one ellipsoid and two spheres using the hybrid TD/IRGN algorithm with violet light.
(a) True geometry:  The ellipsoid is centered at $(5,5,6)$ and oriented along the $y$ axis, with semi-axes $a=0.125$, $b=0.25$, $c=0.125$. The spheres are centered at $(4.75,5,4)$ and $(5.25,5,4)$ with radius $r=0.125$.
(b) Hologram.
(c) Evolution of the cost functional during the optimization process.
(d) and (e) Slices $x=5$ and $y=5$ of the topological derivative (\ref{DT})-(\ref{adjointexplicit}).
(f) Initial guess $\Omega_1$ defined by (\ref{initialguesstd_impl}). Only one
object is detected, elongated and shifted towards the screen.
(g) Approximate object $\Omega_5$ obtained by the IRGNM.
(h) and (i) Slices $x=5$ and $y=5$ of the topological derivative (\ref{DTsimple}), (\ref{forwardomega})-(\ref{adjointomega}) when $\Omega=\Omega_5$.
Cyan contours represent the true object, whereas the approximated object section is
shown in magenta.
(j) Approximate object $\Omega_6$ obtained from $\Omega_5$ using
(\ref{updatedguess2td_impl}). Two new components are detected.
(k,l) Intermediate and final approximate objects $\Omega_{9}$ and $\Omega_{22}$
obtained by the IRGNM. The location of the centers, sizes and orientation are captured.
}
\label{fig7}
\end{figure}

\begin{figure}[h!]
\centering
\hskip 0.0cm (a) \hskip 3.25cm (b) \hskip 3.25cm (c) \\
\includegraphics[width=4cm]{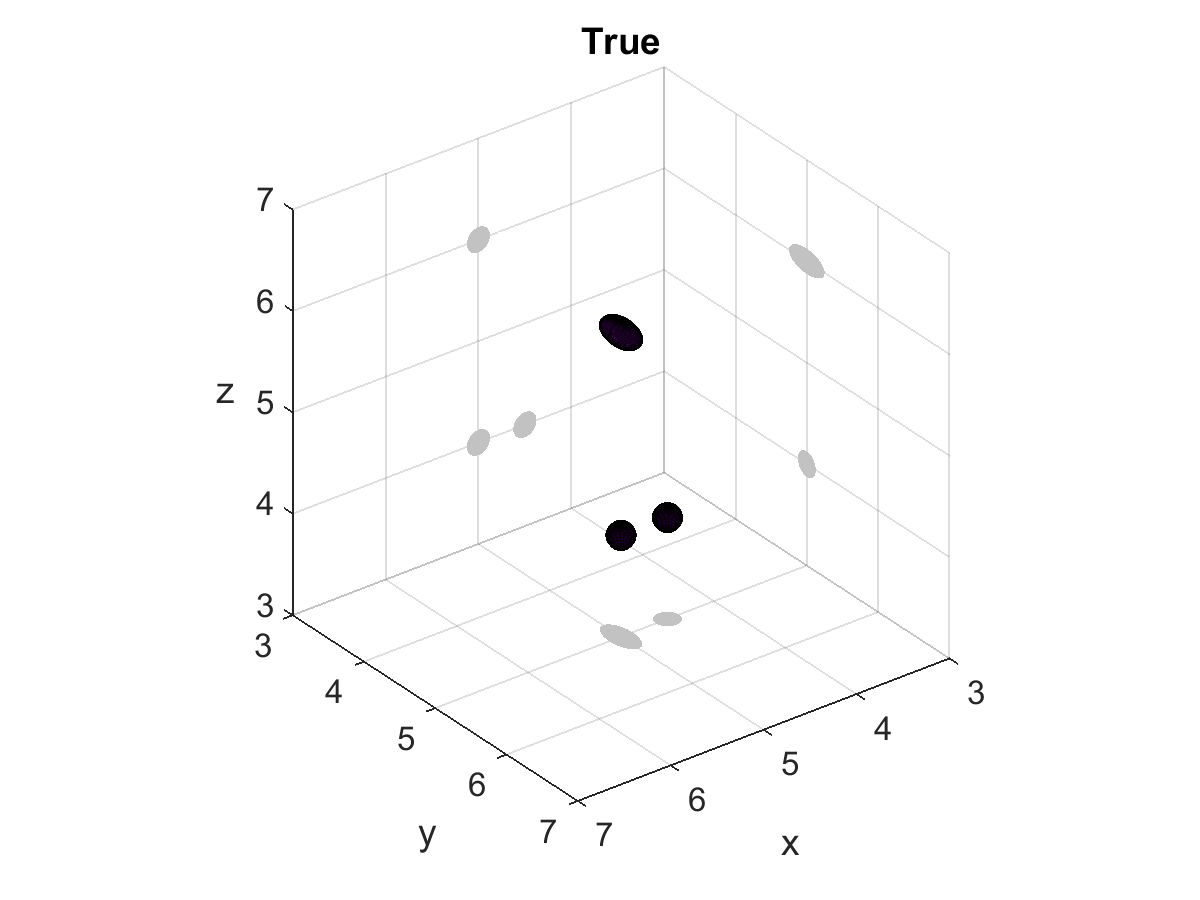} \hskip -2mm
\includegraphics[width=4cm]{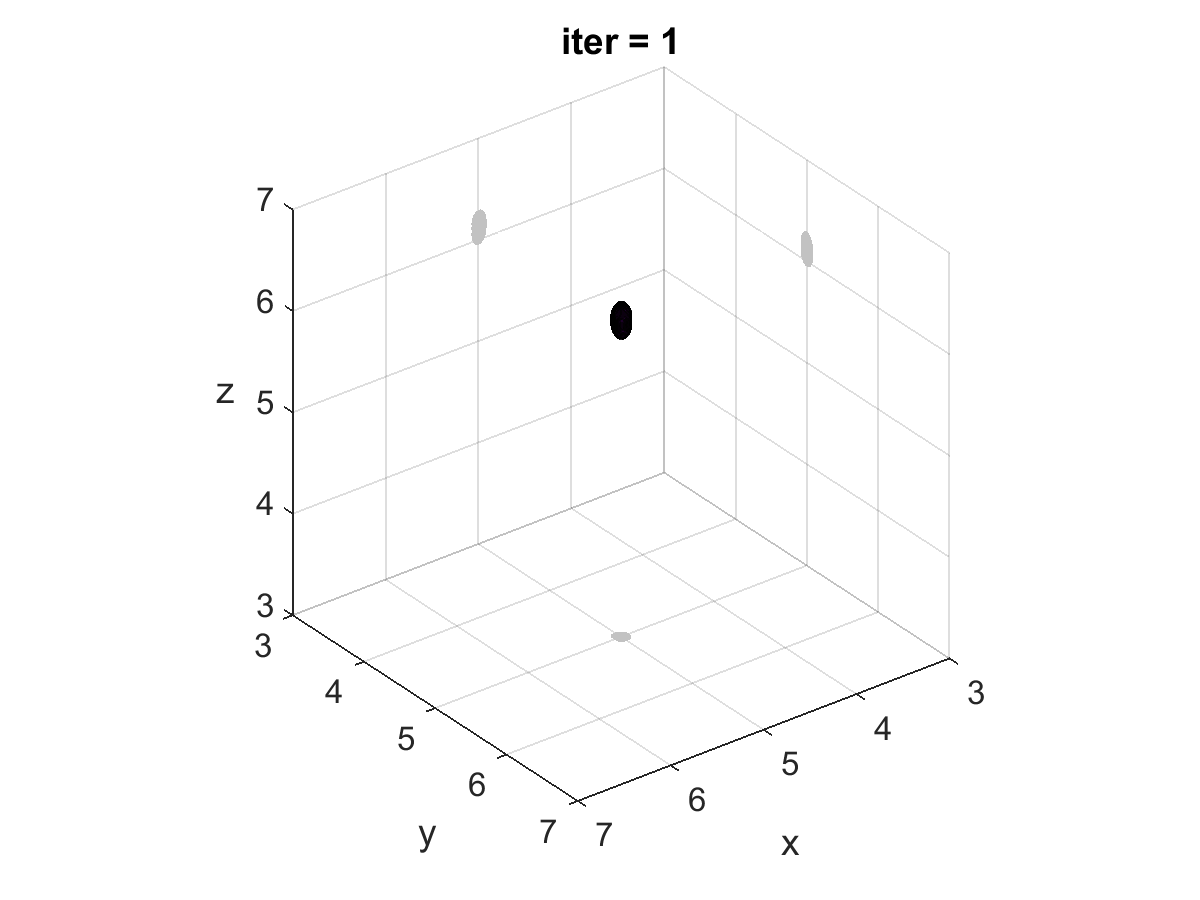} \hskip -2mm
\includegraphics[width=4cm]{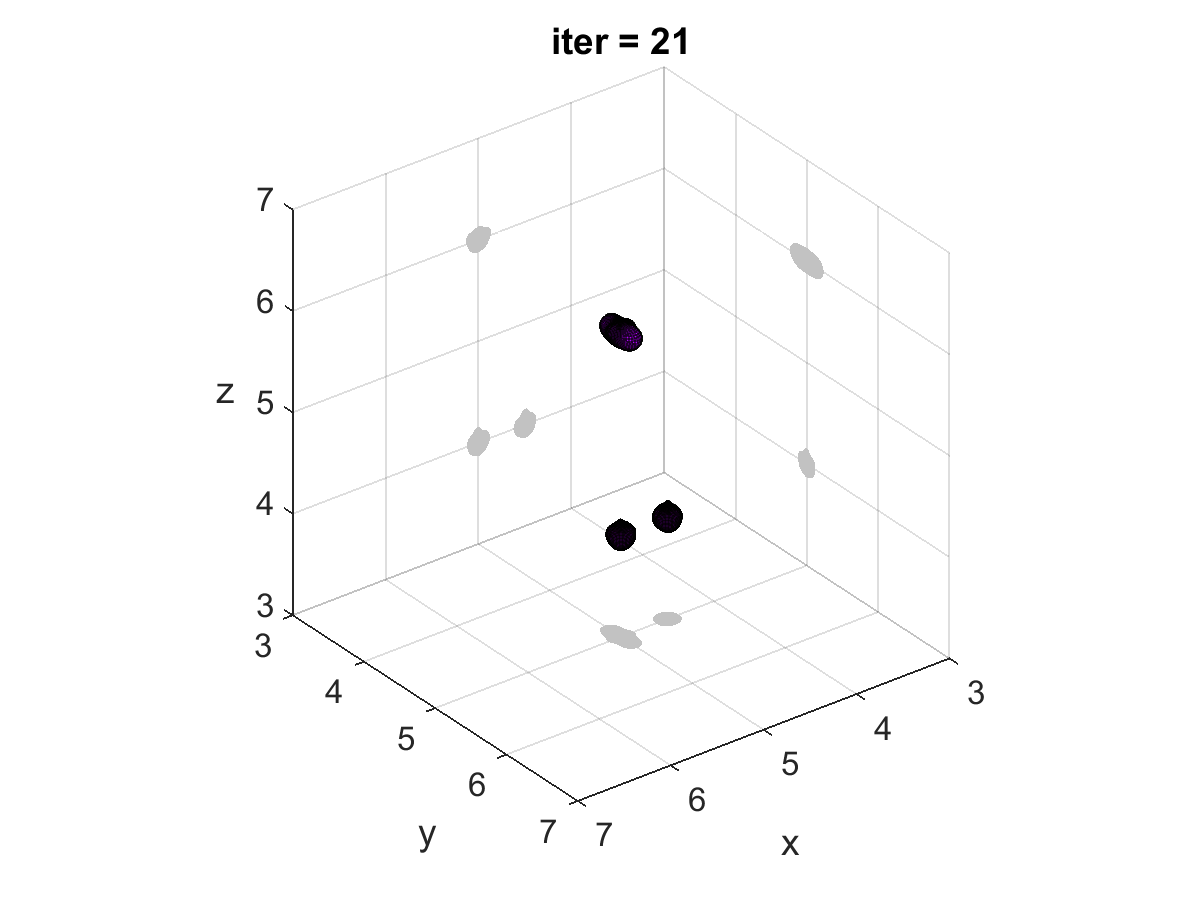}
\caption{\small Counterpart of the panels (a,f,l) in Fig. \ref{fig7} when the spheres are centered at $(5,5,4)$ and $(5.5,5,4)$.}
\label{fig7_bis}
\end{figure}

We next replace the ellipsoid located at $(5,5,4)$ by a peanut. The parameterization of the boundary of a peanut centered at $(c_x,c_y,c_z)$  is:
\begin{equation}\label{parametrizacionpeanut}
(c_x-r_\gamma(\theta)\cos\theta, c_y+2r_\gamma(\theta)\sin\theta\sin\phi,c_z+r_\gamma(\theta)\sin\theta\cos\phi),\end{equation}
with $ \theta\in[0,\pi]$, $\phi\in[0,2\pi]$ and
$r_\gamma(\theta)=\frac{1}{4\sqrt{1+\sqrt{\gamma+1}}}\sqrt{\cos(2\theta)+\sqrt{\gamma+1-\sin^2(2\theta)}}.$
We select $(c_x,c_y,c_z)=(5,5,4)$ and $\gamma=0.5$. The parameter $\gamma>0$ determines the narrowness in the middle of the peanut shape (the closer to zero, the more constricted). Figure \ref{fig6} illustrates the good performance of the hybrid algorithm, which recovers both shapes in  23 iterations for violet light (results for red light are similar in 24 iterations), see also Video 2 in the Supplementary Material.

If we replace the peanut by two spheres, we recover the three objects in $22$ iterations
of the hybrid TD/IRGN method as shown in Figure \ref{fig7}, see also Video 
3 in the Supplementary Material.   Figure \ref{fig7_bis} and Video 4 illustrate convergence when the two spheres are placed asymmetrically. Notice that in this case, recovering the sphere aligned with the ellipsoid should be in principle more complicated than approximating the sphere not screened by it. However, our method overcomes the difficulty and provides an accurate description of both spheres.

\begin{figure}
\centering
\hskip 0.0cm (a) \hskip 3.25cm (b) \hskip 3.25cm (c) \\
\includegraphics[width=4cm]{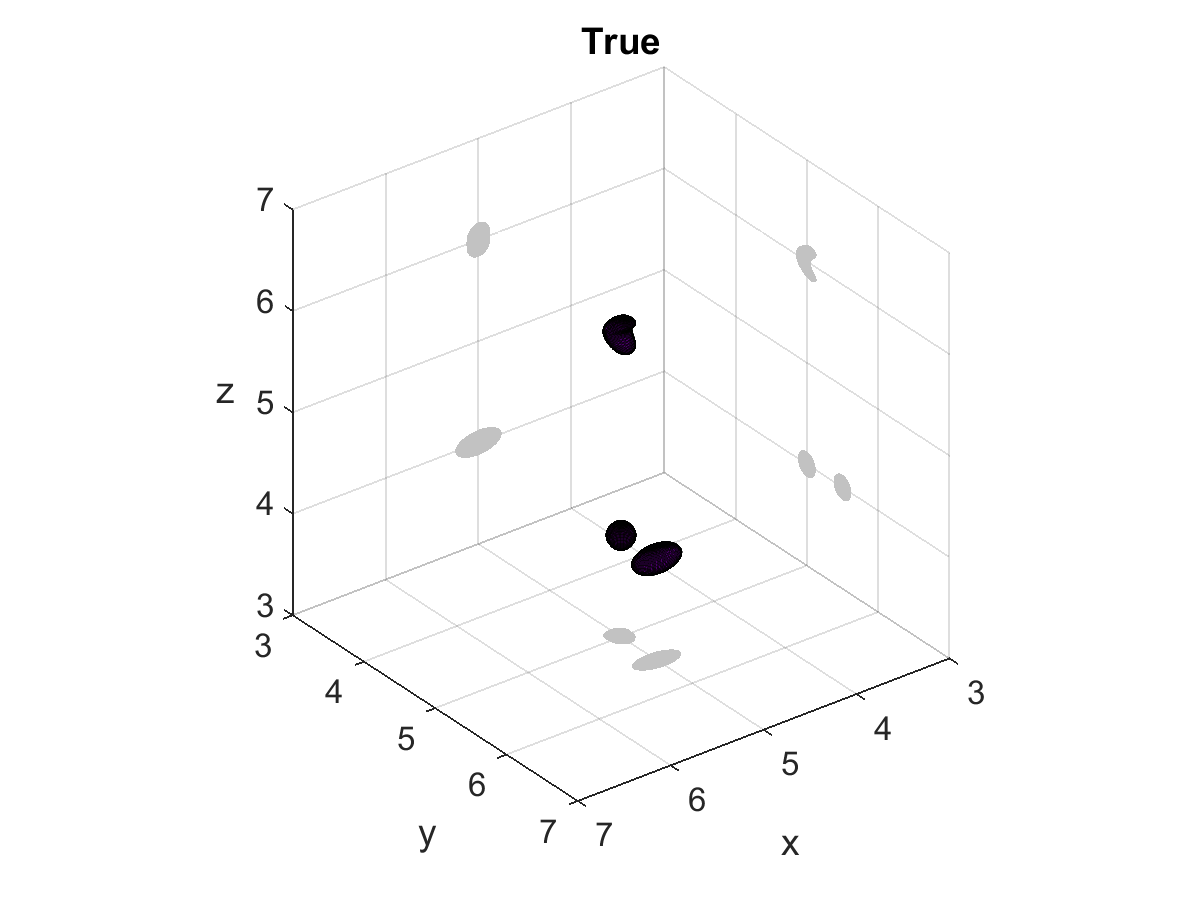}\hskip -2mm
\includegraphics[width=4cm]{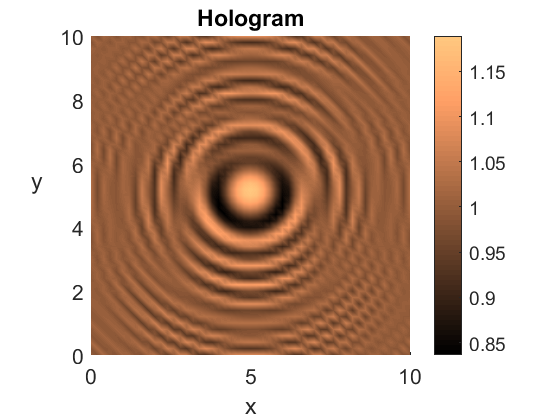}\hskip -2mm
\includegraphics[width=4cm]{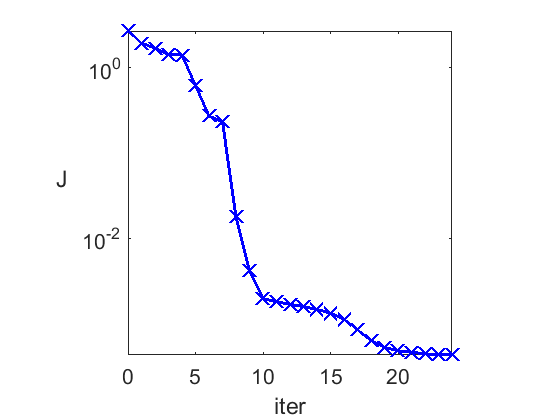} \\
\hskip 0.0cm (d) \hskip 3.25cm (e) \hskip 3.25cm (f) \\
\includegraphics[width=4cm]{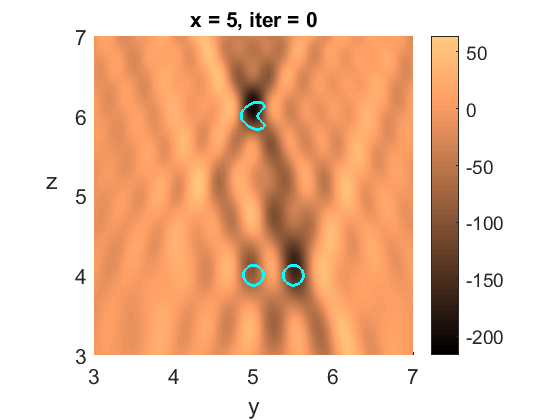}\hskip -2mm
\includegraphics[width=4cm]{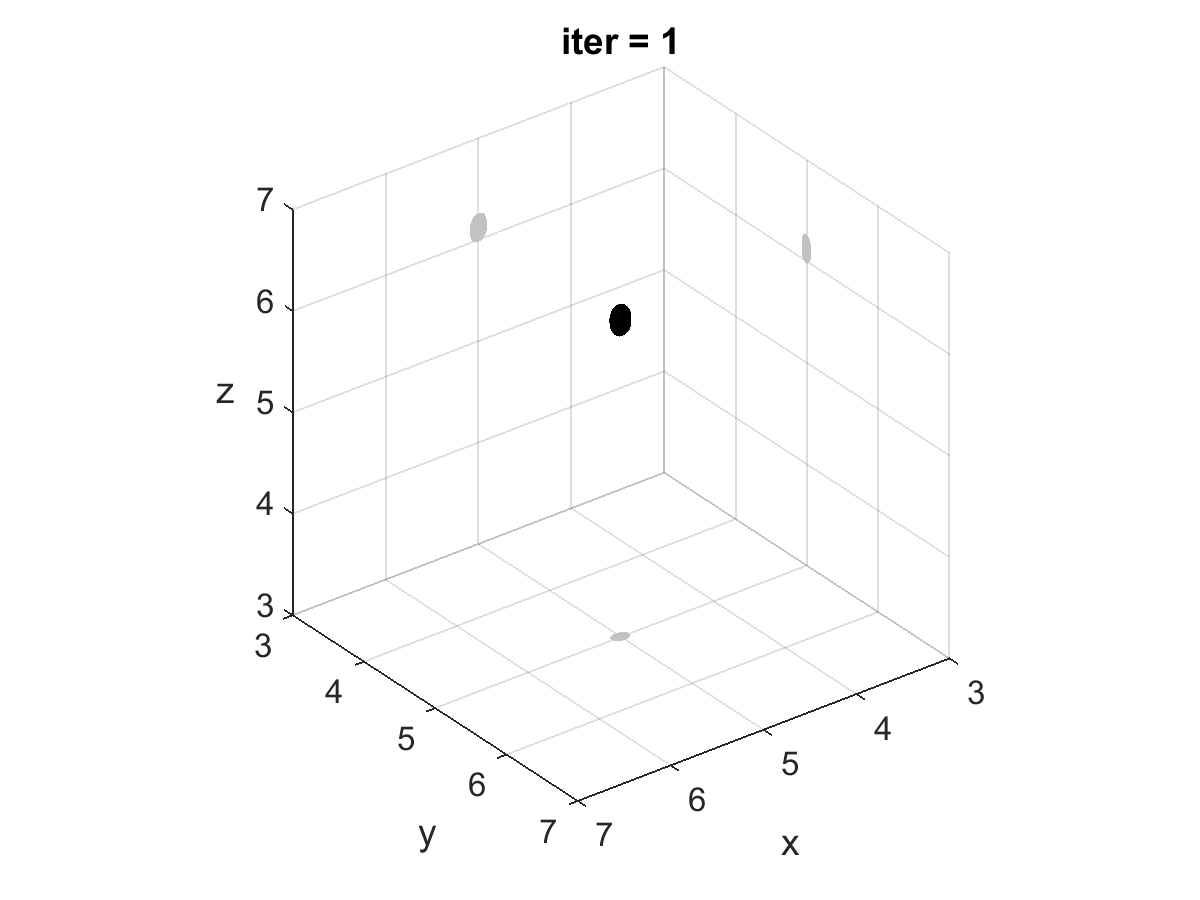}\hskip -2mm
\includegraphics[width=4cm]{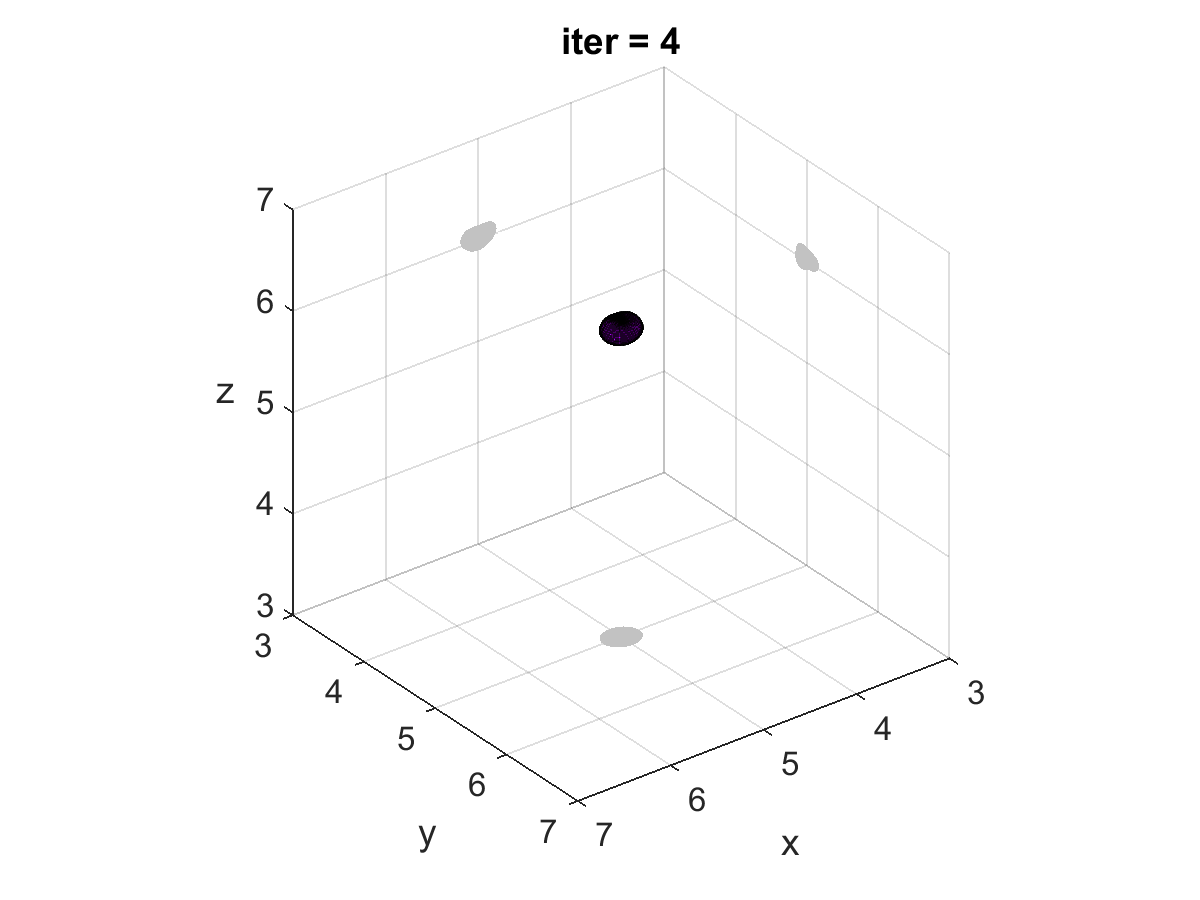} \\
\hskip 0.0cm (g) \hskip 3.25cm (h) \hskip 3.25cm (i) \\
\includegraphics[width=4cm]{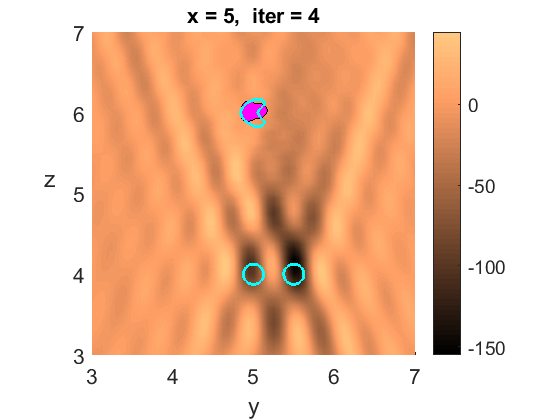}\hskip -2mm
\includegraphics[width=4cm]{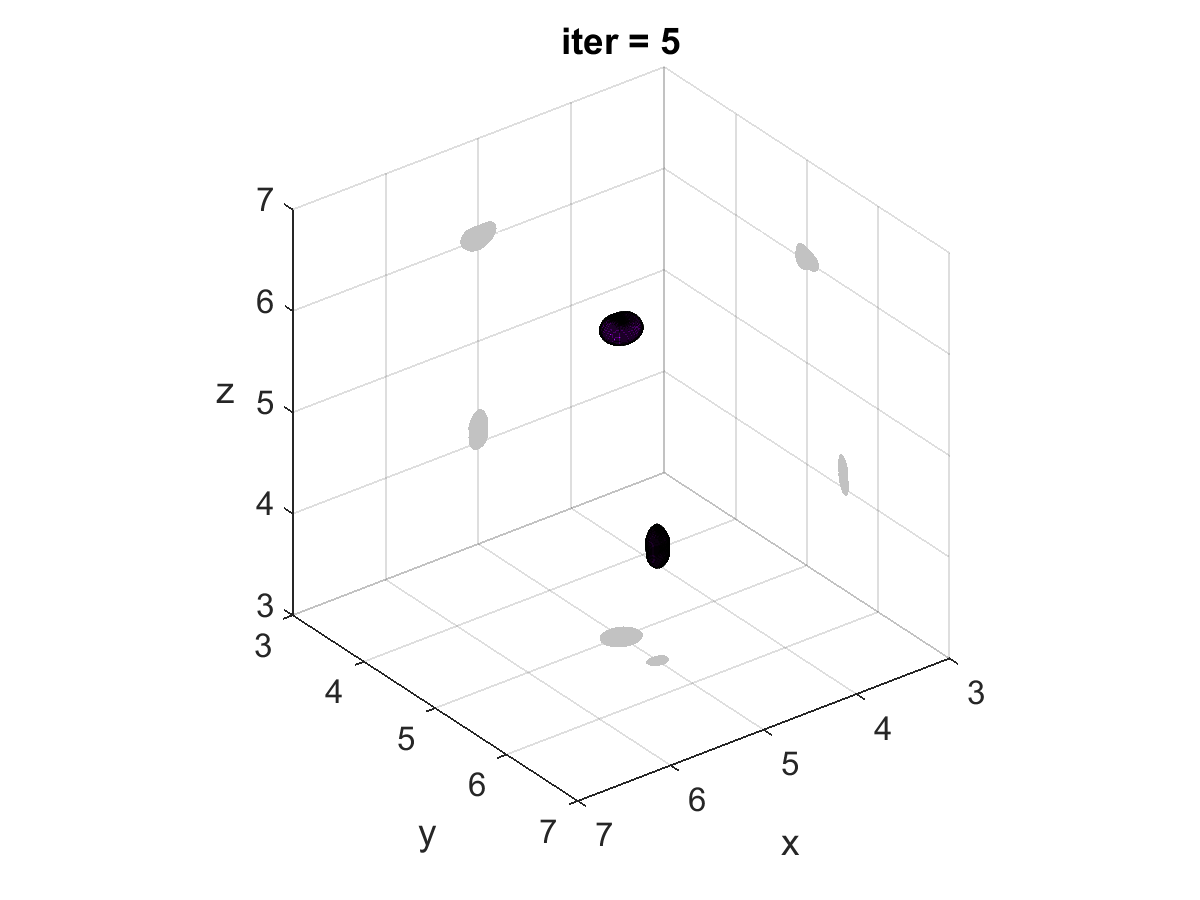}\hskip -2mm
\includegraphics[width=4cm]{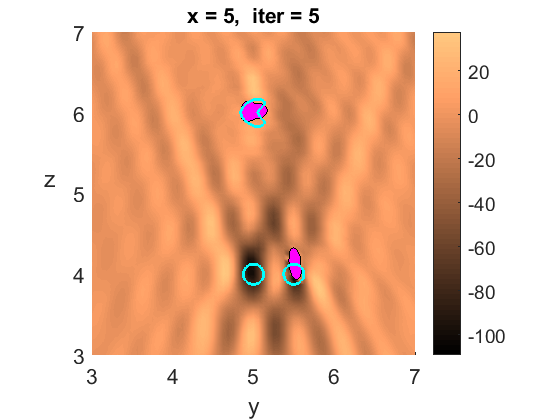} \\
\hskip 0.0cm (j) \hskip 3.25cm (k)  \hskip 3.25cm (l)\\
\includegraphics[width=4cm]{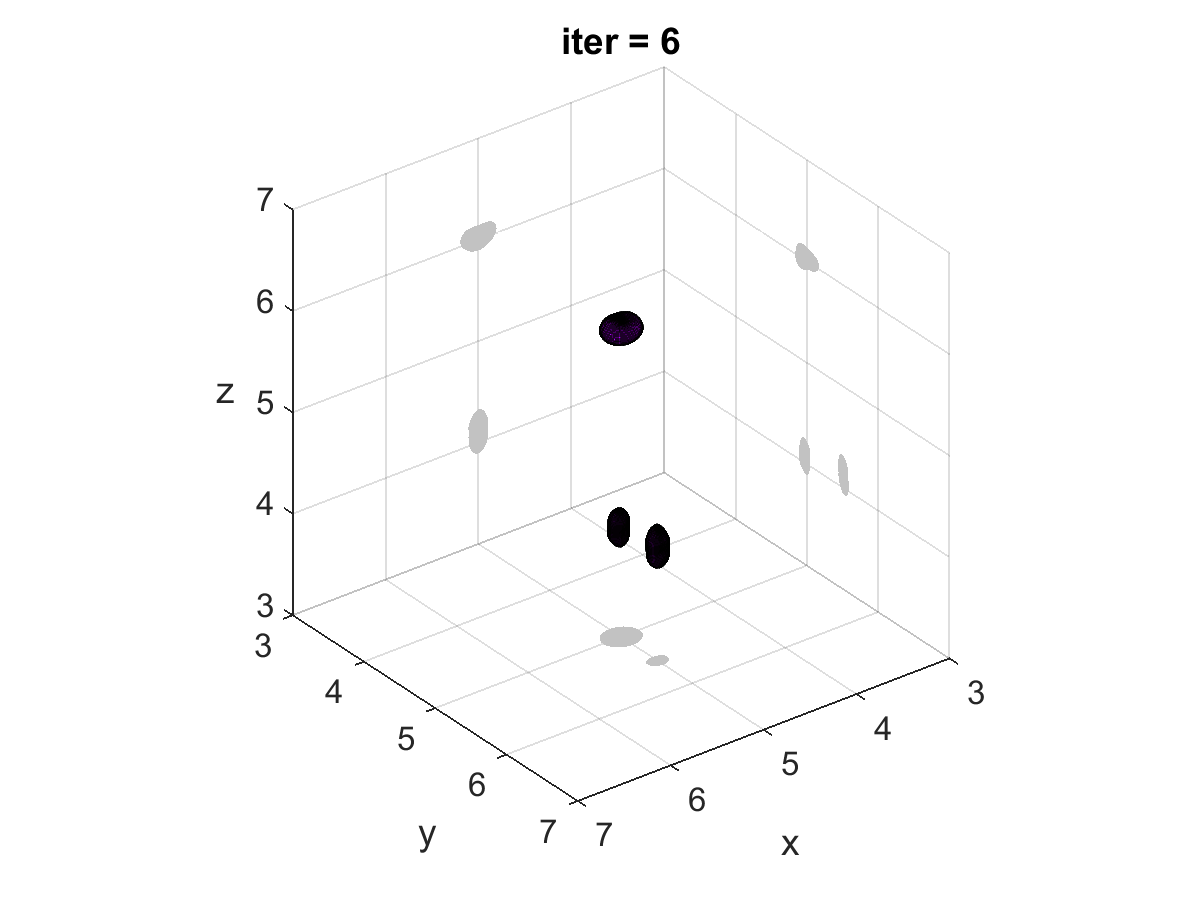}\hskip -2mm
\includegraphics[width=4cm]{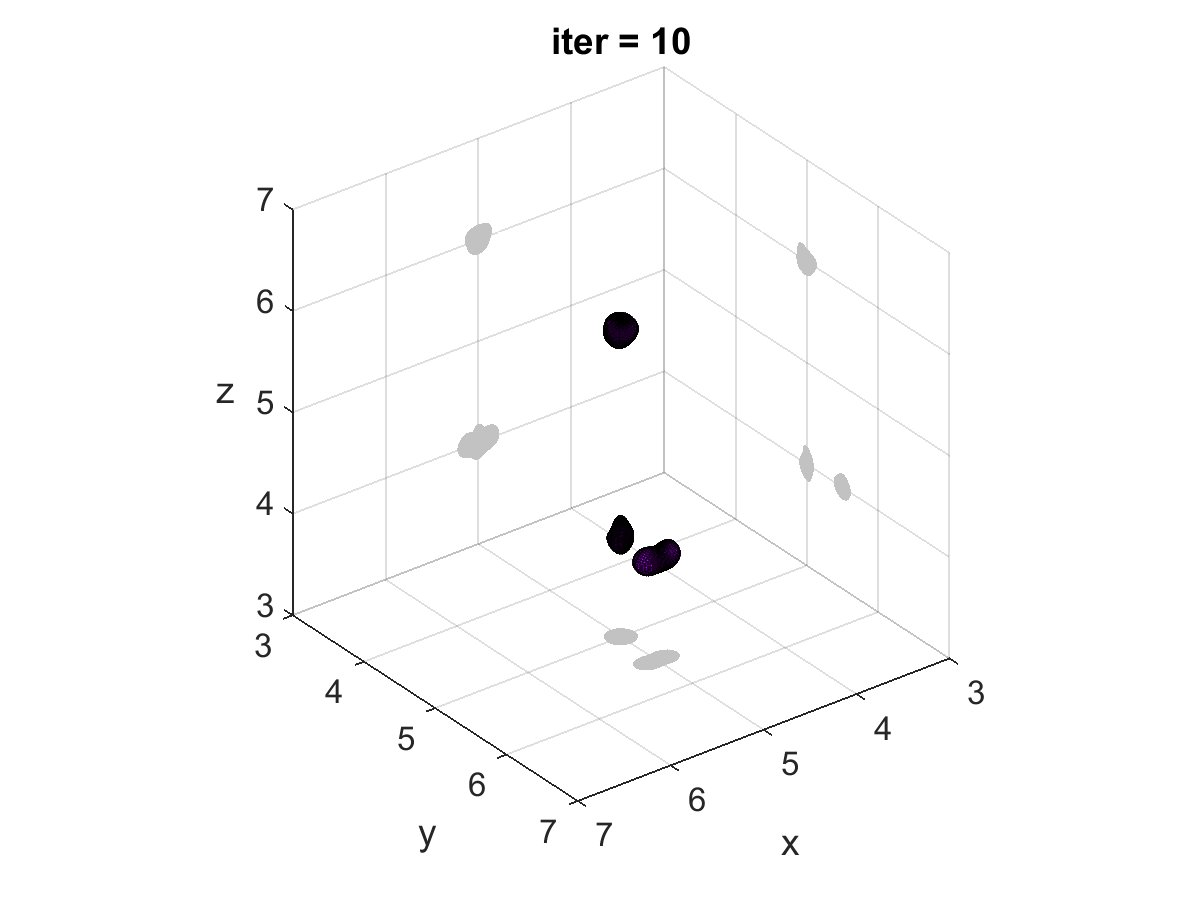}\hskip -2mm
\includegraphics[width=4cm]{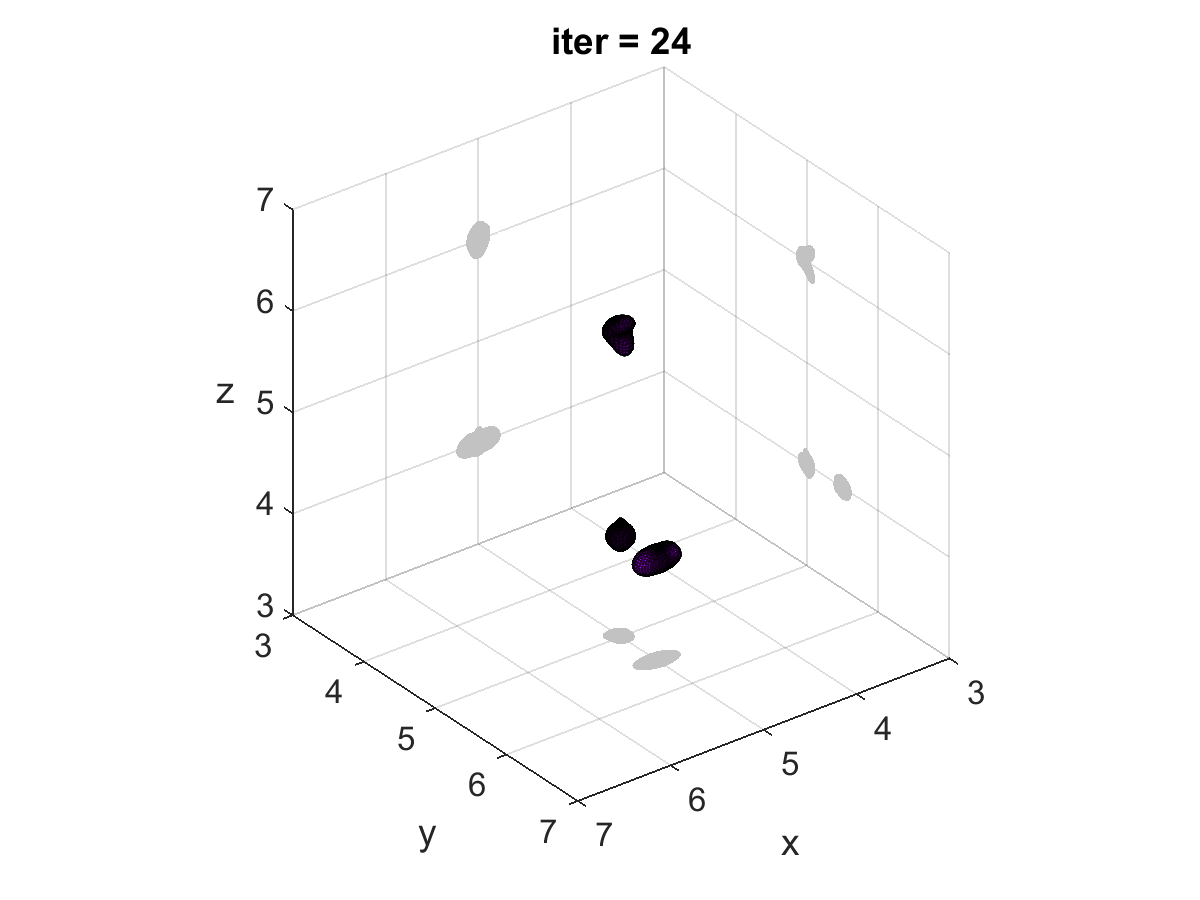}
\caption{ \small
Detection of three objects with different shapes and sizes using the hybrid TD/IRGN 
with violet light.
(a) True geometry:  The bean is centered at $(5,5,6)$ and described by the parameterization (\ref{bean1}),  the sphere is centered at $(5,5,4)$ with radius $0.125$ and the ellipsoid is centered at $(5,5.5,4)$ with  semi-axes $a=0.25$, $b=0.125$ and $c=0.125$.
(b) Hologram.
(c) Evolution of the cost functional during the optimization process.
(d) Slice $x=5$  of the topological derivative (\ref{DT})-(\ref{adjointexplicit}).
(e) Initial guess $\Omega_1$ defined by (\ref{initialguesstd_impl}). Only one
object is detected. 
(f) Approximate object $\Omega_4$ obtained by the IRGNM.
(g) Slice $x=5$ of the topological derivative (\ref{DTsimple}), (\ref{forwardomega})-(\ref{adjointomega}) when $\Omega=\Omega_4$.
Cyan contours represent the true object, whereas the approximated object section is
shown in magenta. 
(h) Approximate object $\Omega_5$ obtained from $\Omega_4$ using
(\ref{updatedguess2td_impl}). A new component is detected.
(i) Slice $x=5$ of the topological derivative when $\Omega=\Omega_5$.
(j) Approximate object $\Omega_6$ obtained from $\Omega_5$ using
(\ref{updatedguess2td_impl}). The third component, corresponding to the sphere, is detected.
(k,l) Intermediate and final approximate objects $\Omega_{10}$ and $\Omega_{24}$ obtained by the IRGNM.
}
\label{fig8}
\end{figure}

Finally, in Figure \ref{fig8} (see also Video 5 in the supplementary material), we illustrate the performance of the hybrid algorithm in a configuration with three objects of different sizes and shapes. One of them is a bean, a non-star shaped object described by the parameterization
\begin{equation}\label{bean1}
(5,5,6)+0.17(A(\theta)\sin\theta\cos\phi, B(\theta)\sin\theta\sin\phi-0.3 C(\theta),\cos\theta),
\end{equation}
with
$A(\theta)\!=\!\sqrt{0.64(1\!-\!0.1C(\theta))}$,  $B(\theta)\!=\!\sqrt{0.64(1\!-\!0.4C(\theta))}$, $C(\theta)=\cos(\pi\cos\theta)$
for $\theta\in[0,\pi]$, $\phi\in[0,2\pi]$.
The first  identified object is the bean, that is closer to the screen. When the algorithm carries out a topological derivative iteration to update the number of objects, the existing approximation resembles an ellipsoid rather than a bean. After this update, only the ellipsoid is detected because is bigger than the sphere, which is not seen. Once this new component is found, the algorithm requires a new topological derivative iteration,  finally determining the correct number of objects. At the 6th iteration (see panel (j)), the three objects resemble ellipsoids, but in a few iterations the true shapes are approximated with accuracy (panel (l)).

\begin{figure}
\centering
\hskip 0.0cm (a) \hskip 3.25cm (b) \hskip 3.25cm (c) \\
\includegraphics[width=4cm]{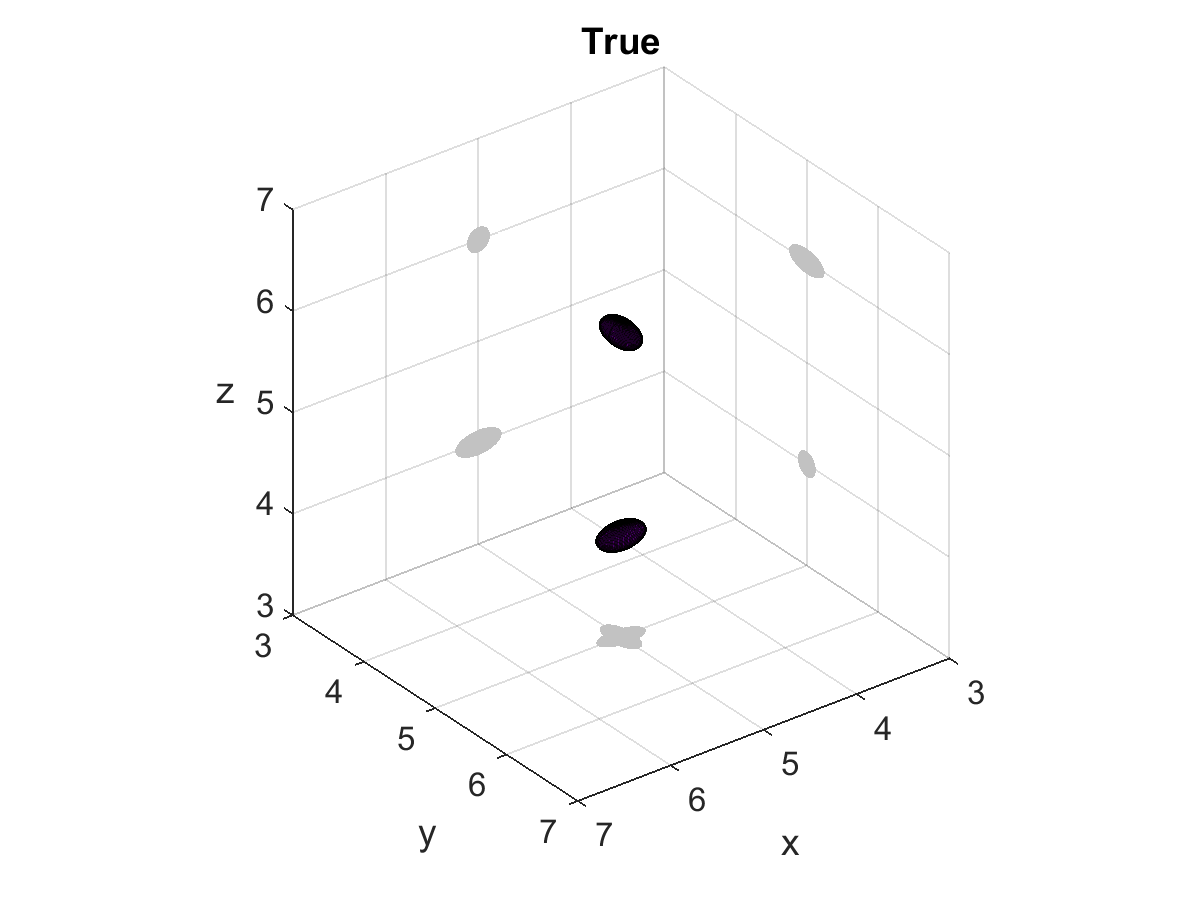}\hskip -2mm
\includegraphics[width=4cm]{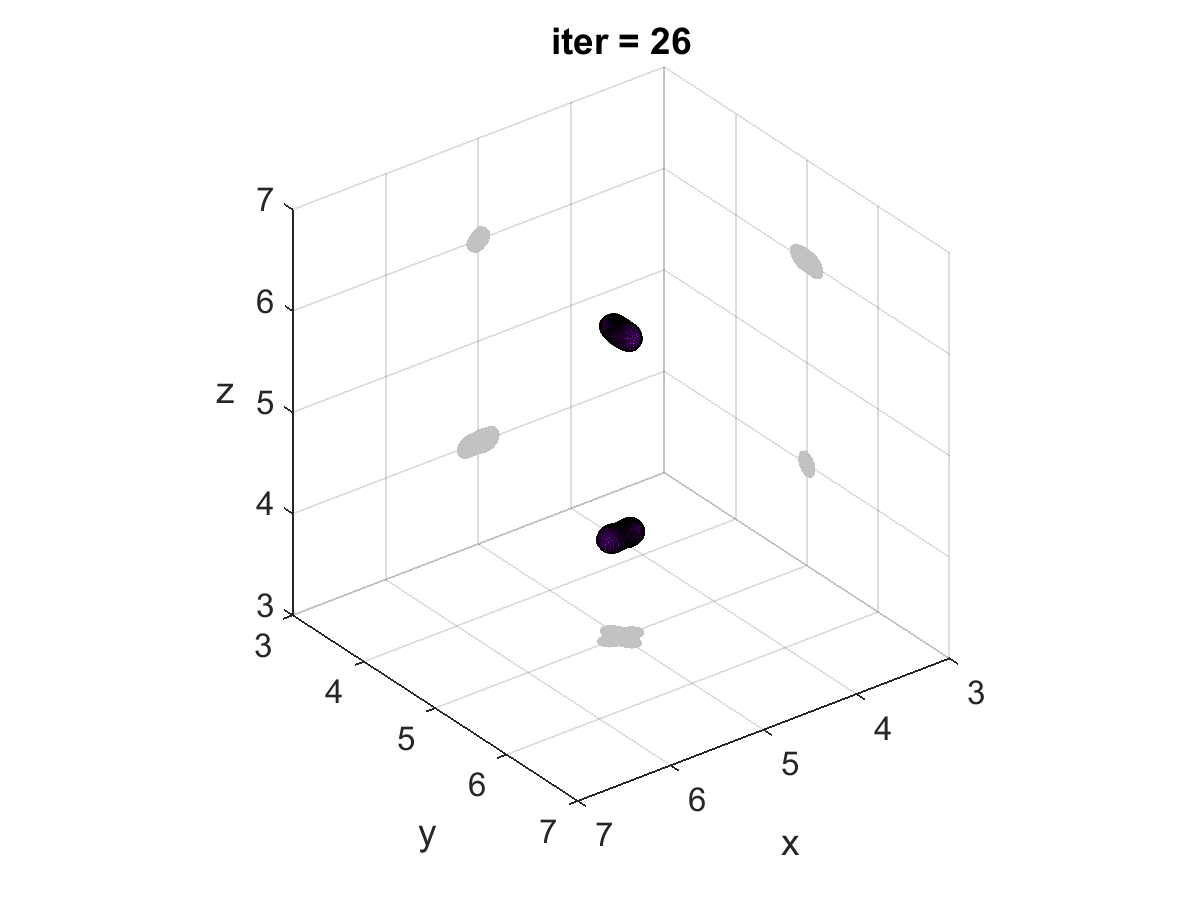}\hskip -2mm
\includegraphics[width=4cm]{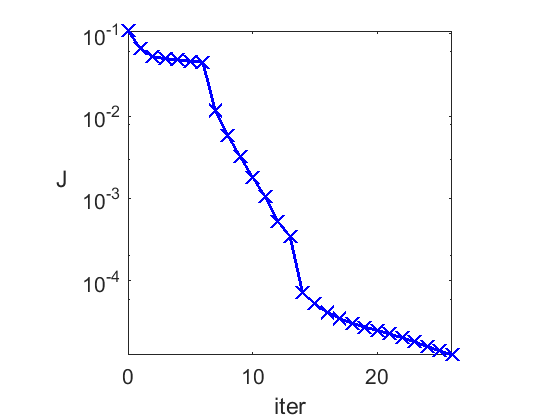} \\
\hskip 0.0cm (d) \hskip 3.25cm (e) \hskip 3.25cm (f) \\
\includegraphics[width=4cm]{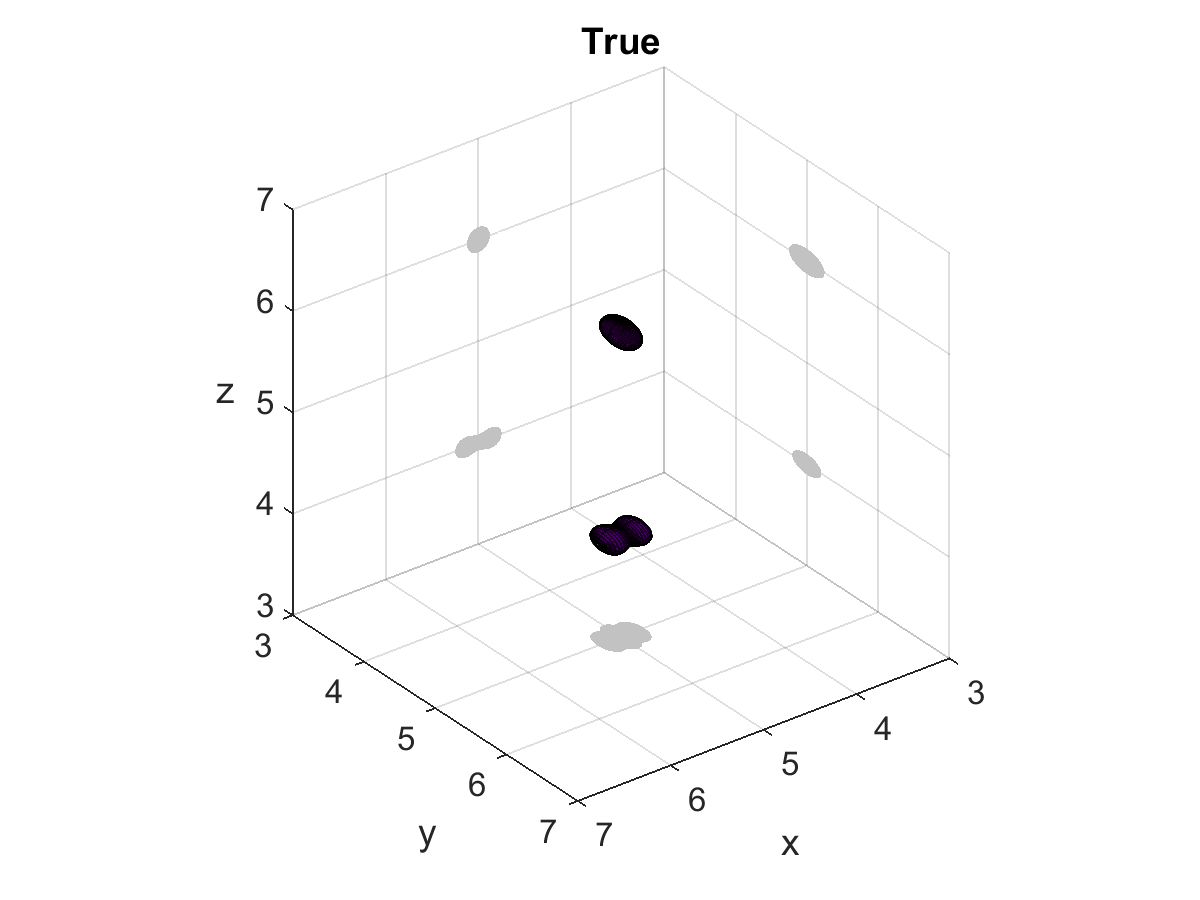}\hskip -2mm
\includegraphics[width=4cm]{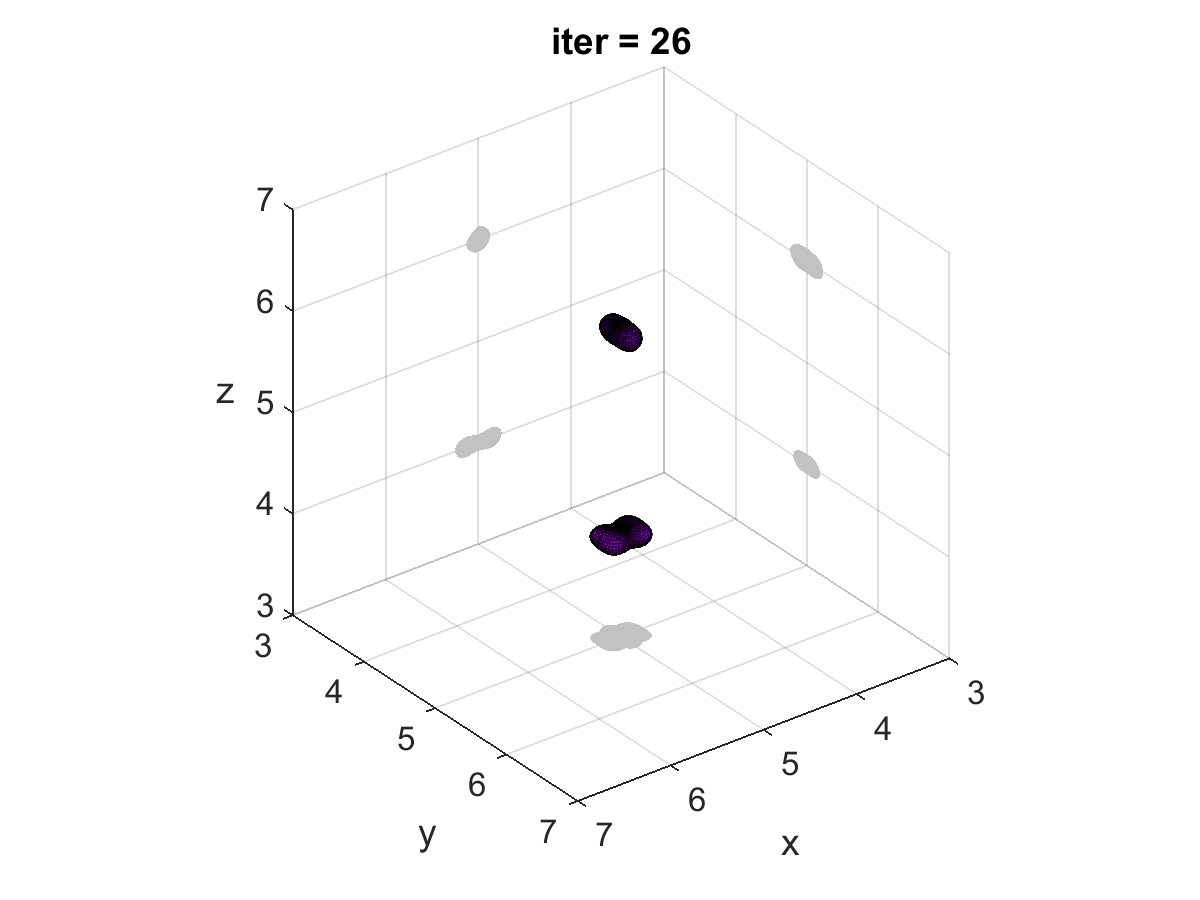}\hskip -2mm
\includegraphics[width=4cm]{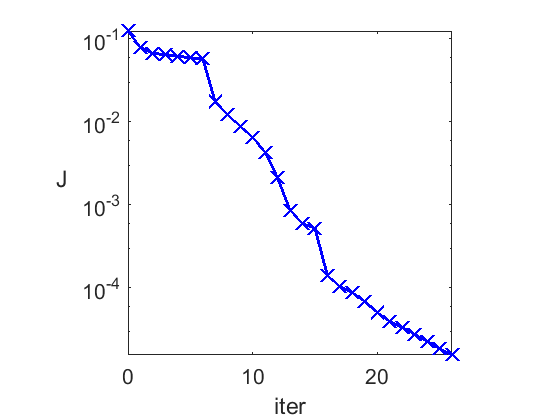} \\
\hskip 0.0cm (g) \hskip 3.25cm (h) \hskip 3.25cm (i) \\
\includegraphics[width=4cm]{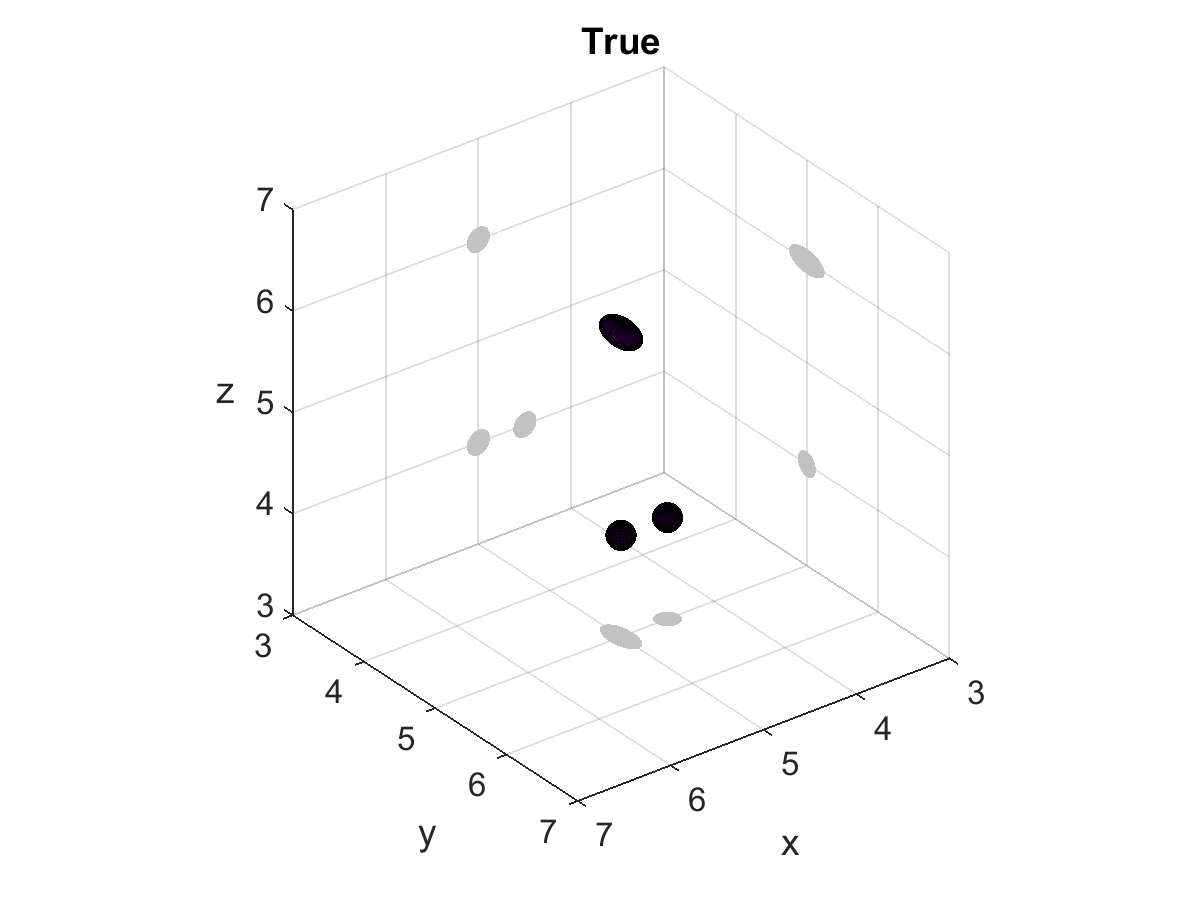}\hskip -2mm
\includegraphics[width=4cm]{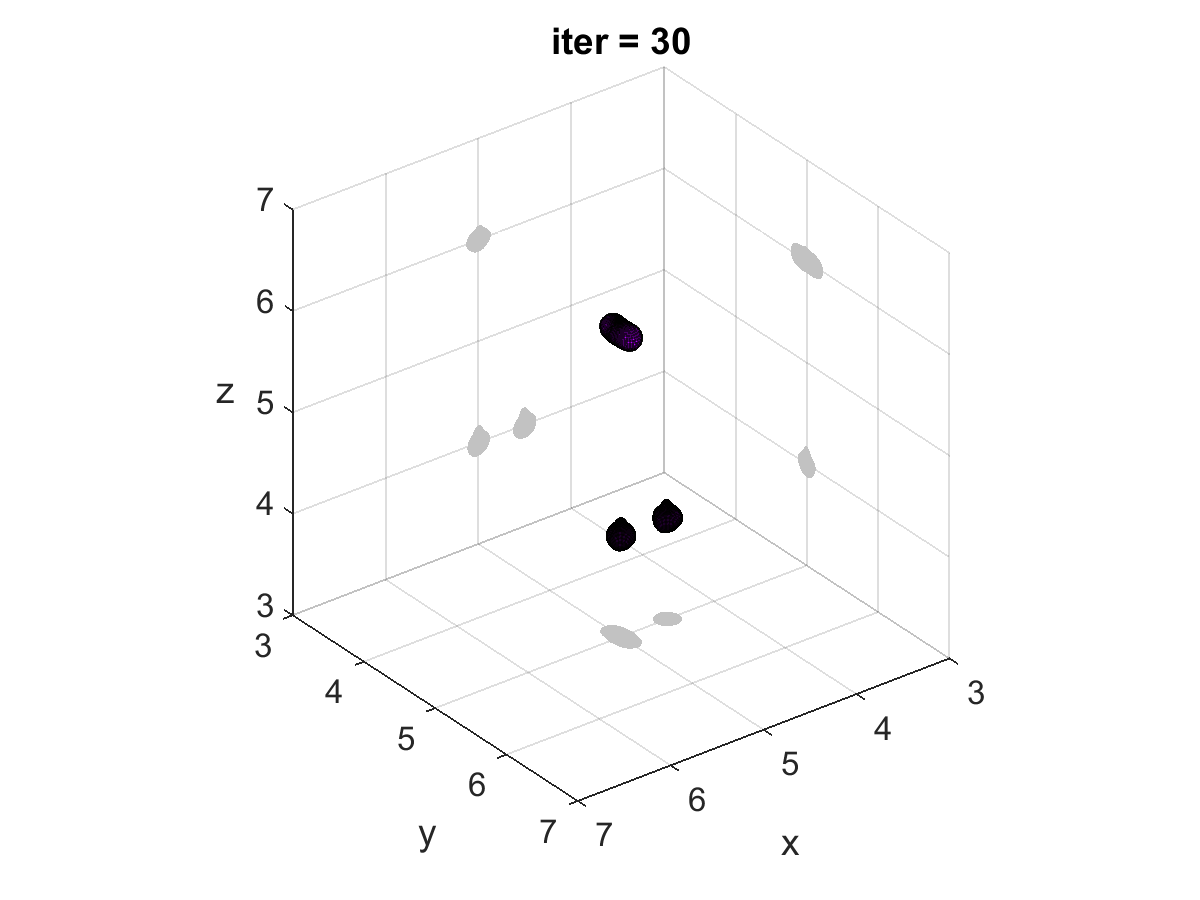}\hskip -2mm
\includegraphics[width=4cm]{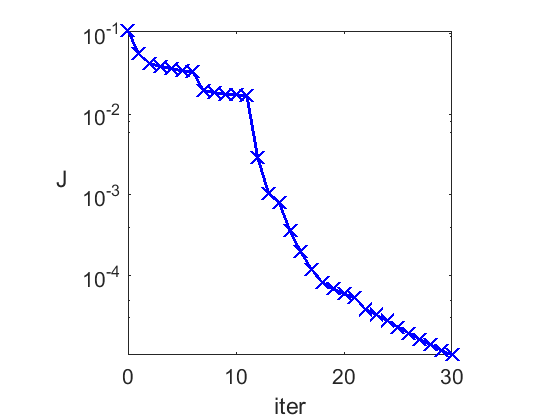} \\
\hskip 0.0cm (j) \hskip 3.25cm (k)  \hskip 3.25cm (l)\\
\includegraphics[width=4cm]{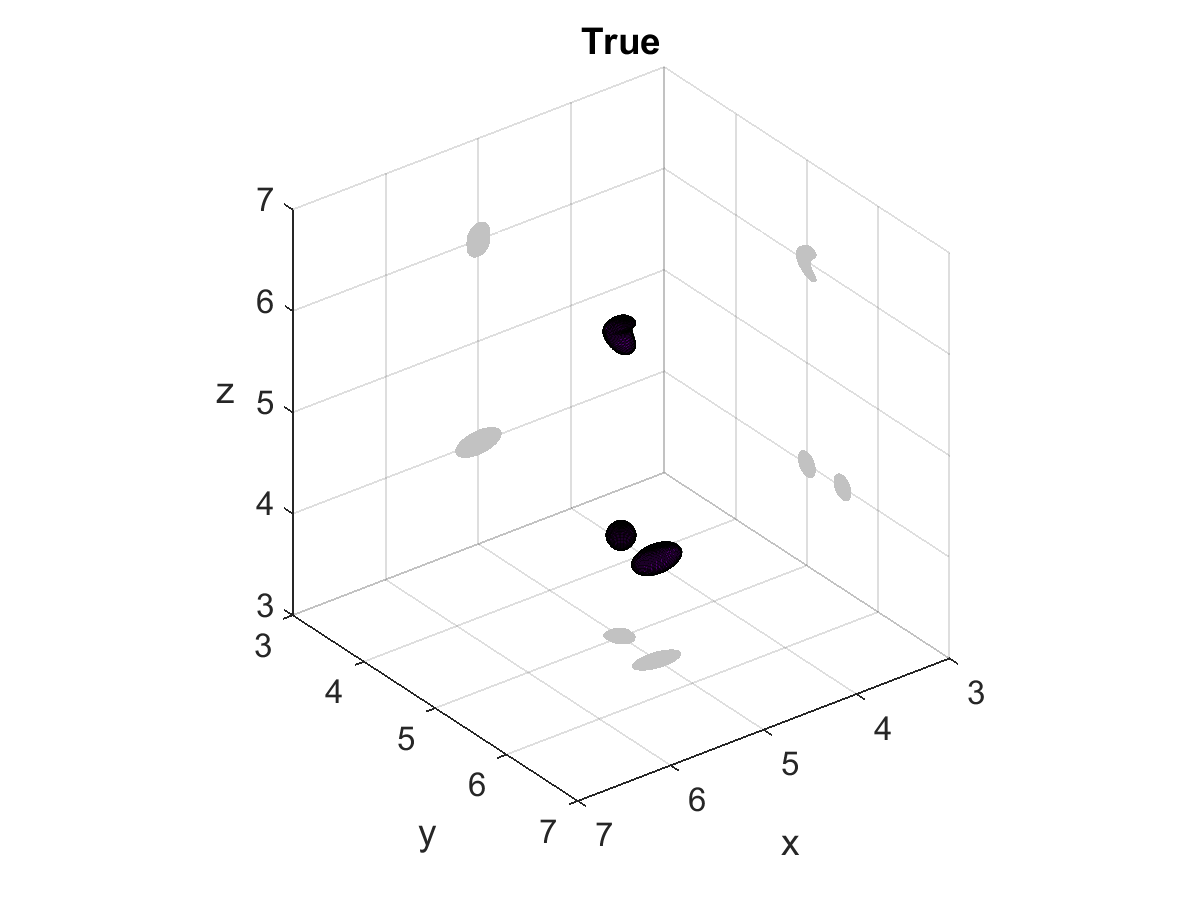}\hskip -2mm
\includegraphics[width=4cm]{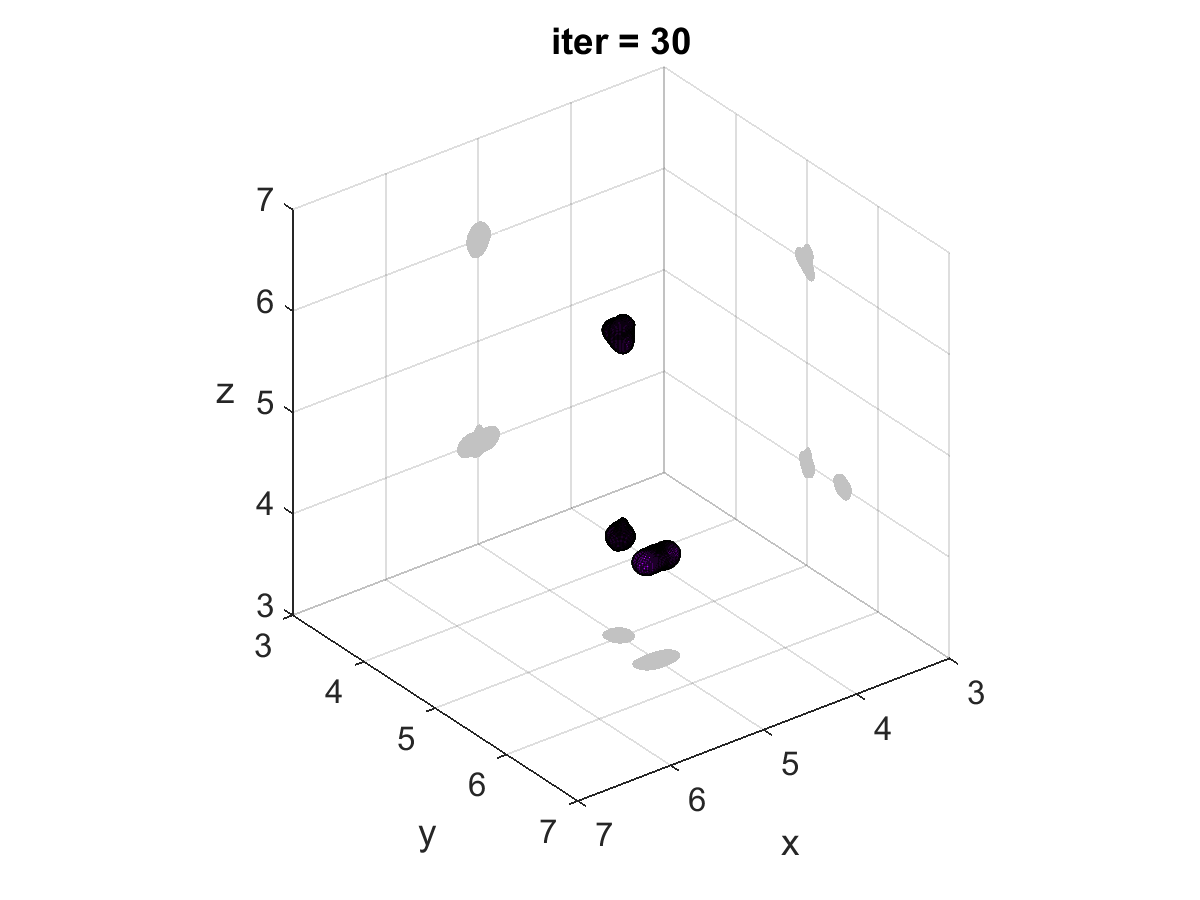}\hskip -2mm
\includegraphics[width=4cm]{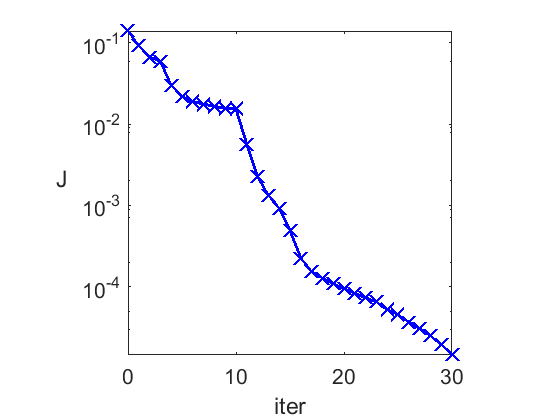}
\caption{ \small Performance of the hybrid TD/IRGN algorithm for different configurations of scatterers when only $121$ detectors are placed on the screen. First column: true geometries. Second column: final reconstructions. Third column: evolution of the cost functional.
}
\label{fig9}
\end{figure}

Remarkably, the quality of the previous approximations is maintained when reducing drastically the number of detectors, in agreement with the observations made
in \cite{dimidukrandom}.  Let us consider a much coarser detector grid formed by just $121$ detectors located at the points ${\bf x}_{k\ell}=(k,\ell,10)$, $k,\ell=0,\dots,10$. Figure \ref{fig9} shows the results obtained for different sets of scatterers using violet light. Comparing Figure \ref{fig9}(b,e,h,k) with Figure \ref{fig5_bis}(e), Figure \ref{fig6}(l), Figure \ref{fig7_bis}(c) and Figure \ref{fig8}(l), respectively, we observe that the quality of the approximations is similar.

\section{Numerical solution of boundary value and linearized problems}
\label{sec:numerical}

The optimization techniques to address the inverse holography
problem described in Sections \ref{sec:priors}-\ref{sec:algorithm}
rely on the availability of adequate solvers
for the associated forward and adjoint problems on one side,
and  the characterization of the Fr\'echet derivatives on the other.
All of them can be rewritten to adopt the same mathematical
structure.

\subsection{Forward and adjoint fields}
\label{sec:bvp}

Expressing the total light field $\mathbf E$ in terms  of transmitted
${\mathbf E}_{\rm tr}$, scattered ${\mathbf E}_{\rm sc}$ and incident
${\mathbf E}_{\rm inc}$ wave fields, and the  conjugate adjoint field in a
similar way where $\overline{\mathbf P}_{\rm inc}$ is the solution of the
conjugate adjoint problem in the whole space (\ref{adjointexplicit}) we find
\begin{eqnarray}
{\mathbf E} = \left\{ \begin{array}{ll}
{\mathbf E}_{\rm sc} + {\mathbf E}_{\rm inc} &  {\rm in} \,
\mathbb R^3 \setminus{\overline{\Omega}}, \\
{\mathbf E}_{\rm tr}                 & {\rm in} \, \Omega, \\
\end{array}\right.
\qquad
\overline{\mathbf P} = \left\{ \begin{array}{ll}
\overline{\mathbf P}_{\rm sc} + \overline{\mathbf P}_{\rm inc} &  {\rm in} \,
\mathbb R^3 \setminus{\overline{\Omega}}, \\
\overline{\mathbf P}_{\rm tr}     & {\rm in} \, \Omega, \\
\end{array}\right.
\end{eqnarray}
systems (\ref{forwardomega}) and (\ref{adjointomega}) governing the forward and
adjoint fields become
\begin{eqnarray} \label{general}
\begin{array}{ll}
\mathbf{curl} \,  (\mathbf{curl} \,  \mathbf W_{\rm sc})  - k_e^2 \mathbf W_{\rm sc} =0
& \mbox{in $\mathbb R^3\setminus\overline{\Omega}$},   \\
\mathbf{curl} \,  (\mathbf{curl} \,  \mathbf W_{\rm tr})  - k_i^2 \mathbf W_{\rm tr} =0
& \mbox{in $\Omega$},   \\
\hat{\mathbf n}  \times \mathbf W_{\rm tr} - \hat{\mathbf n}  \times \mathbf W_{\rm sc}
= \hat{\mathbf n} \times \mathbf W_{\rm inc},
& \mbox{on $\partial \Omega$} \\
\beta \, \hat{\mathbf n} \times \mathbf{curl} \,  \mathbf W_{\rm tr} -
\hat{\mathbf n} \times \mathbf{curl} \,  \mathbf W_{\rm sc} =
\hat{\mathbf n} \times \mathbf{curl} \,  \mathbf W_{\rm inc},
& \mbox{on $\partial \Omega$},    \\
{\rm lim}_{|\mathbf x| \rightarrow \infty} |\mathbf x|  \big|
\mathbf{curl} \,  \mathbf W_{\rm sc} \times \hat{\mathbf x}
-\imath k_e \mathbf W_{\rm sc} \big| =0, &
\end{array} \label{unified}
\end{eqnarray}
for $\mathbf W= \mathbf E$ or $\overline{\mathbf P}.$ In principle, any solver for transmission Maxwell problems can be used.

In our setting, for scatterers whose boundary is defined  by a 
${\cal C}^2$-parameteriza- tion (in particular, by a star-shaped one)
with piecewise constant refractive indexes,
we use fast boundary integral/spectral codes \cite{ganesh,lelouerspectral}.
Maxwell equations are solved by a Muller type boundary integral formulation
\cite{greengard,lelouerspectral,muller}.
The parametrization is exploited to transport the integral equations into unit
spheres and the system is solved by means of Galerkin schemes using
tangential   vector spherical harmonics of low order \cite{lelouerspectral}.
This approach may be faster than  implementing the discrete
dipole approximation  \cite{ddatheory,dimidukdda} or combing boundary
 element (BEM)/ finite element (FEM) in a 3D region
 \cite{siam2016,virginiafem,nedelec}.
Unlike these two latter methods, it is only applicable when a specific
parametrization is available. Fast multipole scattering methods would also
be adequate in the presence of many particles \cite{greengard2}.

\subsection{Linearized equation to correct parameterizations}
\label{sec:linearized}

The least squares problem (\ref{xik+1}) is minimized solving numerically
equation (\ref{xik+1bis})  in $H^s(\mathbb{S}^2)$ using
the conjugate gradient method with initial solution $\boldsymbol\xi_{k+1}=0$
and residual $10^{-2}$.
In each step of the conjugate gradient method, we must apply
the operators ${\cal I}'(\mathbf q_k)$ and ${\cal I}'(\mathbf q_k)^*$
to known vectors. Applying ${\cal I}'(\mathbf q_k)$ to a vector
$\boldsymbol \xi$ amounts to solving system (\ref{characterization}).
System (\ref{characterization}) has the structure (\ref{unified})
with transmission data (\ref{gD})-(\ref{gN}). The techniques
mentioned in Section \ref{sec:bvp} can be used to solve it.

To compute numerically the adjoint (\ref{adjoint3}) there are two possibilities, depending on how we evaluate the Cauchy data $(\hat{\mathbf n} \times \,  \overline{\mathbf E^+_{\boldsymbol h}}, \hat{\mathbf n} \times \mathbf{curl} \,  \overline{\mathbf E^+_{\boldsymbol h}} )$:
\begin{itemize}
\item[(i)] Either we compute the data $(\hat{\mathbf n} \times \,  {\mathbf E^+_{\boldsymbol h}}, \hat{\mathbf n} \times \mathbf{curl} \,  {\mathbf E^+_{\boldsymbol h}} )$ by solving the transmission problem with the incident field $\mathbf E^{inc}_{\boldsymbol h}$ as indicated above. Then we take the conjugate.
\item[(ii)] Or else, we compute the hermitian adjoint (transpose of the conjugate matrix) of the discretized approximation of the operator
$$\begin{pmatrix}g_{D}\\g_{N}\end{pmatrix}\mapsto {\cal I}'(\mathbf q_{\rm ap}) \boldsymbol \xi$$
that provides the vector $(\hat{\mathbf n} \times   \overline{\mathbf E^+_{\boldsymbol h}}, \hat{\mathbf n} \times \mathbf{curl} \,  \overline{\mathbf E^+_{\boldsymbol h}} )$. This is our choice. Indeed, we have
\begin{eqnarray}
\boldsymbol h\cdot {\cal I}'(\mathbf q_{\rm ap})\boldsymbol\xi={\rm Re}\left(\langle g_{D},\hat{\mathbf n} \times   \overline{\mathbf E^+_{\boldsymbol h}}\rangle_{L^2}+\langle g_{N},\hat{\mathbf n} \times \mathbf{curl} \,  \overline{\mathbf E^+_{\boldsymbol h}}\rangle_{L^2}\right). \label{adjoint4}
\end{eqnarray}
\end{itemize}

\section{Conclusions}
\label{sec:conclusions}

We have proposed a fully automatic algorithm to numerically reconstruct
3D objects from the holograms they generate. The method seeks
to minimize the difference when comparing the true holograms
and the holograms generated by approximate objects as predicted
by a Maxwell forward model of light scattering.
The topological derivative of this error functional provides an initial guess of the holographied objects in the absence of a priori information, other than the ambient refractive index and the incident wave. Working with star-shaped parameterizations, we implement an iteratively regularized Gauss-Newton method to successively correct the parametrization by solving linearized problems.  Automatically combined with additional topological derivative based iterations
to create or destroy objects when the decrease in the error functional stagnates, this procedure yields accurate reconstructions of a variety of objects in an experimental holography setting.  This scheme is quite general and can be applied to other inverse electromagnetic problems by changing the error functional, so that it is defined in terms of the adequate data. This usually requires adjusting the sources in the adjoint problems and adjusting the Fr\'echet derivatives.

In general `nonlinear least squares' problems, gradient methods seek to deform initial domains in the direction in which some kind of derivative of the error functional is negative, so that the functional decreases. This kind of methods includes topological derivative (TD), shape derivative (SD) and level set (LS) based optimization.  Instead, Gauss-Newton type methods aim to correct the current approximation by linearizing about it and solving the resulting problem.  Even when no spurious solutions are introduced by these procedures, they all may suffer from stagnation in the direction of  small `gradient'.  Here, the inclusion of Tikhonov regularizing terms in Gauss-Newton methods allows for convergence avoiding this artifact. From a technical point of view, while Gauss-Newton approaches require an explicit expression for a Fr\'echet derivative with respect to the domain, descent methods based on TD, SD or LS rely on explicit expressions for the shape/topological derivatives. Such expressions can usually be obtained introducing auxiliary adjoint problems without the need of characterizing the Fr\'echet derivative. The availability of these latter characterizations allows us to implement the IRGNM considered here.

We have focused on recovering shapes assuming the refractive indexes of the objects known. To obtain both we may combine our algorithm for shape optimization with parameter optimization, as in \cite{siam2018} or explore bayesian techniques \cite{bayesian}.
Moreover, this paper considers object sizes of the same or smaller order  than the employed light wavelength. The design of effective methods for larger objects and for objects  whose components cannot be well approximated by star-shaped descriptions will require further developments.

In principle, the algorithm we propose could be extended to other inverse electromagnetic scattering problems by adjusting the closed-form formulas for the pertinent derivatives, and also to other inverse scattering problems, provided closed-form formulas for the derivatives  as well as solvers for forward/adjoint problems are available (acoustics, elastography...).

\section*{Acknowledgments}

This research has been supported by MINECO grants No.
MTM2017-84446-C2-1-R (AC, MLR) and TRA2016-75075-R (MLR) as well as by
sabbatical funds from Fundaci\'on Caja Madrid and the Salvador de Madariaga 
Program PRX18/00112 (AC).
T.G. Dimiduk and A. Carpio thank V.N. Manoharan for discussions of holography and the Kavli Institute Seminars at Harvard for the interdisciplinary communication environment
that initiated this work. A. Carpio thanks M.P. Brenner for hospitality while visiting Harvard  University and R.E. Caflisch for hospitality while visiting the Courant Institute at NYU.

\appendix

\section{Comparison with scalar approximations}
\label{sec:explicitwhole}

When working with an incident wave  polarized in the $x$ or $y$ directions, such as  
$\mathbf E_{\rm inc} = (E_{\rm inc},0,0)$, $E_{\rm inc} =e^{\imath k_e z},$ a standard approximation sets the non polarized components of the electric field equal to zero and uses a scalar Helmholtz transmission problem to approximate the polarized component $E_1$.
In this framework, the inverse problem (\ref{eqinverse}) is reformulated as finding $\Omega$ such that the solution $E_{\Omega}$ of the scalar forward problem
\begin{eqnarray*}
\begin{array}{l}
\Delta E - k_e^2  E =0  \; \mbox{in}  \; \mathbb R^3\setminus\overline{\Omega},  \qquad
\Delta E - k_i^2  E  =0  \; \mbox{in}  \; \Omega,  \quad  \\
E^- = E^+  \; \mbox{on}  \;  \partial  \Omega,  \qquad
\beta {\partial E ^-\over \partial \hat{\mathbf n}}  =
{\partial E ^+\over \partial \hat{\mathbf n}}
 \; \mbox{on}  \; \partial \Omega,   \\
{\rm lim}_{| \mathbf x | \rightarrow \infty} | \mathbf x |  \big|
{\partial \over \partial | \mathbf x |} (E - E_{\rm inc}) -
\imath k_e (E - E_{\rm inc}) \big| =0,
\end{array}
\label{forwardscalar}
\end{eqnarray*}
satisfies the equations $I_{\rm meas}(\mathbf x_j) = |E_\Omega(\mathbf x_j)|^2$, $j=1,\dots, N$, or is a global minimum of the functional ${1\over 2} \sum_{j=1}^N |I_{\rm meas}(\mathbf x_j) - |E_\Omega(\mathbf x_j)|^2|^2.$ A scalar version of the topological
methods explained in Section \ref{sec:priors} was implemented in \cite{siam2018}.

This strategy may be reasonable in the presence of isolated smooth scatterers, placed far enough from the detector screen. However, it introduces a number of errors of varying magnitude:
\begin{itemize}
\item First,  $I_{\rm meas}(\mathbf x_j) = E_1(\mathbf x_j)^2 + E_2(\mathbf x_j)^2 +E_3(\mathbf x_j)^3$ where $\mathbf E =(E_1,E_2,E_3)$ is the solution of the vector Maxwell system (\ref{forwardadim}) for the true object $\Omega$.  The boundary conditions couple the three components of the electric field at the boundary of the object. However, if we work with well separated objects and the detectors are placed far enough from them, the value of $E_2(\mathbf x_j)^2 +E_3(\mathbf x_j)^2$ is expected to be small.

\item   The topological derivative for the scalar problem in the whole
space is $D_T(\mathbf x, \mathbb R^3)= {\rm Re} [(k_e^2-k_i^2) E \overline{Q}]$ where $E=E_{\rm inc}$ and the scalar adjoint $\overline
Q$ is an outgoing solution of
\begin{eqnarray*} \begin{array}{l}
-\Delta \overline Q - k_e^2 \overline Q =  \sum_{j=1}^N d_{j} \delta_{\mathbf x_j}, \quad d_{j} = 2 (I_{\rm meas}(\mathbf x_j) -  |E(\mathbf x_j)|^2 )\overline E(\mathbf x_j).
\end{array} \end{eqnarray*}
Particularizing formula (\ref{DT}) for the vector Maxwell problem with
polarized light and $\beta=1$, we find $D_T(\mathbf x, \mathbb R^3)={3 k_e^2 \over k_i^2+ 2 k_e^2} {\rm Re} [(k_e^2-k_i^2) E \overline{P}_1]$, where $\mathbf P = (P_1,P_2,P_3)$
is given by (\ref{adjointempty}).
To quantify the difference between both  formulas for topological derivatives, we need to relate $Q$ and $P_1$. We find a term of the form $k_e^{-2} \sum_{j=1}^N \nabla {\rm div} G_{k_e} * d_{j} \delta_{\mathbf x_j}$, where $G_{k_e} $ is the Green function of the Helmholtz equation. This error term decays with the distance  to the detectors.

Indeed, conjugating system (\ref{adjointempty}) we obtain:
\begin{eqnarray} \label{conjadjointempty}
\begin{array}{ll}
\mathbf{curl} \, (  \mathbf{curl} \,  \overline{\mathbf P} )
- k_e^2 \overline{\mathbf P} =
2 \sum_{j=1}^N (I_{\rm meas} - |\mathbf E_{\rm inc}|^2) \overline{\mathbf E}_{\rm inc} \,
\delta_{\mathbf x_j} &
\quad \mbox{in $\mathbb R^3$}, \\[1ex]
{\rm lim}_{|\mathbf x| \rightarrow \infty} |\mathbf x|  \big|
\mathbf{curl} \,  \overline{\mathbf P} \times \hat{\mathbf x}
- \imath k_e \overline{\mathbf P} \big| =0.    &
\end{array}
\end{eqnarray}
For $j=1,\ldots, N$, we take the divergence $\mathbf{curl} \, ( \mathbf{curl} \,  \overline{\mathbf P}_j ) - k_e^2 \overline{\mathbf P}_j = {\mathbf d_j} \delta_{\mathbf x_j}$, where $\mathbf d_j= 2(I_{\rm meas}(\mathbf x_j)  - |\mathbf E(\mathbf x_j) |^2) \overline{\mathbf E(\mathbf x_j) }$. Since
${\rm div} \,  (\mathbf{curl} \,  \mathbf A)=0$ for any vector $\mathbf A$, we find
$ {\rm div} \,  {\overline{\mathbf P}_j} =
- {1\over k_e^2} {\rm div} \,  {\mathbf d}_j \delta_{\mathbf x_j}. $
Making use of the vector identity
$\mathbf{curl} \, ( \mathbf{curl} \,  \overline{\mathbf P}_j ) = \nabla ( {\rm div} \,  \overline{\mathbf P}_j) - \Delta {\overline{\mathbf P}_j} $
we have \begin{eqnarray*}
\begin{array}{r}
-\Delta {\overline{\mathbf P}_j} - k_e^2 {\overline{\mathbf P}_j} = \delta_{\mathbf x_j} {\mathbf d}_j
+  {1\over k_e^2} \nabla ({\rm div} \,  {\mathbf d}_j \delta_{\mathbf x_j}).
\end{array}
\end{eqnarray*}
We can solve the equations by convolution with the Green function of Helmholtz equation:
\begin{eqnarray}
\label{difference}
\begin{array}{r}
\overline{\mathbf P}_j = G_{k_e}  * {\mathbf d}_j  \delta_{\mathbf x_j} +
{1\over k_e^2}  G_{k_e}  * \nabla ({\rm div} \,  {\mathbf d} _j\delta_{\mathbf x_j}).
\end{array}
\end{eqnarray}
This expression quantifies the difference between the Green function of
Maxwell equation  and Green functions of Helmholtz
equations. Indeed, notice that the right hand side can be rewritten as
$G_{k_e}  * {\mathbf d}_j  \delta_{\mathbf x_j} + {1\over k_e^2}  G_{k_e}  *
[\mathbf{curl} \,  \mathbf{curl} \,  {\mathbf d}_j \delta_{\mathbf x_j}
+ \Delta {\mathbf d}_j \delta_{\mathbf x_j}].$
Interchanging the derivatives in the convolution we find:
\begin{eqnarray*}
\label{adjointexplicit4}
\begin{array}{r}
\overline{\mathbf P}_j(\mathbf x) =
{1\over k_e^2}  \mathbf{curl} \,  \mathbf{curl} \,
G_{k_e}(\mathbf x - \mathbf x_j) {\mathbf d}_j(\mathbf x_j).
\end{array}
\end{eqnarray*}
for $\mathbf x \neq \mathbf x_j$.
Summing over $j$, this  implies that formula (\ref{adjointexplicit}) for the adjoint
holds pointwise as long as we do not reach the screen where the hologram is recorded.

\item When we consider the topological derivatives in the presence of approximated objects $\Omega_{\rm ap}\neq \emptyset$, the
differences between the vector and the scalar approach increase, since
the expression for the topological derivative (\ref{DTsimple}) becomes discontinuous across the boundary and the coupling of the field components at the boundary in  (\ref{forwardomega}) and (\ref{adjointomega}) produces nonzero components for
the forward and adjoint fields.
\end{itemize}

\section{Fr\'echet, shape and topological derivatives for holography}
\label{sec:derivatives}

In this section we justify the formulas for the Fr\'echet, shape and
topological derivatives for  the hologram and the cost functional
(\ref{costH}) employed throughout the paper.

Given an object $\Omega \subset \mathbb R^3$ with boundary
$\Gamma=\partial \Omega$, we consider variations of $\Omega$
along vector fields $\boldsymbol \xi.$ For $\tau >0$,
we introduce the family of deformations
\begin{eqnarray}\begin{array}{l}
\varphi_{\tau \boldsymbol \xi}(\mathbf x)=
\mathbf x + \tau \boldsymbol \xi(\mathbf x),
\quad \mathbf x \in \mathbb R^3,
\label{deformation}
\end{array}\end{eqnarray}
and consider the deformed domains
$\varphi_{\tau \boldsymbol \xi}(\Omega)=\Omega_{\tau \boldsymbol \xi}.$
Let us quantify first the variations of the hologram under such deformations.

\subsection{Fr\'echet derivative of the hologram}
\label{sec:frechet}

The Fr\'echet derivative extends the concept of differential
to general spaces of infinite dimension.
Given two Banach spaces $X$, $Y$ and  a function
${\cal F}: D({\cal F}) \subset X \longrightarrow Y$, its Fr\'echet derivative
${\cal F}': X \longrightarrow Y$ is a linear bounded operator satisfying
${\cal F}(x+ \xi) = {\cal F}(x) + {\cal F}'(x) \xi +  o (\xi)$
for $\xi \in X$ as $\| \xi \|_X \rightarrow 0$, for any $x \in X$. In other
words, it satisfies
$ {\rm lim}_{\|\xi \|_X \rightarrow 0} { \| {\cal F}(x+ \xi)  - {\cal F}(x) -
{\cal F}'(x) \xi \|_Y \over  \|\xi \|_X} =0. $
It is related to the directional Gateaux derivative
\begin{eqnarray}\begin{array}{l}
D_\xi{\cal F}(x)
= {\rm lim}_{\tau \rightarrow 0} {{\cal F}(x+ \tau \xi) - {\cal F}(x)
\over \tau } \end{array} \label{gateaux} \end{eqnarray}
by ${\cal F}'(x) \xi =  D_\xi {\cal F}(x).$

Let us introduce the operator that maps the deformation of the
object boundary to  the solution of the forward problem evaluated at
the detectors:
\begin{eqnarray}\begin{array}{ll}
{\cal F}: & D({\cal F}) \subset C^k(\Gamma, \mathbb R^3) 
\longrightarrow {\cal M}_{N\times 3}(\mathbb C) \\
& {\boldsymbol \xi} \longrightarrow (\mathbf E_{\Omega_{\boldsymbol \xi}}
(\mathbf x_1),\ldots, \mathbf E_{\Omega_{\boldsymbol \xi}}(\mathbf x_N))^t,
\end{array} \label{mapholo} \end{eqnarray}
where  $k \geq 2$, ${\mathbf E}_{\Omega_{\boldsymbol \xi}}$ is the solution of
(\ref{forwardadim}) with $\Omega={\Omega_{\boldsymbol \xi}},$
$\mathbf x_j$, $j=1,\ldots,N,$ are detectors placed on a screen.
Then, we can write the corresponding hologram as
${\cal I}({\boldsymbol \xi})_j=
| \mathbf E_{\Omega_{\boldsymbol \xi}} (\mathbf x_j) |^2 =
{\cal F}({\boldsymbol \xi})_j \cdot \overline{{\cal F}({\boldsymbol \xi})_j},$
$j=1,\ldots,N.$   \\

{\bf Theorem 1.}{\it
The Fr\'echet derivative of the operator ${\cal I}({\boldsymbol \xi})$  and its adjoint operator are given by (\ref{frechetholo}) and (\ref{adjoint3}), respectively.}

{\bf Proof.} The Gateaux derivative $\dot {\mathbf E}$ of the solution of (\ref{forwardadim}) in the direction $\boldsymbol \xi$ is characterized as the solution of (\ref{characterization}) in  \cite{lelouerderivative}, Theorem 6.6. This provides the Fr\'echet derivative of the operator ${\cal F}$ defined by (\ref{mapholo}). Then,
\begin{eqnarray*}\begin{array}{l}
{\cal I}'(0) \boldsymbol \xi = 2  \left( {\rm Re} [ \overline{{\cal F}(0)_j}
\cdot ({\cal F}'(0)\boldsymbol \xi)_j] \right)_{j=1}^N
= 2 \left(  {\rm Re} [
\overline{\mathbf  E_{\Omega_{\boldsymbol \xi}}(\mathbf x_j) }
\cdot \dot{\mathbf E}(\mathbf x_j)] \right)_{j=1}^N  .
\end{array}
 \end{eqnarray*}
We deduce that
$${{\cal I}'}^*(0)\boldsymbol h =  2{{\cal F}'}^*(0) \overline{{\cal F}(0)}\boldsymbol h ,  $$
where ${{\cal F}'}^*(0)$ is characterized in \cite{lelouerspectral},  Proposition 6.
The only difference with respect to the use of full measurements (phase+intensity) \cite{lelouerspectral} is the multiplication of the incident adjoint field by $2\boldsymbol h$. $\square$  
\\

 Using the same formula we get similar results inspired by \cite[Eq. (1.3)]{HohageSchormann} for the scalar case.

{\bf Theorem 1bis. }{\it 
Using the scalar approximation for polarized waves, the Fr\'echet derivative of the operator 
${\cal I}({\boldsymbol \xi})$ } is given by 
\begin{eqnarray*} 
{\cal I}'(\mathbf q_{\rm ap}) \boldsymbol \xi = \begin{pmatrix}2 {\rm Re} \left[
\overline{  E_{\Omega_{\rm ap}}(\mathbf x_1)}  
\dot{ E}(\mathbf x_1)\right]\\\vdots\\2 {\rm Re} \left[
\overline{  E_{\Omega_{\rm ap}}(\mathbf x_N)}  
\dot{ E}(\mathbf x_N)\right]\end{pmatrix}, 
\end{eqnarray*}
where $\dot { E}$ is the solution of 
\begin{eqnarray*}
\begin{array}{l}
\Delta \dot{E} - k_e^2  \dot{E} =0  \; \mbox{in}  \; \mathbb R^3\setminus\overline{\Omega},  \qquad
\Delta \dot{E} - k_i^2  \dot{E}  =0  \; \mbox{in}  \; \Omega,  \quad  \\
\dot{E}^+ - \dot{E}^-=g_{D}  \; \mbox{on}  \;  \partial  \Omega,  \qquad
 {\partial \dot{E} ^+\over \partial \hat{\mathbf n}}   -
\beta{\partial \dot{E} ^-\over \partial \hat{\mathbf n}} =g_{N} 
 \; \mbox{on}  \; \partial \Omega,   \\
{\rm lim}_{| \mathbf x | \rightarrow \infty} | \mathbf x |  \big|
{\partial \over \partial | \mathbf x |} \dot{E}  - 
\imath k_e \dot{E}  \big| =0, 
\end{array} 
\end{eqnarray*}
with object $\Omega=\Omega_{\rm ap}$ and transmission data 
\begin{eqnarray*}
g_D= - ({\boldsymbol \xi} \cdot {\hat{\mathbf n}} ) \left( 
{\partial {E} ^+\over \partial \hat{\mathbf n}}-{\partial {E} ^-\over \partial \hat{\mathbf n}}
 \right), \\
g_N=  ({\boldsymbol \xi} \cdot {\hat{\mathbf n}}) 
\left( k_e^2    E^+
- k_i^2 \beta \, \  E^- \right)  + {\rm div}_{\partial\Omega} \left( ({\boldsymbol \xi} \cdot 
{\hat{\mathbf n}}) (\nabla E^+ - \beta  \nabla  E^-)  \right), 
\end{eqnarray*}
$ E=  E_{\Omega_{\rm ap}}$ being the solution of
(\ref{forwardscalar}) with $\Omega= \Omega_{\rm ap}$ and $\hat{\mathbf n}$
the outer unit normal.
The $L^2$-adjoint operator  is defined  by
\begin{eqnarray*}
\begin{aligned}
&    \boldsymbol\xi^*(\hat{\mathbf x})=  r^* (\hat{\mathbf x}) \hat{\mathbf x}={\cal I}'(\mathbf q_{\rm ap})_{|L^2}^*\boldsymbol h, \\
 r^*=\hat{\mathbf x}\cdot{\cal I}'(\mathbf q_{\rm ap})_{|L^2}^*\boldsymbol h=&\; r^2_{\rm ap}{\rm Re}\Big[-(1-\beta^{-1}) {\partial \overline{ E^+_{\boldsymbol h}}\over \partial \hat{\mathbf n}}\cdot {\partial   \overline{ E^+}\over \partial \hat{\mathbf n}}-({k_{e}^2}-{\beta k_{i}^2})  \overline{ E^+_{\boldsymbol h}}\cdot  \overline{  E^+}    \\
&   \hspace{1cm} +(1-\beta){\nabla}_{\partial\Omega}  \overline{ E^+_{\boldsymbol h}} \,\cdot\,{\nabla}_{\partial\Omega}\overline{  E^+} \Big]  \circ\mathbf q_{\rm ap},
\end{aligned}
\end{eqnarray*}
where ${ E^+_{\boldsymbol h}}$ is the solution of the transmission problem \eqref{forwardscalar} with object $\Omega=\Omega_{\rm ap}$ and incident field
$$ E^{inc}_{\boldsymbol h}(\mathbf x)=\sum_{j=1}^N\frac{e^{ik_e|\mathbf x- \mathbf x_{j}|}}
{4\pi |\mathbf x-\mathbf x_{j}|}2\boldsymbol h(\mathbf x_{j})\overline{ E_{\Omega_{\rm ap}}(\mathbf x_{j})}.$$

\subsection{Shape derivative of the cost functional}
\label{sec:sdcost}

When instead of the hologram we differentiate the cost functional (\ref{costH}) with respect to deformations along vector fields, we obtain the so-called shape derivative. Let us fix a vector field $\boldsymbol \xi$.
For any region ${\cal R}$, the deformed domain $\varphi_\tau({\cal R})$ is the image of ${\cal R}$ by the deformation (\ref{deformation}).
Evaluating ${ J}$ on the deformed regions, we obtain a scalar function $J(\tau)={J}(\varphi_\tau({\cal R}))$  of the deformation parameter $\tau$, which can be differentiated with respect to it. The shape derivative along the vector field $\boldsymbol \xi$ is  precisely this derivative:
\begin{eqnarray}\label{shapederivativedefinition}
\langle  D{J}({\cal R}), \boldsymbol \xi \rangle  :=
\frac{d}{d\tau}\,{J}(\varphi_\tau({\cal R}))
\bigg|_{\tau=0}.
\end{eqnarray}

{\bf Theorem 2.}
{\it Let us assume that $\Omega$ is a $C^2$ domain and the coefficients $k_e,k_i,\beta$ are piecewise constant in the different components. Then the shape derivative $\langle DJ(\mathbb R^3 \setminus \overline{\Omega}), \boldsymbol \xi  \rangle$ of functional (\ref{costH}) is given by:
\begin{eqnarray}
\begin{array}{l}
{\rm Re} \bigg[ \int_{\partial \Omega} (\boldsymbol \xi  \! \cdot \! \hat{\mathbf n})
(k_e^2 \!-\! k_i^2 \beta)
\left[ (\hat{\mathbf n} \! \times \! \mathbf E^-) \! \cdot \!(\hat{\mathbf n}
\! \times \! \overline{\mathbf P}^-) \!+\! {k_i^2  \beta \over k_e^2 }
(\hat{\mathbf n} \! \cdot  \!\mathbf E^-) \! \cdot \! (\hat{\mathbf n}
\! \cdot \! \overline{\mathbf P}^-) \right]  \\
\ -(\boldsymbol \xi  \! \cdot \! \hat{\mathbf n}) (1\!- \!\beta)
\left[\beta (\hat{\mathbf n} \! \times \! \mathbf{curl} \, \mathbf E^-) \! \cdot \!
(\hat{\mathbf n} \! \times \! \mathbf{curl} \, \overline{\mathbf P}^-) \!+\!
(\hat{\mathbf n} \! \cdot \! \mathbf{curl} \, \mathbf E^-) \! \cdot \!
(\hat{\mathbf n} \! \cdot \! \mathbf{curl} \, \overline{\mathbf P}^-)
\right] \bigg] d S,
\end{array}
\nonumber
\end{eqnarray}
where $\mathbf E$ and $\overline{\mathbf P}$ are the solutions of the
forward and conjugate adjoint problems (\ref{forwardomega}) and (\ref{adjointomega}).
}

{\bf Proof.} Given the vector field $\boldsymbol \xi $, we must evaluate
(\ref{shapederivativedefinition}).
Differentiating with respect to $\tau$ the function
$J(\mathbb R^3 \setminus \overline{\phi_\tau(\Omega)})$
with $J$ given by (\ref{costH}) we find a vector version of
the scalar formula in \cite{siam2018}:
\begin{eqnarray*}
\begin{array}{l}
{d J(\mathbb R^3 \setminus \overline{\phi_\tau(\Omega)})
\over d \tau}=
{\rm Re} \left[
\sum_{j=1}^N 2\left(|\mathbf E_\tau(\mathbf x_j)|^2  -  
I_{\rm meas}(\mathbf x_j)
\right)  \overline{\mathbf E}_\tau(\mathbf x_j)  \cdot {d \mathbf E_\tau
\over d \tau}(\mathbf x_j)   \right],
\end{array}
\end{eqnarray*}
where $\mathbf E_\tau$ is a solution of the forward problem
(\ref{forwardomega}) with object $\Omega_\tau=\varphi_\tau(\Omega).$
Evaluating at $\tau=0$, we obtain
\begin{eqnarray}
\begin{array}{l}
{d J(\mathbb R^3 \setminus \overline{\phi_\tau(\Omega)}) \over
d \tau} \Big|_{\tau =0}=
{\rm Re} \left[
\sum_{j=1}^N 2\left(|\mathbf E(\mathbf x_j)|^2  -
I_{\rm meas}(\mathbf x_j)
\right)  \overline{\mathbf E}(\mathbf x_j)  \cdot \dot{\mathbf E}(\mathbf x_j)
\right],
\end{array}
\label{sdinicial}
\end{eqnarray}
since $\mathbf E_0 = \mathbf E$ is the solution of  (\ref{forwardomega})
with object $\Omega$.

We set $\dot{\mathbf E} = {d \mathbf E_\tau \over d \tau} \Big|_{\tau=0}$,
which is the  Gateaux derivative in the direction $\boldsymbol \xi$ of the
solutions of the forward problems (\ref{forwardomega}) with deformed
domains $\Omega_\tau$.
For the transmission Maxwell system it is characterized as the solution
of (\ref{characterization}), see  \cite{lelouerderivative}. Then,
elliptic regularity and Sobolev's embeddings ensure continuity of the solution
and continuity of derivatives away from the  interface $\partial \Omega$
\cite{gilbart,grisvard,nedelec}.  Therefore, we can evaluate it at the detectors.

We  rewrite the right hand side in identity (\ref{sdinicial})  as follows.
Integrating the equations for the adjoint field and using Green's formula
we obtain the identities:
\begin{eqnarray}  \label{inner}
\hskip -1cm \begin{array}{l}
\int_{\Omega}
(\beta \, \mathbf{curl} \,  \mathbf{curl} \,  \overline{\mathbf P}
- \beta \, k_i^2 \overline{\mathbf P}) \cdot \dot{\mathbf E} -
\int_{\partial \Omega}  \beta  \ \hat{\mathbf n}
\times \mathbf{curl}\, \overline{\mathbf P}^-
\cdot \dot{\mathbf E}^-   \\
= \int_{\Omega}
(\beta \, \mathbf{curl} \,  \mathbf{curl} \,  \dot{\mathbf E}
- \beta \, k_i^2 \dot{\mathbf E} )  \cdot \overline{\mathbf P}  -
\int_{\partial \Omega}  \beta  \,  \hat{\mathbf n}
\times \mathbf{curl}\,  \dot{\mathbf E}^- \cdot \overline{\mathbf P}^-
\end{array}
\end{eqnarray}
and
\begin{eqnarray}  \label{outer}
\begin{array}{l}
\int_{B_R\setminus \overline{\Omega} }
( \mathbf{curl} \,  \mathbf{curl} \,  \overline{\mathbf P}
\!- \!k_e^2 \overline{\mathbf P}) \! \cdot \! \dot{\mathbf E} \!+\!
\int_{\partial \Omega}   \hskip -1mm
\hat{\mathbf n} \! \times \! \mathbf{curl}\,
\overline{\mathbf P}^+ \!\! \cdot \! \dot{\mathbf E}^+
\!\!-\!\! \int_{\partial B_R}  \hskip -2.5mm \imath k_e
L_{k_e}\!(\hat{\mathbf n}_R \! \times \! \overline{\mathbf P})
\! \cdot \! \dot{\mathbf E}
 \\
=\int_{B_R\setminus{\overline{\Omega}}}
( \mathbf{curl} \,  \mathbf{curl} \,  \dot{\mathbf E}
\!-\! k_e^2 \dot{\mathbf E} )  \! \cdot \! \overline{\mathbf P}  \!+\!
\int_{\partial \Omega}  \hskip -1mm
\hat{\mathbf n} \! \times \! \mathbf{curl}\,
\dot{\mathbf E}^+ \!\! \cdot \! \overline{\mathbf P}^+
\!\!-\!\! \int_{\partial B_R} \hskip -2.5mm \imath k_e
L_{k_e}\!(\hat{\mathbf n}_R \! \times \! \dot{\mathbf E} )
\! \cdot \! \overline{\mathbf P}.
\end{array}
\end{eqnarray}
Here, $B_R=B(\mathbf 0,R)$ is a sphere with radius $R$ large enough
to contain the objects and detectors and $L_{k_e}$ stands
for the Dirichlet to Neumann operator for Maxwell equations
\cite[Chapter 10]{monk}.
To be able to use the transmission boundary conditions for $\dot{\mathbf E}$
we combine (\ref{inner})-(\ref{outer}) to get:
\begin{eqnarray} \label{combination}
\hskip -1cm \begin{array}{l}
\sum_{j=1}^N 2\left(|\mathbf E(\mathbf x_j)|^2  - I_{\rm meas}(\mathbf x_j)
\right)  \overline{\mathbf E}(\mathbf x_j)  \cdot \dot{\mathbf E}(\mathbf x_j)
= \\
 - \int_{\partial \Omega}   \hat{\mathbf n} \times
(\mathbf{curl}\, \dot{\mathbf E}^+
- \beta \, \mathbf{curl}\, \dot{\mathbf E}^-) \cdot \overline{\mathbf P}^-
- \int_{\partial \Omega}  \hat{\mathbf n} \times
\mathbf{curl}\,  \dot{\mathbf E}^+
\cdot (\overline{\mathbf P}^+ - \overline{\mathbf P}^-)
\\
+ \int_{\partial \Omega}  \beta \, \hat{\mathbf n} \times \mathbf{curl}\,
\overline{\mathbf P}^- \cdot (\dot{\mathbf E}^+ -\dot{\mathbf E}^-)
- \int_{\partial \Omega}   \hat{\mathbf n} \times
(\beta \,  \mathbf{curl}\, \overline{\mathbf P}^-  -
 \mathbf{curl}\, \overline{\mathbf P}^+) \cdot \dot{\mathbf E}^+.
\end{array}
\end{eqnarray}
Following  \cite{lelouerrapun}, we analyze the 
boundary terms in this
identity using the boundary conditions satisfied by $\dot{\mathbf E}$,
$\mathbf E$, $\overline{\mathbf P}$, together with
classical vector identities and differential calculus relations
on surfaces \cite{nedelec}, to finally obtain the desired expression using
(\ref{sdinicial}), see \cite{lelouerrapun} for details. $\square$ \\

This Theorem is stated for piecewise constant coefficients and $C^2$
domains. However, piecewise $C^1$ coefficients and Lipschitz domains
suffice \cite{costabel,hettlich,nedelec}.

Shape derivatives can be exploited to optimize the cost functional
(\ref{costH}) since they provide vector fields $\boldsymbol \xi$  to
 deform an initial guess $\Omega_{\rm ap}$
in such a way that the cost functional  decreases. The vector field
$\boldsymbol \xi$ is selected to ensure that the shape derivative
of (\ref{costH}) is negative. The process can be repeated, generating
a sequence of reconstructions along which the cost functional
diminishes.

When $\beta=1$, the shape derivative of the cost functional  (\ref{costH})
at the scatterer  $\Omega_{\rm ap}$ is given by
\begin{eqnarray}
\begin{array}{r}
\langle DJ(\mathbb R^3 \setminus \overline{\Omega}_{\rm ap}),
\boldsymbol \xi \rangle =
{\rm Re} \Big[ \int_{\partial \Omega_{\rm ap}}
(\boldsymbol \xi \cdot \hat{\mathbf n}) (k_e^2 - k_i^2)
\Big( (\hat{\mathbf n} \times \mathbf E^-) \cdot (\hat{\mathbf n} \times \overline{\mathbf P}^-)    
\\
 + {k_i^2 \over k_e^2 } (\hat{\mathbf n} \cdot \mathbf E^-)
\cdot (\hat{\mathbf n} \cdot \overline{\mathbf P}^-)
\Big) d S \Big]
\end{array}
 \label{shapederivative}
\end{eqnarray}
where $\mathbf E$ and $\mathbf P$ are the solutions of the
forward and adjoint problems (\ref{forwardomega}) and (\ref{adjointomega})
with object $\Omega_{\rm ap}$.  Choosing
$\boldsymbol \xi =  \xi_n \hat{\mathbf n}$, $\hat{\mathbf n}$ being the unit normal vector on $\partial \Omega_{\rm ap},$ with
\begin{eqnarray} \begin{array}{l}
 \xi_n= \boldsymbol \xi \cdot \hat{\mathbf n} =  (k_i^2 - k_e^2 ) \,
{\rm Re} \Big[ (\hat{\mathbf n} \times \mathbf E^-) \cdot
(\hat{\mathbf n} \times \overline{\mathbf P}^-)
+ {k_i^2  \over k_e^2 }(\hat{\mathbf n} \cdot \mathbf E^-)
\cdot (\hat{\mathbf n} \cdot \overline{\mathbf P}^-)
\Big] \label{shapecorrector}
\end{array} \end{eqnarray}
we ensure
$\langle DJ(\mathbb R^3 \setminus \overline{\Omega}_{\rm ap}),
\boldsymbol \xi \rangle <0$.  A new approximation is defined as
\begin{eqnarray}\begin{array}{l}
\Omega_{\rm new} = \Omega_{\rm ap} + \tau \boldsymbol \xi(\Omega_{\rm ap})
= \{ \mathbf x + \tau \boldsymbol \xi(\mathbf x), \;
\mathbf x \in \Omega_{\rm ap} \}.
\end{array}\label{shapenew}
\end{eqnarray}
Then,  $J(\mathbb R^3 \setminus \overline{\Omega}_{\rm new})
<J(\mathbb R^3 \setminus \overline{\Omega}_{\rm ap})$
for $\tau >0$ small.

When enforcing star-shaped parameterizations we should fit a star-shaped parametrization
to the resulting object.
Alternatively, we can adapt an strategy to preserve the parametrization
used in \cite{caubet}. We define a set of directions associated to the
parametrization:
\begin{eqnarray*} \begin{array}{l}
\mathbf V_1=(1,0,0), \, \mathbf V_2=(0,1,0),  \mathbf V_3=(0,0,1), \\[1ex]
\mathbf V_{n,m}={\mathbf x \over |\mathbf x|}Y_n^m({\mathbf x \over |\mathbf x|}),
\quad m=-n,...,n, \, n=0,...,n_{max}.
\end{array} \end{eqnarray*}
Then, the parameters defining the star-shaped boundary
of the new approximation $\Omega_{\rm new}$ would be:
\begin{eqnarray*} \begin{array}{l}
c_\ell^{\rm new} = c_\ell^{\rm ap} - \tau_\ell
\langle DJ(\mathbb R^3 \setminus \overline{\Omega}_{\rm ap}),\mathbf V_\ell
\rangle,   \quad
\gamma_{n,m}^{\rm new} = \gamma_{n,m}^{\rm ap} - \tau_{n,m} \langle
DJ(\mathbb R^3 \setminus \overline{\Omega}_{\rm ap}),\mathbf V_{n,m}
\rangle,  \end{array}
\end{eqnarray*}
with $\tau_\ell, \tau_{n,m}>0$ small enough.  We have tested both procedures,
but the approximations stagnate without converging, as it happens for the
topological derivative based iterations.

\subsection{Topological derivative of the cost functional}
\label{sec:tdcost}

Once an explicit formula for the shape derivative of a cost functional is available,
we can obtain an explicit formula for the topological derivative taking an
additional limit \cite{feijoo}. For every $\mathbf x \in {\cal R},$ we set
$B_\varepsilon=B({\mathbf x}, \varepsilon)$ and
${\cal R}_\varepsilon= {\cal R} \setminus \overline{B_\varepsilon},$
we have  \cite{feijoo}:
\begin{eqnarray}\label{topder}
D_T({\mathbf{x}},{\cal R})=
\lim_{\varepsilon\to 0}
{J({\cal R}_\varepsilon) - J({\cal R}) \over {4\over 3} \pi \varepsilon^3}
= \lim_{\varepsilon\to 0} \frac { \langle D{J}({\cal R}_\varepsilon),
\boldsymbol \xi \rangle} {4 \pi \varepsilon^2} \nonumber \\
= \lim_{\varepsilon\to 0}\left(\frac { 1}
{4 \pi \varepsilon^2}  \,\frac{d}{d\tau} {J}(\varphi_\tau(
{\cal R}_\varepsilon))\bigg|_{\tau=0}\right),
\end{eqnarray}
for every $\mathbf x \in {\cal R}$ \footnote{Reference \cite{feijoo} uses
$-{4 \pi \varepsilon^2} $. This  changes the sign of
$D_T({\mathbf{x}},{\cal R})$ at each point.}.
The vector field $\boldsymbol \xi$ is an extension to  $\mathbb{R}^3$ of
$\boldsymbol \xi= \hat {\bf n}({\mathbf{z}}),\, {\mathbf{z}}\in
\Gamma_{\varepsilon}=\partial B_\varepsilon({\mathbf{x}}),$ where
the normal $\hat {\bf n}({\bf z})={ {\bf z}-{\bf x} \over |{\bf z}-{\bf x}| }$
points outside\footnote{Reference \cite{feijoo} and later work
often select the inner normal vector to the ball instead of the exterior
one, setting ${\bf V}= -{\bf n}({\mathbf{z}})$ at the boundary,
which is equivalent.} the ball, and vanishes out of a narrow
neighborhood of $\partial B_\varepsilon$.  \\

{\bf Theorem 3.}
{\it Let us assume that $\Omega$ is a $C^2$ domain and the coefficients $k_e,k_i,\beta$ are constant. Then the topological derivative of functional (\ref{costH}) in $\mathbb R^3$ is given by (\ref{DT}) with forward and adjoint fields governed by
(\ref{forwardempty}) and (\ref{adjointempty}).}

{\bf Proof.}
Expressions of the form (\ref{DT}) for cost functionals involving the full wave 
field were first established in \cite{masmoudi} using asymptotic expansions.
We obtain it here exploiting the relation with shape derivatives (\ref{topder}), 
as in \cite{siam2018,ip2008,feijoo,lelouerrapun}, 
To find an expression for the topological derivative in terms of adjoint and
forward fields we must first evaluate the shape derivative. We set here
${\cal R}= \mathbb R^3$. By Theorem 2 in  Appendix \ref{sec:sdcost},
$\langle DJ(\mathbb R^3 \setminus \overline{B_\varepsilon}), \boldsymbol \xi
\rangle $
is given by
\begin{eqnarray}  \label{shapederivativeepsilon}
\begin{array}{l}
 {\rm Re} \bigg[ \int_{\partial B_\varepsilon}
(k_e^2 \!- \! k_i^2 \beta)
\left[ (\hat{\mathbf n}_\varepsilon \! \times \! \mathbf E_\varepsilon^-) \!\cdot \!
(\hat{\mathbf n}_\varepsilon \! \times \! \overline{\mathbf P}_\varepsilon^-)
\!+ \! {k_i^2  \beta \over k_e^2  }
(\hat{\mathbf n}_\varepsilon \! \cdot \! \mathbf E_\varepsilon^-) \!\cdot \!
(\hat{\mathbf n}_\varepsilon \! \cdot \! \overline{\mathbf P}_\varepsilon^-)
\right]\\
-(1\!-\! \beta ) \! \left[ \beta
(\hat{\mathbf n}_\varepsilon \! \times \! \mathbf{curl} \, \mathbf E_\varepsilon^-)
\! \cdot \! (\hat{\mathbf n}_\varepsilon \! \times \! \mathbf{curl} \, \overline{\mathbf P}_\varepsilon^-) \!+\!
(\hat{\mathbf n}_\varepsilon \! \cdot \! \mathbf{curl} \, \mathbf E_\varepsilon^-) \! \cdot \!
(\hat{\mathbf n}_\varepsilon \! \cdot \! \mathbf{curl} \, \overline{\mathbf P}_\varepsilon^-)
\right] \!\! \bigg] d S,
\end{array}
\end{eqnarray}
where $\mathbf E_\varepsilon$ and $\mathbf P_\varepsilon$ are the solutions of the forward and adjoint problems (\ref{forwardomega}) and (\ref{adjointomega}) with object $B_\varepsilon$. These solutions admit explicit forms given by the series expansions in   \ref{sec:sphere}. From them, we obtain the following asymptotic behaviors when $|\boldsymbol \chi |=1$ as in \cite{siam2018,lelouerrapun}:
\begin{eqnarray} \hskip -1cm \begin{array}{r}
\hat{\mathbf n}_\varepsilon \times \mathbf E_\varepsilon^-(\mathbf x 
+ \varepsilon \boldsymbol \chi )
\rightarrow 3 \left({k_i^2 \beta \over k_e^2 } + 2 \right)^{-1}
\boldsymbol \chi  \times  \mathbf E_{\rm inc}(\mathbf x),  \\
\hat{\mathbf n}_\varepsilon \cdot
\mathbf E_\varepsilon^-(\mathbf x + \varepsilon \boldsymbol \chi )
\rightarrow 3 \left( {k_i^2 \beta  \over k_e^2 } + 2 \right)^{-1}
\boldsymbol \chi  \cdot  \mathbf E_{\rm inc}(\mathbf x), \\
\hat{\mathbf n}_\varepsilon \times \mathbf{curl} \, \mathbf
E_\varepsilon^-(\mathbf x + \varepsilon  \boldsymbol \chi )
\rightarrow 3 \left(1 + 2 \beta \right)^{-1}
\boldsymbol \chi  \times \mathbf{curl} \, \mathbf E_{\rm inc}(\mathbf x),  \\
\hat{\mathbf n}_\varepsilon \cdot \mathbf{curl} \,
\mathbf E_\varepsilon^-(\mathbf x + \varepsilon  \boldsymbol \chi )
\rightarrow  3 \left( 1 + 2 \beta \right)^{-1}
\boldsymbol \chi  \cdot \mathbf{curl} \, \mathbf E_{\rm inc}(\mathbf x),
\end{array} \label{asymptotic}
\end{eqnarray}
when $\varepsilon \rightarrow 0.$ To justify this asymptotic behavior,
we use the series expansion (\ref{series2}) with coefficients (\ref{coef1})
for $R=\varepsilon$ at $\mathbf z= \mathbf x + \varepsilon \boldsymbol
\chi ,$ evaluating $\alpha_{n,m}$,$\beta_{n,m}$ from  the series expansion
of the incident wave (\ref{coef6}). Since the spheres are centered at a
point $\mathbf x$ instead of $\mathbf 0$, $\mathbf M_{n,m}^{(1)}(k_i, \mathbf x)$
and $\mathbf N_{n,m}^{(1)}(k_i,\mathbf x)$ are replaced by
$\mathbf M_{n,m}^{(1)}(k_i,\mathbf z-\mathbf x)$
and $\mathbf N_{n,m}^{(1)}(k_i, \mathbf z- \mathbf x)$   
\cite{Chew,Mackowski, lelouerrapun}.

The behavior of the conjugate adjoint fields $ \overline{\mathbf P}^-_{\varepsilon}$ is similar to (\ref{asymptotic}), replacing $\mathbf E_{\rm inc}$ by $\overline{\mathbf  P}_{\rm inc}.$
Combining the asymptotic behaviors (\ref{asymptotic}) with the identities
\[
\int_{\partial B_1} (\boldsymbol \chi(\mathbf z) \cdot \mathbf a) (\boldsymbol \chi(\mathbf z) \cdot \mathbf b)
d S_{\mathbf z} ={4 \over 3} \pi \, \mathbf a \cdot \mathbf b, \quad
\int_{\partial B_1} (\boldsymbol \chi(\mathbf z) \times \mathbf a) (\boldsymbol \chi(\mathbf z) \times \mathbf b)
d S_{\mathbf z} ={8 \over 3} \pi \, \mathbf a \cdot \mathbf b,
\]
where $\boldsymbol \chi(\mathbf z) = {\mathbf z -\mathbf x \over |\mathbf z -\mathbf x|} $,
$ |\mathbf z -\mathbf x| =1$, and
$
\int_{\partial B_\varepsilon} a(\mathbf z) d S_{\mathbf z} =
\varepsilon^2 \int_{\partial B_1} a(\mathbf x + \varepsilon \boldsymbol \chi) d S_{\boldsymbol \chi}.
$
we obtain expression (\ref{DT}) when taking the limit (\ref{topder}) using
formula (\ref{shapederivativeepsilon}).  $\square$ \\

{\bf Theorem 4.}
{\it Let us assume that $\Omega$ is a $C^2$ domain, the coefficients $k_e,k_i$ are piecewise constant functions and the coefficient $\beta=1$. Then the topological derivative of functional (\ref{costH}) in $\mathbb R^3$ is given by (\ref{DTsimple}) with forward and adjoint fields governed by (\ref{forwardomega}) and (\ref{adjointomega}).}

{\bf Proof.} 
We can adapt the arguments in either \cite{siam2018} for scalar holography or in \cite{lelouerrapun2} for full Maxwell measurements.
Consider first points $\mathbf x \in {\cal R}= \mathbb R^3 \setminus \Omega$. Since $\boldsymbol \xi$ vanishes on $\partial \Omega$, the only difference with the proof of Theorem 2 arises when justifying the limits (\ref{asymptotic}).
Let us first observe that (\ref{asymptotic}) hold when $\mathbf E_{\rm inc}$ is replaced by a general field $\mathbf W_{\rm inc}$ which can be expanded as (\ref{coef6}) and $\mathbf E_\varepsilon$ is replaced by the solution $\mathbf W_\varepsilon$ of (\ref{forwardomega2}) with $R=\varepsilon.$ We set $\mathbf W_{\rm inc}=\mathbf E$ equal to the solution of the forward problem (\ref{forwardomega}) with object $\Omega$.

The proof is concluded by showing that to first order in $\varepsilon$ we have $\mathbf  W_{\varepsilon} \sim \mathbf E_{\varepsilon}$ on
$\partial B_\varepsilon(\mathbf x).$
To do so, let us quantify the difference between the solutions $\mathbf E_\varepsilon$ of (\ref{forwardomega}) with object $\Omega \cup B_\varepsilon(\mathbf x)$  and the total fields $\tilde {\mathbf E}_\varepsilon$:
\begin{eqnarray*}
\tilde {\mathbf E}_\varepsilon = \left\{ \begin{array}{ll}
{\mathbf W}_{\varepsilon, \rm sc} + {\mathbf W}_{\rm inc} &  {\rm in} \,
\mathbb R^3 \setminus{\overline{B_\varepsilon(\mathbf x)}}, \\
{\mathbf W}_{\varepsilon, \rm tr}  & {\rm in} \, B_\varepsilon(\mathbf x), \\
\end{array}\right.
\end{eqnarray*}
$\mathbf W_\varepsilon$ being the solution of (\ref{forwardomega2}) with $R=\varepsilon.$ The functions $\tilde {\mathbf E}_\varepsilon$ satisfy
\begin{eqnarray}
\begin{array}{lcl}
\mathbf{curl} \,  (\mathbf{curl} \,  \tilde {\mathbf E}_\varepsilon)  - k_e^2
\tilde {\mathbf E}_\varepsilon =0, & \mbox{in} & \mathbb R^3\setminus\overline{\Omega
\cup B_\varepsilon(\mathbf x)}, \\
\mathbf{curl} \,  (\mathbf{curl} \,  \tilde {\mathbf E}_\varepsilon)  - k_i^2
\tilde {\mathbf E}_\varepsilon =0, & \mbox{in} & B_\varepsilon(\mathbf x),   \\
\mathbf{curl} \,  (\mathbf{curl} \,  \tilde {\mathbf E}_\varepsilon)  - k_i^2
\tilde {\mathbf E}_\varepsilon = (k_e^2  - k_i^2) {\mathbf W}_{\varepsilon,\rm sc} , & \mbox{in} & \Omega,   \\
\hat{ \mathbf n}  \times \tilde {\mathbf E}_\varepsilon^- =
\hat{ \mathbf n} \times \tilde {\mathbf E}_\varepsilon^+,  \quad
\hat{ \mathbf n} \times \mathbf{curl} \,  \tilde {\mathbf E}_\varepsilon^- =
\hat{ \mathbf n} \times \mathbf{curl} \,  \tilde {\mathbf E}_\varepsilon^+,
& \mbox{on} & \partial B_\varepsilon(\mathbf x),  \\
\hat{ \mathbf n}  \times \tilde {\mathbf E}_\varepsilon^- =
\hat{ \mathbf n} \times \tilde {\mathbf E}_\varepsilon^+,  \quad
\hat{ \mathbf n} \times \mathbf{curl} \,  \tilde {\mathbf E}_\varepsilon^- =
\hat{ \mathbf n} \times \mathbf{curl} \,  \tilde {\mathbf E}_\varepsilon^+,
& \mbox{on} & \partial \Omega,    \\
{\rm lim}_{|\mathbf x| \rightarrow \infty} |\mathbf x|  \big|
\mathbf{curl} \,  (\tilde {\mathbf E}_\varepsilon - \mathbf E_{\rm inc}) \times \hat{\mathbf x} - \imath k_e ( \tilde {\mathbf E}_\varepsilon - \mathbf E_{\rm inc})
\big| =0. &&
\end{array} \label{forwardomega4}
\end{eqnarray}
Notice that both $\mathbf W_{\rm inc} = \mathbf E$ and  $\mathbf
W_{\varepsilon,\rm sc}$ satisfy the transmission boundary conditions
on $\partial \Omega$, the first one by construction and the second one by regularity. Now, $\mathbf W_{\varepsilon,\rm sc}(\mathbf z)$ is given by a series of the form (\ref{series1}) setting in the coefficients $R=\varepsilon$ and using $\mathbf M^{(3)}_{n,m}(k_e,\mathbf z - \mathbf x)$, $\mathbf N^{(3)}_{n,m}(k_e,\mathbf z - \mathbf x)$  \cite{Chew,Mackowski}.
Exploiting the asymptotic behavior of the spherical Bessel functions
\cite{coltonkress} as $\varepsilon \rightarrow 0$,
$j_n(k \varepsilon) \sim  (k \varepsilon)^n {2^n n! \over (2n+1)!}$,
$y_n(k \varepsilon) \sim -(k \varepsilon)^{-(n+1)} {(2n-1)! \over
2^{n-1} (n-1)!}$, as well as the relations $h^{(1)}_n= j_n+ \imath y_n$, and $z_n(k \varepsilon)' = -z_{n+1}(k \varepsilon)
+ n (k \varepsilon)^{-1} z_n(k \varepsilon)$ for $z_n=j_n,h_n^{(1)},$
we see that the coefficients $a_n(\varepsilon)$
and $b_n(\varepsilon)$ given by (\ref{coef1})
behave like $\varepsilon^{2n+1}$. Therefore,
the source term $(k_e^2-k_i^2) \mathbf W_{\varepsilon,\rm sc}(\mathbf z)=O(\varepsilon^3)$ and  $\tilde{\mathbf E}_{\varepsilon} \sim
\mathbf E_{\varepsilon}$ at zero order in $\varepsilon$.

Let us consider now $\mathbf x \in \Omega$. The proof of identity (\ref{topder})  in \cite{feijoo} uses expansions of the form (\ref{expansion}). If we revisit that proof using expansion (\ref{expansion2}) instead and setting ${\cal R}= \mathbb R^3 \setminus \overline{\Omega}$, we find  that
$D_T({\mathbf{x}},{\cal R})= - \lim_{\varepsilon\to 0} \frac { 1}
{4 \pi \varepsilon^2}  \,\frac{d}{d\tau} {J}(\varphi_\tau(
{\cal R}\cup B_\varepsilon))\big|_{\tau=0} $ for $\mathbf x \in \Omega$.  Now, the roles of $k_i$ and $k_e$ are exchanged when computing the shape derivative $\frac{d}{d\tau} {J}(\varphi_\tau( {\cal R}\cup B_\varepsilon))\big|_{\tau=0}$ following the proof of Theorem 1. Therefore the formula for the shape derivative has the opposite sign and the two minus signs cancel each other,
yielding (\ref{DTsimple}) for $\beta=1.$ \footnote{When $\beta \neq 1$, similar arguments show that the formula for the topological derivative inside $\Omega$ becomes
$ 3 \, {\rm Re}
\left[
- {k_i^2  (k_i^2  - k_e^2 \beta^{-1}) \over   (k_e^2 \beta^{-1} +2k_i^2 )}
\mathbf E(\mathbf x) \cdot \overline{\mathbf P}(\mathbf x) +
{1 -\beta^{-1} \over 1 + 2 \beta^{-1}}
\mathbf{curl} \,  \mathbf E(\mathbf x) \cdot \mathbf{curl} \,
\overline{\mathbf  P}(\mathbf x)
\right].$
}. The passage to the limit remains similar. $\square$ \\

When $k_i$ and $k_e$ are piecewise $C^1$ in space, the formula in Theorem 2 persists, replacing  $k_i$ and $k_e$ by $k_i(\mathbf x)$ and $k_e(\mathbf x)$.
When taking limits in Theorems 3 and 4, we  expand $k(\mathbf x + \varepsilon \boldsymbol \xi) =k(\mathbf x) + \varepsilon r(\mathbf x) $, $r$ being a bounded function.  Similar results hold replacing  $k_i$ and $k_e$ by $k_i(\mathbf x)$ and $k_e(\mathbf x)$ again, revisiting
the arguments in \cite{ip2008}.

%
%
%

\section{Spherical  harmonic expansions for spheres}
\label{sec:sphere}

In this Appendix, we obtain the series expansions for the forward and
adjoint fields employed to calculate the topological
derivative in Appendix \ref{sec:tdcost}.
We consider the transmission problems (\ref{unified})
when  $\Omega = B_R= B(\mathbf 0, R)$
is a sphere centered at $(0,0,0)$ with radius $R>0$.
The boundary $\partial \Omega= \partial B_R$ is the surface
$|\mathbf x |=R$, with normal vector $\hat{\mathbf n} = \hat{\mathbf x}
= {\mathbf x \over |\mathbf x |}$.  Both problems take the form:
\begin{equation}  \label{forwardomega2}
\begin{array}{llll}
\mathbf{curl} \,  (\mathbf{curl} \,   \mathbf W_{\rm sc})
- k_e^2  \mathbf W_{\rm sc} = 0
& \mbox{in $\mathbb R^3\setminus\overline{B_R}$},   \\
\mathbf{curl} \,  (\mathbf{curl} \,   \mathbf W_{\rm tr})
- k_i^2  \mathbf W_{\rm tr} =0
& \mbox{in $B_R$},   \\
\hat{\mathbf x}  \times  \mathbf W_{\rm tr}
- \hat{\mathbf x}  \times  \mathbf W_{\rm sc}
= \hat{\mathbf x} \times  \mathbf W_{\rm inc}
& \mbox{on $\partial B_R$},   \\
\beta \, \hat{\mathbf x} \times \mathbf{curl} \,   \mathbf W_{\rm tr} -
\hat{\mathbf x} \times \mathbf{curl} \,   \mathbf W_{\rm sc} =
\hat{\mathbf x} \times \mathbf{curl} \,   \mathbf W_{\rm inc}
& \mbox{on $\partial B_R$},    \\
{\rm lim}_{|\mathbf x| \rightarrow \infty} |\mathbf x|  \big|
\mathbf{curl} \,  \mathbf W_{\rm sc} \times \hat{\mathbf x}
-\imath k_e  \mathbf W_{\rm sc} \big| =0, & &
\end{array}
\end{equation}
for specific choices of $\mathbf W_{\rm inc}.$
We set $\mathbb S^2= \partial B_1.$

Any solution of (\ref{forwardomega2}) admits a series expansion in terms of spherical 
harmonic functions \cite{borhen,coltonkress}. Different choices for this basis coexist in 
the literature. Initial calculations of Mie series for the scattering of plane waves
by spheres used functions of the form $P_n^{m}(\cos(\theta)) \cos(m \psi),$
$P_n^{m}(\cos(\theta)) \sin(m \psi),$ $m=0,1,...$, $n=m,...$, where $P_n^m$
are associated Legendre polynomials \cite{borhen}.
For many computations it is useful to select a basis that is invariant by complex conjugation, 
such that $\overline{Y}_n^m={Y}_n^{-m}$. We will work with \cite{coltonkress,monk}:
\begin{eqnarray} \begin{array}{l}
Y_n^m(\hat{\mathbf x}) = \sqrt{{2n+1 \over 4\pi} {(n- |m|)! \over (n+|m|)!}}
P_n^{|m|}(\cos(\theta)) e^{\imath m \psi},
\end{array} \end{eqnarray}
where $r>0, \theta \in [0, \phi], \psi \in [0, 2 \phi]$ are the spherical
coordinates of a point $\mathbf x \in \mathbb R^3$ and  $P_n^{|m|}(\cos(\theta))$  are the associated Legendre polynomials for $m=-n,...,n$  and $n=0,1,2...$.
These spherical functions constitute an orthonormal  basis in $L^2(\mathbb S^2)$. Denoting  by $j_n$ the spherical Bessel functions of the first kind and by $h^{(1)}_n$ the spherical Hankel functions, the sets of functions
\begin{eqnarray} \begin{array}{l}
\mathbf M^{(1)}_{n,m}(\mathbf x) = \mathbf{curl} \,  (\mathbf{x} j_n(k r)
Y_n^m(\hat{\mathbf x})), \quad
\mathbf N^{(1)}_{n,m}(\mathbf x) =  {1\over \imath k} \mathbf{curl} \,
\mathbf M^{(1)}_{n,m}(\mathbf x),
\end{array} \label{interior} \end{eqnarray}
are solutions of
$ \mathbf{curl} \,  \mathbf{curl} \,  \mathbf W - k^2 \mathbf W =0$
in $\mathbb R^3$  for $m=-n,...,n$ and $n=0,1,2...$  and the functions
\begin{eqnarray} \begin{array}{l}
\mathbf M^{(3)}_{n,m}(\mathbf x) = \mathbf{curl} \,  (\mathbf{x} h^{(1)}_n(k r)
Y_n^m(\hat{\mathbf x})), \quad
\mathbf N^{(3)}_{n,m}(\mathbf x) =  {1\over \imath k} \mathbf{curl} \,
\mathbf M^{(3)}_{n,m}(\mathbf x),
\end{array} \label{exterior} \end{eqnarray}
are solutions of
$ \mathbf{curl} \,  \mathbf{curl} \,  \mathbf W - k^2 \mathbf W =0$
in $\mathbb R^3 \setminus \{\mathbf 0\}$ satisfying the Silver-M\"uller
radiation condition for $m=-n,...,n$ and $n=0,1,2...$.
We can use $\mathbf M^{(1)}_{n,m}$ and $\mathbf N^{(1)}_{n,m}$
to seek series expansions of solutions of interior problems inside
spheres and $\mathbf M^{(3)}_{n,m}$ and $\mathbf N^{(3)}_{n,m}$
for exterior problems outside.
More precisely, the unique solution of (\ref{forwardomega2}) is given by
\cite{lelouerthesis}:
\begin{eqnarray}
\mathbf W_{\rm sc}(\mathbf x)= \sum_{n=1}^\infty \sum_{m=-n}^n [
a_{n,m} \mathbf  M^{(3)}_{n,m}(k_e, \mathbf x) +
b_{n,m} \mathbf  N^{(3)}_{n,m}(k_e, \mathbf x)], \label{series1} \\
\mathbf W_{\rm tr}(\mathbf x)= \sum_{n=1}^\infty \sum_{m=-n}^n [
c_{n,m} \mathbf  M^{(1)}_{n,m}(k_i, \mathbf x)
+ d_{n,m} \mathbf  N^{(1)}_{n,m}(k_i, \mathbf x)], \label{series2}
\end{eqnarray}
with coefficients
\begin{eqnarray} \hskip -1cm \begin{array}{l}
a_{n,m}= \alpha_{n,m}  a_n, \, b_{n,m}= \beta_{n,m}  b_n, \,
c_{n,m}= \alpha_{n,m}  c_n, \, d_{n,m}= \beta_{n,m}  c_n, \\[1ex]
a_n\!=\! - {j_n(k_i R) [j_n(k_e R) + k_e R j_n'(k_e R)] -
\beta 
j_n(k_e R) [j_n(k_i R) + k_i R j_n'(k_i R)]
\over j_n(k_i R)  [h^{(1)}_n(k_e R) + k_e R (h^{(1)}_n)'(k_e R)]
- \beta 
h_n^{(1)}(k_e R) [j_n(k_i R) + k_i R j_n'(k_i R)] },  \\[1ex]
b_n\!=\! - { {k_i^2 \beta \over k_e^2 } j_n(k_i R) [j_n(k_e R) +
k_e R j_n'(k_e R)]  - j_n(k_e R) [j_n(k_i R) + k_i R j_n'(k_i R)]  \over
{k_i^2 \beta \over k_e^2 } j_n(k_i R)  [h^{(1)}_n(k_e R) +
k_e R (h^{(1)}_n)'(k_e R)] - h_n^{(1)} (k_e R) [j_n(k_i R) +
k_i R j_n'(k_i R)]}, \\[1ex]
c_n\!=\!  { k_e R \left[ (h^{(1)}_n)'(k_e R)) j_n(k_e R)
- h_n^{(1)}(k_e R)   j_n'(k_e R))  \right]
\over j_n(k_i R)  [h^{(1)}_n(k_e R) + k_e R (h^{(1)}_n)'(k_e R)]
- \beta 
h_n^{(1)}(k_e R) [j_n(k_i R) + k_i R j_n'(k_i R)] },   \\[1.5ex]
d_n\!=\!  {  k_i R  \left[ (h^{(1)}_n)'(k_e R)
j_n(k_e R) - h_n^{(1)}(k_e R)  j_n'(k_e R) \right]
\over {k_i^2 \beta \over k_e^2 } j_n(k_i R)  [h^{(1)}_n(k_e R) +
k_e R (h^{(1)}_n)'(k_e R)] - h_n^{(1)} (k_e R) [j_n(k_i R)
+ k_i R j_n'(k_i R)] },
\end{array} \label{coef1} \end{eqnarray}
where $\alpha_{n,m}$ and $\beta_{n,m}$ are the coefficients of the series
expansion of $\mathbf W_{\rm inc}:$
\begin{eqnarray} \begin{array}{l}
\mathbf W_{\rm inc}(\mathbf x)= \sum_{n=1}^\infty \sum_{m=-n}^n [
\alpha_{n,m} \mathbf M^{(1)}_{n,m}(k_e, \mathbf x)
+ \beta_{n,m} \mathbf  N^{(1)}_{n,m}(k_e, \mathbf x)].
\end{array} \label{coef6}
\end{eqnarray}

These transfer coefficients are the same for the spherical harmonics used in
\cite{borhen}, with a change of sign in $a_n$ and $b_n$ because of a
sign difference in the coefficients employed in \cite{borhen} for the expansion
of the scattered waves. They are obtained as follows.
We take the ${\bf curl}$ of the series (\ref{series1})-(\ref{series2}) and use
$\mathbf{curl \, M}^{(j)} = \imath k \mathbf N^{(j)}$,
$\mathbf{curl \, N}^{(j)} = - \imath k \mathbf M^{(j)}$, $j=1,3$ to
express them again in terms of the basis functions. We then
impose the transmission conditions (\ref{forwardomega2}) on
$|\mathbf x|=R$ computing the cross product with the normal vectors.
Notice that on the surface of a sphere $|\mathbf x| = R$,
\begin{eqnarray}  
\hskip -2mm \begin{array}{l}
 \mathbf M_{n,m}^{(1)}(k, R \boldsymbol \chi) \!=\!
j_n(k R) \mathbf{curl}_{\mathbf S^2}
Y^m_n(\boldsymbol \chi), \\
\mathbf N_{n,m}^{(1)}(k,R \boldsymbol \chi) \!=\!
{\boldsymbol \chi \over \imath k R} n(n\!+\!1)
j_n(k R)  Y^m_n(\boldsymbol \chi)
\!+\! {1\over \imath k R} [j_n(k R) \!+\! k R j_n'(k R)]
 \nabla_{\mathbb S^2}   Y_n^m(\boldsymbol \chi),
\end{array} \nonumber
\end{eqnarray}
where $\mathbf{curl}_{\mathbb S^2}$ and $\nabla_{\mathbb S^2}$ represent
surface curl and gradients on the sphere $|\boldsymbol \chi| = 1$. Similar
identities hold for $\mathbf M^{(3)}$ and $\mathbf N^{(3)}$  replacing $j_n$
by $h_n^{(1)}.$ We observe that,  in the transmission conditions,  the 
terms involving  ${\boldsymbol \chi}Y_n^m(\boldsymbol \chi)$
vanish.   The functions $\mathbf{curl}_{\mathbb S^2} Y_n^m$ and
$\nabla_{\mathbb S^2} Y_n^m$ form an orthogonal basis in 
$L^2_{\mathbf t}(\mathbb S^2)$ (tangent fields), satisfying 
$\hat {\mathbf n} \times \nabla_{\mathbb S^2}  Y_n^m
= - \mathbf{curl}_{\mathbb S^2} Y_n^m$ and
$ \hat {\mathbf n} \times \mathbf{curl}_{\mathbb S^2} Y_n^m
= \nabla_{\mathbb S^2}   Y_n^m.$} 
Setting the coefficients of the resulting expansions equal to zero,
we obtain the expressions  (\ref{coef1}), where $a_n,b_n,c_n,d_n$
are the solutions of the systems of equations:
\begin{eqnarray*} \hskip -2mm \begin{array}{l}
  c_n  j_n(k_i R) - a_n  h_n^{(1)}(k_e R) = j_n(k_e R),  \\[1ex]
  c_n  \beta [j_n(k_i R) \!+\! k_i R j_n'(k_i R)] \!-\! a_n
[h^{(1)}_n(k_e R) \!+\! k_e R (h^{(1)}_n)'(k_e R)]  
=  j_n(k_e R) \!+\! k_e R j_n'(k_e R),   \\ [1ex]
  d_n   k_i^{-1} [j_n(k_i R) \!+\! k_i R j_n'(k_i R)]
\!-\!  b_n k_e^{-1} [h^{(1)}_n(k_e R) \!+\! k_e R (h^{(1)}_n)'(k_e R)]   
=  k_e^{-1} [j_n(k_e R) \!+\! k_e R j_n'(k_e R)],  \\[1ex]
  d_n  k_i \beta j_n(k_i R)  - b_n k_e  h_n^{(1)}(k_e R)
= k_e  j_n(k_e R).
\end{array} \end{eqnarray*}

\noindent Thus, to obtain explicit expressions for the forward and adjoint fields in the
presence of spheres we only need to determine the coefficients
$\alpha_{n,m}$ and $\beta_{n,m}$ of the series expansion (\ref{coef6})
of $\mathbf E_{\rm inc}$ and of $\overline{\mathbf P}_{\rm inc}$ given by (\ref{adjointempty}).

\subsection{Series expansion of forward fields for spheres}
\label{sec:forwardsphere}

For a plane wave advancing in the direction of the axis $z$ and
polarized along the axis $x$, that is,
$\mathbf E_{\rm inc}(\mathbf x) =  (e^{\imath k_e z},0,0),$
the Mie series expansion for $\mathbf E_{\rm inc}$ is obtained from the
expansion in \cite{borhen} setting
$\sin(m \theta) = {e^{\imath m \theta}- e^{-\imath m \theta} \over 2 \imath}$
and
$\cos(m \theta) = {e^{\imath m \theta}+ e^{-\imath m \theta} \over 2}$:
\begin{eqnarray}
\begin{array}{rrr} \hskip -2mm
\mathbf E_{\rm inc}(\mathbf x) \!=\!  - {\displaystyle \sum_{n=1}^\infty}
\! \imath^n \! \sqrt{4 \pi {2n+1 \over n(n+1)}}
\Big[  {\mathbf M_{n,1}^{(1)}(k_e,\mathbf x)  - \mathbf M_{n,-1}^{(1)}(k_e,\mathbf x) \over 2 \imath}
\!+\!{\mathbf N_{n,1}^{(1)}(k_e,\mathbf x)  + \mathbf N_{n,-1}^{(1)}(k_e,\mathbf x) \over 2}
\Big].
\end{array}
\label{expansionincident}
\end{eqnarray}
With these coefficients we obtain the corresponding Mie solution (\ref{series1})-(\ref{coef6}).
The above formulas apply to spheres centered at $(0,0,0)$. 
If the sphere is centered at a point $\mathbf y$, the formulas must be conveniently shifted.

\subsection{Series expansion of adjoint fields for spheres}
\label{sec:adjointsphere}

The field $\overline{\mathbf P}_{\rm inc}$ given by (\ref{adjointempty}) involves terms of the form
\begin{eqnarray} \begin{array}{l}
{1\over k_e^2}
\mathbf{curl} \,  \mathbf{curl} \,  \left( \mathbf d_j {e^{\imath k_e |\mathbf x - \mathbf x_j|}
\over 4 \pi |\mathbf x - \mathbf x_j| }
\right), \end{array} \label{termj}
\end{eqnarray}
where $\mathbf d_j =
2(I_{\rm meas}(\mathbf x_j)  - |\mathbf E(\mathbf x_j) |^2) \overline{\mathbf E(\mathbf x_j) },$  where $\mathbf E$ is the solution to (\ref{forwardempty}). 
When the incident wave is a plane wave, the coefficients (\ref{coef6}) are given in   \ref{sec:forwardsphere}.
The Mie expansion of the functions (\ref{termj})
 is given in \cite{coltonkress}, page 222
\footnote{In \cite{coltonkress}, the notation is
$\mathbf M_{n,m} = \mathbf M_{n,m}^{(1)}$,
$\mathbf N_{n,m} = \mathbf M_{n,m}^{(3)}$.
}. Summing over $j$:
\begin{eqnarray*}
\begin{array}{rrr}
\overline{\mathbf P}_{\rm inc}(\mathbf x)
=  
\sum_{n=1}^\infty { \imath k_e \over n(n+1)}
\sum_{m=-n}^n \Big[
\mathbf M_{n,m}^{(1)}(\mathbf x)
\sum_{j=1}^N
\big( \mathbf M_{n,-m}^{(3)}(\mathbf x_j) \cdot \mathbf d_j \big) \\
-  \mathbf N_{n,m}^{(1)}(\mathbf x) \sum_{j=1}^N
\big( \mathbf N_{n,-m}^{(3)}(\mathbf x_j) \cdot \mathbf d_j \big)
\Big] .
\end{array}
\label{expansionadjoint2}
\end{eqnarray*}
With these coefficients we obtain the corresponding Mie solution
(\ref{series1})-(\ref{coef6}). Such formulas apply to spheres centered at $(0,0,0).$
When the sphere is centered at a point $\mathbf y$, we
replace everywhere the functions $\mathbf  M_{n,m}^{(1)}(\mathbf x)$,
$\mathbf  N_{n,m}^{(1)}(\mathbf x)$, $M_{n,m}^{(3)}(\mathbf x_j)$,
$\mathbf  N_{n,m}^{(3)}(\mathbf x_j)$ by
$\mathbf  M_{n,m}^{(1)}(\mathbf x- \mathbf y)$,
$\mathbf  N_{n,m}^{(1)}(\mathbf x- \mathbf y)$,
$\mathbf  M_{n,m}^{(3)}(\mathbf x_j- \mathbf y)$,
$\mathbf  N_{n,m}^{(3)}(\mathbf x_j- \mathbf y)$. The resulting formula
holds when $|\mathbf x_j- \mathbf y|> | \mathbf x - \mathbf y|,$
for $j=1,...,N$.  An alternative derivation follows \cite{Chew,Mackowski}.

\bibliographystyle{model1-num-names}

\end{document}